\documentclass[11pt]{article}

\usepackage[utf8]{inputenc}
\usepackage[T1]{fontenc}
\usepackage{amsmath}
\usepackage{amssymb}
\usepackage{amsthm}
\usepackage{mathtools}
\usepackage[margin=1in]{geometry}
\usepackage{booktabs}
\usepackage{array}
\usepackage[hidelinks]{hyperref}
\usepackage{tikz}

\title{Nonnegative Rank Bounds for the Algebraic Monotone Calculus}
\author{Fedor Part\thanks{JetBrains Research, \texttt{fedor.part@gmail.com}}}
\date{}

\begin{document}

\newcommand{\AMC}{\mathsf{AMC}}
\newcommand{\SHAMC}{\Sigma\Pi\Sigma\text{-}\mathsf{AMC}}
\newcommand{\AMCb}{\Sigma\Pi\Sigma\text{-}\mathsf{AMC}_{b}}
\newcommand{\TS}{\mathsf{TS}}
\newcommand{\rkp}{\mathrm{rk}_+}
\newcommand{\rkt}{\mathrm{rk}^{\otimes}_+}
\newcommand{\rktu}{\mathrm{rk}^{\otimes}_{+,b}}
\newcommand{\flt}[3]{M_{#1 \mid #2}(#3)}
\newcommand{\ex}{\mathrm{ex}}
\newcommand{\PM}{\mathsf{PM}}
\newcommand{\OF}{\mathsf{OF}}
\newcommand{\amcle}{\preccurlyeq}
\newcommand{\viol}{\mathrm{viol}}
\newcommand{\Lift}{\mathrm{Lift}}
\newcommand{\Flat}{\mathrm{Flat}}
\newcommand{\SA}{\mathsf{SA}}
\newcommand{\supp}{\mathsf{supp}}

\newtheorem{theorem}{Theorem}[section]
\newtheorem{prop}[theorem]{Proposition}
\newtheorem{lem}[theorem]{Lemma}
\newtheorem{corollary}[theorem]{Corollary}
\theoremstyle{definition}
\newtheorem{definition}[theorem]{Definition}
\theoremstyle{remark}
\newtheorem{remark}[theorem]{Remark}

\newcommand{\ol}{\overline}
\newcommand{\lta}{\langle}
\newcommand{\rta}{\rangle}
\newcommand{\mc}{\mathcal}

\maketitle

\begin{abstract}
In Hrube\v{s}' Algebraic Monotone Calculus $\AMC$ \cite{Hrubes20} a DNF tautology $A$ is proved
by a monotone arithmetic circuit computing $\chi(A) - \epsilon\cdot\chi(1)$ for some
$\epsilon>0$, where $\chi(A)$ is the set-multilinear polynomial whose coefficient at
$\mathbf x^{\sigma}$ counts the terms of $A$ satisfied by $\sigma$, and
$\chi(1) = \prod_{i\leq n}(x_i^{0}+x_i^{1})$. Proof size is circuit size, and $\epsilon$ is
existentially quantified, so a lower bound must hold at arbitrarily small $\epsilon$. We prove
such bounds for bounded-depth fragments by two nonnegative rank measures of
$\chi(A)-\epsilon\cdot\chi(1)$: its monotone tensor rank, and the nonnegative rank of a
flattening of its coefficient tensor.

At depth three, a proof decomposes the coefficient tensor into nonnegative rank-one tensors.
Restricting the summands to nonnegative multiples of subcube indicator tensors yields conical
juntas and a system polynomially equivalent to Sherali--Adams when size counts conjunctions
over the twin variables $x_i,\bar x_i$.
For arbitrary nonnegative rank-one summands, we adapt Rothvo\ss{}' hyperplane separation
method to monotone tensor rank and prove that Tseitin formulas on constant-degree expanders of
sufficiently large girth
require proofs of size $2^{\Omega(\epsilon n)}$ for $\epsilon\in(0,1]$ with
$\epsilon=\omega(\log n/n)$.
This cannot reach every $\epsilon>0$: at shift $2^{-\Theta(n\log n)}$ Tseitin acquires
polynomial-size depth-three proofs, which separates the general depth-three fragment from the subcube one
exponentially. We compose Tseitin formulas with a constant-size index gadget to obtain lower
bounds for every $\epsilon\geq2^{-\Omega(n)}$, converting a Sherali--Adams \emph{degree}
bound into a depth-three \emph{size} bound. Depth three is essential to all of this: we also
show that Tseitin has monotone proofs of size $O(n^{3})$ and depth $O(\log n)$ at \emph{every}
$\epsilon\in(0,1]$.

Beyond depth three, we bound the nonnegative rank of a flattening by $L^{2^{t}}$ for a monotone
formula with $L$ leaves and \emph{mixed-product depth} $t$. Here $t$ is the maximum number,
along a root-to-leaf path, of product gates whose two inputs each involve both sides of the
bipartition; the Tseitin proofs just mentioned have $t=0$ for every bipartition, so this bound
is empty for Tseitin. We give two constructions turning hard matrices into hard DNFs.
Approximate extension complexity yields bounds for perfect matching and odd factor instances.
Linear programming integrality gaps for planted constraint satisfaction yield an explicit
family of planted Tseitin DNFs hard at inverse-polynomial shift.
\end{abstract}

\section{Introduction}

In \cite{Hrubes20} Hrube\v{s} discovered a family of reductions showing that sufficiently
strong \emph{monotone} lower bounds imply unconditional lower bounds for general, non-monotone
computation: on arithmetic circuit size, on Boolean circuit size, and --- for a proof
system whose proofs are monotone circuits and which polynomially simulates Resolution --- on
proof size. All of them ask for
\emph{$\epsilon$-sensitive} lower bounds: one bounds the monotone complexity not of a single
hard object but of a perturbation of an easy one, the perturbation governed by a parameter
$\epsilon>0$ tending to $0$. In the two instances this paper is concerned with, the arithmetic
and the proof-complexity one, the perturbation is a difference
\[
    F \;-\; \epsilon\cdot G
\]
with $F$ alone monotonically easy. They differ in where the
hardness resides and in the quantification over $\epsilon$: in the arithmetic instance $F$ is fixed and easy while $G$ is
arbitrary and carries the hardness, and it suffices to have monotone lower bounds along
\emph{some} sequence of $\epsilon$ tending to $0$; in the proof complexity instance both $F$ and
$G$ are monotone-easy, and the lower bound is required for \emph{every} $\epsilon > 0$, since
a proof may use any single $\epsilon$. We briefly describe the two instances.

\paragraph{General arithmetic circuits.}
The arithmetic instance of the reduction is
\cite[Theorem~2.1]{Hrubes20}: if a polynomial $f \in \mathbb{R}[x_1, \dots, x_n]$ of degree $d$
has an arithmetic circuit of size $s$, then for every sufficiently small $\epsilon > 0$ the
polynomial
\[
    g_\epsilon \;:=\; (x_1 + \cdots + x_n + 1)^d \;+\; \epsilon\cdot f
\]
has a \emph{monotone} arithmetic circuit of size $O(sd^2 + n\log n)$. This is $F - \epsilon G$
with $F$ the universal polynomial $U := (x_1 + \cdots + x_n + 1)^d$ and $G = -f$ an arbitrary,
possibly non-monotone polynomial; since $U$
contains every monomial of degree at most $d$ with coefficient at least $1$, once $\epsilon$
is small enough for these coefficients to absorb those of $\epsilon f$, every coefficient of
$g_\epsilon$ is indeed nonnegative, so $g_\epsilon$ is a legitimate monotone target. Up to the
factor $d^2$ the reduction is an exact equivalence: the identity $f = (g_\epsilon - g_0)/\epsilon$
turns a monotone circuit for $g_\epsilon$ at any single $\epsilon > 0$ into a general circuit
for $f$ of essentially the same size, so the monotone size of $g_\epsilon$ and the general
circuit size of $f$ coincide for all sufficiently small $\epsilon$, up to the factor $d^2$. In
particular, monotone lower bounds on $g_{\epsilon_i}$ along any sequence $\epsilon_i\to0$
yield an unconditional lower bound for $f$, one of the $\epsilon_i$ eventually falling below the
relevant threshold; a bound at a single fixed $\epsilon$ does not suffice, since the easiness of
$g_\epsilon$ for easy $f$ is guaranteed only below an $\epsilon_0$ depending on the coefficients
of $f$, and $\epsilon_0$ is not known in advance. The identity also shows that the required lower bounds are genuine:
if $f$ is hard, then $g_\epsilon$ is monotone-hard at \emph{every} $\epsilon > 0$. An equivalent
formulation, via
the standard observation that every circuit of size $s$ can be simulated with a single subtraction
as $f = f^{+} - f^{-}$ with $f^{\pm}$ monotone of monotone circuit size $O(s)$: the negative
part $f^{-}$ can always be taken to be a scalar multiple of the fixed universal polynomial $U$.

The measure $\epsilon \mapsto \mathrm{size}_{+}(g_\epsilon)$ is thus highly
\emph{discontinuous} at $\epsilon = 0$: $g_0 = U$ has monotone circuits of size
$O(n + \log d)$, yet whenever $f$ is hard the monotone size of $g_\epsilon$ jumps, for every
$\epsilon > 0$, to essentially the \emph{general} circuit size of $f$ --- an arbitrarily small
perturbation of the easy polynomial $U$ blows up its monotone complexity, and the hardness is
created entirely by the $\epsilon$-perturbation. This discontinuity is both the power and the
price of the approach. The power: it converts a lower bound in the \emph{monotone} model ---
a restricted setting in which strong unconditional lower bounds are known --- into a lower
bound for \emph{general} arithmetic circuits, where no such techniques exist. The price: most
classical monotone lower bound methods are inapplicable here, as they depend only on the set of
monomials that occur --- the exponential monotone lower bound for the permanent of Jerrum and
Snir \cite{JerrumSnir82}, for instance, is of this kind --- but
$\supp(g_\epsilon) = \supp(U)$ is the full set of monomials of degree at most
$d$, the support of an easy polynomial. An $\epsilon$-sensitive lower bound must therefore be
genuinely coefficient-sensitive and, moreover, discontinuous in the coefficients.

Previous $\epsilon$-sensitive lower bounds use communication complexity for monotone arithmetic
circuits \cite{ChattopadhyayDattaMukhopadhyay21,ChattopadhyayDattaGhosalMukhopadhyay22} and
extend to separations among monotone circuits, branching programs, formulas and bounded-depth
formulas \cite{ChattopadhyayGhosalMukhopadhyay22}. Here we study the tautology encodings arising
from $\AMC$.

\paragraph{The proof complexity setting: Algebraic Monotone Calculus.}
Hrube\v{s} \cite{Hrubes20} applied the same idea to \emph{proof complexity}, defining the
Algebraic Monotone Calculus ($\AMC$). The lines of $\AMC$ are formal inequalities
$f \amcle g$ between monotone polynomials, meaning that every monomial occurring in $f$ occurs
also in $g$, that is, the support containment
$\supp(f) \subseteq \supp(g)$; accordingly the axioms are $f \amcle f$ and
$a \amcle b$ for reals $a \geq 0$, $b > 0$, and the rules are transitivity together with
compatibility with the two operations, from $f_1 \amcle g_1$ and $f_2 \amcle g_2$ infer
$f_1 \circ f_2 \amcle g_1 \circ g_2$ for $\circ \in \{+, \times\}$. The size of a proof
$f_1 \amcle g_1, \dots, f_m \amcle g_m$ is the size of a smallest monotone arithmetic circuit
computing all the polynomials $f_1, g_1, \dots, f_m, g_m$ simultaneously. Nothing so far
mentions $\epsilon$: the lines assert only support containment. The parameter enters through the
semantics, since $\supp(f) \subseteq \supp(g)$ holds iff $g - \epsilon\cdot f$ is monotone for
some $\epsilon > 0$ iff $f \amcle g$ is provable. Quantitatively, the $\AMC$ proof size of
$f \amcle g$ equals, up to a constant factor, the minimum size of a monotone circuit computing
$f$, $g$ and $g - \epsilon\cdot f$ simultaneously, over all $\epsilon > 0$
\cite[Proposition~2.6, Theorem~2.7]{Hrubes20}.

For a DNF $A$, the coefficient of the set-multilinear monomial
$x^\sigma = \prod_i x_i^{\sigma(i)}$ in its characteristic polynomial $\chi(A)$ counts the
terms of $A$ satisfied by $\sigma$. Here $(x_i^0,x_i^1)$ is the variable pair for the Boolean
variable $x_i$, while $\chi(1) = \prod_i (x_i^0 + x_i^1)$ collects all such monomials, one per
assignment; thus $A$ is a tautology precisely when
$\supp(\chi(1)) \subseteq \supp(\chi(A))$, and a proof of $A$ is an $\AMC$ proof
of $\chi(1) \amcle \chi(A)$. A proof of size $s$ gives a monotone circuit of size $O(s)$ for
$\chi(A) - \epsilon\cdot\chi(1)$ for some $\epsilon>0$. Conversely, a monotone circuit of size $s$ for this
difference gives a proof of size $O(s+mn)$, since $\chi(A)$ and $\chi(1)$ have \emph{monotone}
circuits of size $O(mn)$ when $A$ has $m$ terms: both are monotone-easy, and all the difficulty
is in the perturbation. Thus a lower bound on the $\AMC$ complexity of $A$ reduces to a
monotone circuit lower bound for
$\chi(A) - \epsilon\cdot\chi(1)$ for \emph{all} $\epsilon > 0$. Since $\chi(A)$
and $\chi(1)$, and hence their difference, have small \emph{general} arithmetic circuits, such a
bound exhibits polynomials $\chi(A) - \epsilon\cdot\chi(1)$ in $\mathsf{VP}$ with no
polynomial-size monotone circuit. Separations of monotone $\mathsf{VP}$ from $\mathsf{VP}$ as
such are classical \cite{Valiant80}; what is demanded here is that the bound hold for every
$\epsilon > 0$, for targets whose support, for every sufficiently small positive $\epsilon$,
is that of the easy polynomial $\chi(1)$. On the proof side such a bound has unconditional
consequences, because $\AMC$ is a strong system: it polynomially simulates Resolution --- a Resolution refutation with $k$ lines yields an $\AMC$
proof of size $O((m + k)n)$ \cite[Proposition~5.8]{Hrubes20} --- so an $\AMC$ lower bound, and
hence a monotone lower bound on $\chi(A) - \epsilon\cdot\chi(1)$ for every $\epsilon > 0$,
implies a Resolution size lower bound. As we show in Section~\ref{sec:SA}, considerably more is
true already for the weakest fragment considered here: it simulates Sherali--Adams, and
therefore Resolution, when size counts displayed conjunctions. A distinctive
feature of the proof complexity setting is that here the ``positive part'' $F = \chi(A)$
varies with the instance while $G = \chi(1)$ is \emph{fixed}, whereas in the
circuit-complexity applications of the reduction it is the other way around: $F$ is a fixed,
simple polynomial (such as a power $(1+\sum_i x_i)^d$) and the varying, hardness-carrying part
is $G$. The interest of $\AMC$ as a proof system is that it may be much stronger than
Resolution, and here the relation to the monotone calculus $\mathsf{MLK}$ of Atserias, Galesi
and Pudl\'ak \cite{AtseriasGalesiPudlak02} is the relevant benchmark. $\mathsf{MLK}$ proves
implications $A \to B$ between monotone Boolean formulas, and an $\mathsf{LK}$ proof of size
$s$ of a monotone sequent can be turned into one containing only monotone formulas, of size
$s^{O(\log s)}$ and with polynomially many lines \cite{AtseriasGalesiPudlak02}; on monotone
sequents $\mathsf{MLK}$ therefore quasipolynomially simulates Frege. Replacing $\wedge, \vee$ in a monotone circuit $A$ by
$\times, +$ yields a monotone polynomial $F_A$, and $\supp(F_A) \subseteq
\supp(F_B)$ implies that $A \to B$ is a tautology; the converse fails --- $x \wedge y
\to y$ is a tautology while $\supp(F_{x \wedge y}) = \{xy\}$ and
$\supp(F_y) = \{y\}$ --- so $\AMC$ is, over such translations, a genuine weakening of
$\mathsf{MLK}$: it is blind to the Boolean identities $x^2 = x$ and $x \leq 1$. Hrube\v{s}
\cite{Hrubes20} indicates --- as a remark, without proof --- how Frege could be reached in
principle by restoring exactly these: $\AMC$ augmented with the Boolean axioms
$f \amcle 1$ and $f \amcle f^2$ simulates $\mathsf{MLK}$. Whether the Boolean axioms actually
help, and whether $\AMC$ itself polynomially simulates Frege, are open --- both under this
translation and under the encoding used in the rest of the paper, where a proof of a DNF
tautology $A$ is a proof of $\chi(1) \amcle \chi(A)$.

\paragraph{Nonnegative rank and extension complexity.}
The \emph{nonnegative rank} of a matrix
$M \in \mathbb{R}_{\geq 0}^{m \times n}$ with
nonnegative entries is the smallest $r$ such that $M$ can be written as a sum of
$r$ nonnegative rank-one matrices, equivalently $M = UV$ for nonnegative matrices
$U \in \mathbb{R}_{\geq 0}^{m \times r}$ and $V \in \mathbb{R}_{\geq 0}^{r \times n}$.
The relevance of this measure to optimization was established in the seminal work of
Yannakakis \cite{Yannakakis91}, who showed that the smallest number of facets of any linear
program projecting onto a polytope $P$ --- the \emph{extension complexity} of $P$ --- equals the
nonnegative rank of its slack matrix. Lower bounds on extension complexity therefore rule
out small linear programming relaxations, and proving them amounts to bounding the
nonnegative rank of an explicit matrix from below. After a long hiatus this program produced
the first superpolynomial bounds, for the correlation, cut and travelling salesman polytopes
\cite{FioriniMassarPokuttaTiwaryWolf15}, and culminated in the breakthrough of Rothvo\ss{}
\cite{Rothvoss17}, who proved that the perfect matching polytope has exponential extension
complexity; Braun, Fiorini, Pokutta and Steurer \cite{BraunFioriniPokuttaSteurer12} and Braun
and Pokutta \cite{BraunPokutta15} extended such bounds to \emph{approximate} extended
formulations.

The same measure is the common currency of several lower bound programs. In \emph{circuit}
complexity, Hrube\v{s} \cite{Hrubes20} attaches to every Boolean function $f$ an explicit
nonnegative matrix $M(f)$, with rows indexed by the $1$-inputs of $f$ and columns by its
$0$-inputs, and shows that $R(f) := \min_{\epsilon>0}\rkp(M(f)-\epsilon J)$, with $J$ the
all-ones matrix, satisfies $R(f) = O(C(f)+n)$ for the Boolean circuit complexity $C(f)$
\cite[Thm.~2.3]{Hrubes20}: a superpolynomial lower bound on $R(f)$ is a superpolynomial circuit
lower bound. The same quantity measures the cost of separating $f^{-1}(1)$ from $f^{-1}(0)$ by a
linear program, to within an additive term linear in $n$: writing $\mathrm{sep}(f)$ for the least
extension complexity of a separating polyhedron, the \emph{linear separation complexity},
$R(f)-2n-1 \leq \mathrm{sep}(f)\leq R(f)$ \cite[Thm.~2.5]{Hrubes20}.
For monotone $f$, the corresponding monotone separation complexity $\mathrm{sep}_{+}(f)$ uses
monotone separating polyhedra and the matrix $M_{+}(f)_{y,x}=\sum_i x_i(1-y_i)$, with
$x\in f^{-1}(1)$ and $y\in f^{-1}(0)$ \cite[Thm.~3.16]{Hrubes20}. Up to polynomial factors,
this monotone quantity is the weak monotone linear programming gate complexity of
de Oliveira Oliveira and Pudl\'ak \cite{PudlakOliveira17}. An explicit monotone function with
superpolynomial $\mathrm{sep}_{+}$ is sought in \cite[Open Problem 3]{Hrubes20} and
\cite[Open Problem 1]{PudlakOliveira17}; nonconstructive lower bounds for separation complexity
are known \cite{Hrubes20,HrubesTalebanfard21}. In \emph{formula} complexity, splitting
the variable blocks of a monotone set-multilinear polynomial into two groups and reading its
coefficients as a matrix sends every product term of a monotone $\Sigma\Pi\Sigma$ formula to a
nonnegative rank-one matrix, so the nonnegative rank of that matrix bounds the number of terms
from below. Moreover $\lceil\log_2\rkp(M)\rceil$ is exactly the randomized communication
complexity of computing the entries of the nonnegative matrix $M$ \emph{in expectation}
\cite{FaenzaFioriniGrappeTiwary15}.
Information-theoretic estimates enter more directly: a nonnegative factorization of $M$ with
$r$ terms, read as a mixture of $r$ product distributions on the normalised matrix, bounds
$\log r$ from below by Wyner's common information \cite{Wyner75}, the route taken by Braun
and Pokutta \cite{BraunPokutta15}. In \emph{proof} complexity the measure is carried by Sherali--Adams: by
Fleming, G\"o\"os, Grosser and Robere \cite{FlemingGoosGrosserRobere22} a refutation of an
unsatisfiable CNF $C$ normalises to a single conical junta
$\viol_C - \epsilon = \sum_j \eta_jQ_j$, with $\eta_j > 0$ and the $Q_j$ conjunctions, where
$\viol_C(\sigma)$ is the number of clauses of $C$ falsified by $\sigma$ --- so that degree
becomes width, size becomes a number of subcubes, which is again a nonnegative rank of
$\viol_C-\epsilon$ with its rank-one terms constrained to that shape, and $\epsilon > 0$ is
quantified existentially; and \cite{PudlakOliveira17} use extension complexity in the same
spirit to prove monotone feasible interpolation for Lov\'asz--Schrijver and its mixed variant.

\subsection{Our contributions}\label{sec:contributions}

In this paper we study the strength of bounded-depth formula fragments of $\AMC$ using
nonnegative rank measures. We obtain the following results.

\subsubsection{Depth three: \texorpdfstring{$\SHAMC$}{Sigma Pi Sigma-AMC}, Sherali--Adams and size bounds}

At depth three the minimum circuit and formula sizes differ by at most a factor $O(n)$, so we
take $\SHAMC$ proofs to be formulas and measure their size by the number of product terms. Let
$f$ be monotone, homogeneous and set-multilinear over blocks $X_1,\dots,X_k$, and put
$n:=\sum_i|X_i|$. A monotone $\Sigma\Pi\Sigma$ formula computing $f$ is then set-multilinear
itself --- monotonicity is what forces this, a formula with cancellation being free to compute a
set-multilinear polynomial through gates that are not set-multilinear --- so it \emph{is} a
nonnegative tensor
decomposition $f=\sum_{i\leq r}\ell_{1,i}(X_1)\otimes\cdots\otimes\ell_{k,i}(X_k)$ with each
$\ell_{j,i}$ a nonnegative linear form on $X_j$; the least such $r$ is the \emph{monotone
tensor rank} $\rkt(f)$. Flattening $f$ along a bipartition of the blocks into $Y$ and $Z$ sends
each rank-one tensor to a nonnegative rank-one matrix, so
$\rkt(f)\geq\rkp(\flt{Y}{Z}{f})$: a nonnegative-rank bound for a single flattening bounds the
whole formula.

In the encoding of a DNF $A$ the blocks are the pairs $\{x_i^0,x_i^1\}$, one per variable of
$A$, and a rank-one term is
$\eta\prod_i(p_ix_i^1+(1-p_i)x_i^0)$, a nonnegative multiple of a product distribution with
\emph{biases} $p_i\in[0,1]$; a $\SHAMC$ proof of size $r$ is a representation
$\viol_C-\epsilon=\sum_{i\leq r}\eta_i\mu_i$, where $\viol_C(\sigma)$ counts the clauses of
the $m$-clause CNF $C=\neg A$ falsified by $\sigma$. In the \emph{bounded-bias} fragment
$\AMCb$ the biases lie in $\{0,\tfrac12,1\}$, each term is a subcube indicator, and a proof is
a \emph{conical junta} representation of $\viol_C-\epsilon$ --- the Sherali--Adams normal form
of \cite{FlemingGoosGrosserRobere22}.

\begin{theorem}[Informal; Theorem~\ref{thm:sa-to-amcb} and
Corollary~\ref{cor:amcb-sa-equivalent}]
$\AMCb$ polynomially simulates Sherali--Adams: a degree-$d$ refutation of $C$, displayed as an
explicit list of conjunction multipliers of total length $s$, normalises to a single conical
junta $\viol_C-\epsilon = K$ with at most $ds+m$ conjunctions, which is an $\AMCb$ proof of $A$
with that many terms. Conversely Sherali--Adams
over the twin variables $x_i,\bar x_i$, where a conjunction counts as a single monomial,
polynomially simulates $\AMCb$: an $r$-term $\AMCb$ proof is a Sherali--Adams refutation with
$r+m$ conjunction occurrences. Since $d\leq n$, the two systems are polynomially equivalent.
\end{theorem}

The first direction partitions the complement of an $r$-literal conjunction into $r$
\emph{first-mismatch} cells rather than $2^{r}-1$ complete ones, which is what removes the
factor $2^{d}$ of \cite{FlemingGoosGrosserRobere22}. The second is polynomial only in the twin
variables: after substituting $\bar x_i = 1-x_i$ a single conjunction expands into as many as
$2^{d}$ ordinary monomials, and in that measure the equivalence fails.

Two further facts place the fragment. Hrube\v{s}' simulation of Resolution
\cite[Prop.~5.8]{Hrubes20} can be carried out inside it: a Resolution refutation with $k$ lines
gives an $\AMCb$ proof with at most $2nk+m$ terms (Corollary~\ref{cor:resolution}). And an
$r$-term $\AMCb$ proof is a sums-of-squares refutation with $r+m$ monomials in the twin-variable
measure of Atserias and Hakoniemi, so their bounds \cite{AtseriasHakoniemi19} give
$2^{\Omega(n)}$ for $\AMCb$ on Tseitin and on random $3$-SAT, at every $\epsilon>0$ at once
(Corollary~\ref{cor:all-eps-amcb}). In the other direction $\mathsf{PHP}^{n+1}_{n}$ has an
$\AMCb$ proof with $O(n^{4})$ terms (Theorem~\ref{thm:tensor-PHP-UB}).

None of this reaches arbitrary bias, and that we prove directly, by hyperplane separation. Write
$\TS_G$ for the Tseitin contradiction of a graph $G$ of odd total charge: the DNF, in the edge
variables of $G$, whose terms are the local assignments violating one of its parity constraints
(Section~\ref{sec:tseitin-setup}). One exhibits a signed test tensor, a \emph{witness}, scoring
high against the target $\chi(\TS_G)-\epsilon\cdot\chi(1)$ but low against every rank-one term
lying entrywise below it, and the ratio of the two scores bounds the number of terms from below. This is Rothvo\ss{}'
method \cite{Rothvoss17}, which we adapt to monotone tensor rank and to the regime $\epsilon<1$.
Our witness is calibrated so that no subcube term scores much against it, and removing the bias
restriction is the substantial step. A general product term is not a subcube, but each of its
coordinates is a mixture of two simpler alternatives, pin it to its more likely value or leave
it unbiased:
\[
   p\,x^{1}+(1-p)\,x^{0}
   \;=\;
   (2p-1)\cdot x^{1}
   \;+\;
   (2-2p)\cdot\tfrac{1}{2}\bigl(x^{0}+x^{1}\bigr)
   \qquad\bigl(p\geq\tfrac12\bigr),
\]
a convex combination in which the pinned branch carries the weight $2p-1$. Tensoring these
exhibits the whole term as a mixture of subcubes, which the witness scores by averaging. The
subcubes that could spoil the estimate are held down by the expansion of $G$, and the average
survives because the term lies entrywise below the target;
Section~\ref{sec:tseitin-refinement} carries this out.

\begin{theorem}[Informal; Theorem~\ref{thm:tseitin-lb-arbitrary}]
Let $G$ be connected, bridgeless and $d$-regular, $d\geq3$, on $n$ vertices, with edge expansion
$h$, girth exceeding $4/h$ and odd total charge. For $\epsilon\in(0,1]$ with
$\epsilon n\geq50e^4$,
\[
   \rkt\bigl(\chi(\TS_G)-\epsilon\cdot\chi(1)\bigr)
   \;\geq\;\frac{\epsilon}{2(n-\epsilon)}
      2^{\epsilon\min\{h,1\}n/(50e^4)}.
\]
The same lower bound holds for every depth-three certificate at that shift.
\end{theorem}

On constant-degree expanders this is $2^{\Omega(n)}$ at constant $\epsilon$ and superpolynomial
throughout $\epsilon=\omega(\log n/n)$ (Corollary~\ref{cor:tseitin-depth-three-regimes}).
The proof gives an $\epsilon$-sensitive tensor-rank bound directly for the Tseitin target.
It cannot be pushed to all $\epsilon>0$, and the obstruction is an upper bound:
telescoping the inconsistency of $\TS_G$ along an ordering of the vertices, and decomposing a
weakly biased parity into few nonnegative products by finite differences, gives a polynomial-size
proof once $\epsilon$ is exponentially small.

\begin{theorem}[Informal; Theorem~\ref{thm:tseitin-tiny-shift} and
Corollary~\ref{cor:tiny-shift-separation}]
For fixed degree and every $\epsilon\leq2^{-\Theta(n\log n)}$,
$\rkt\bigl(\chi(\TS_G)-\epsilon\cdot\chi(1)\bigr) = O(n^{3})$. With
Corollary~\ref{cor:all-eps-amcb} this separates $\AMCb$ from $\SHAMC$ exponentially at depth
three.
\end{theorem}

Depth three is what the whole of this point is about, and the restriction is not cosmetic: at
every $\epsilon\in(0,1]$ the same target has a monotone formula of size $O(n^{3})$ and depth
$O(\log n)$, so the exponential bounds above are bounds on tensor rank and not on monotone proof
size as such (Theorem~\ref{thm:mu0-upper}).

So no $\epsilon$-uniform superpolynomial bound for $\SHAMC$ can be proved on Tseitin by any
method, and the quantifier over $\epsilon$ must be settled on a different formula. A
constant-size index gadget supplies one, and in doing so converts a Sherali--Adams \emph{degree}
bound into a depth-three \emph{size} bound: giving each base variable a private table addressed
by selector bits plants a copy of the base instance on every address tuple at once, so the copy
to be tested is chosen only after the purported decomposition.

\begin{theorem}[Informal; Theorem~\ref{thm:selector-lift} and
Corollary~\ref{cor:exp-small-shift}]
Let $I_0$ be an unsatisfiable CSP on $m$ variables with $L$ constraints and Sherali--Adams degree
exceeding $d$, and let $A$ be the DNF encoding the composition $I_0\circ\mathrm{IND}_q^{\,m}$,
with one copy of the index gadget $\mathrm{IND}_q(x,y) = y_x$ per base variable. Then
$\rkt(\chi(A)-\epsilon\cdot\chi(1)) \geq \epsilon\,q^{d+1}/\bigl(L\sum_{j\leq d}\binom mj\bigr)$.
Taking $\TS_G$ as the base, whose Sherali--Adams degree is linear \cite{Grigoriev01}, gives an
explicit family of tautological DNFs with $O(N)$ terms of constant width and size
$2^{\Omega(N)}$ for every $\epsilon\geq2^{-\Omega(N)}$ at once.
\end{theorem}

What the gadget buys is the range: for $\TS_G$ itself the direct argument stops at
$\epsilon = \Omega(\log n/n)$, while the lifted family is hard for every
$\epsilon\geq2^{-\Omega(N)}$. The bound is still not uniform in $\epsilon$, and being a
tensor-rank bound it says nothing about $\rkp$.

\subsubsection{Higher mixed-product depth: the flattening bound, and what it yields}

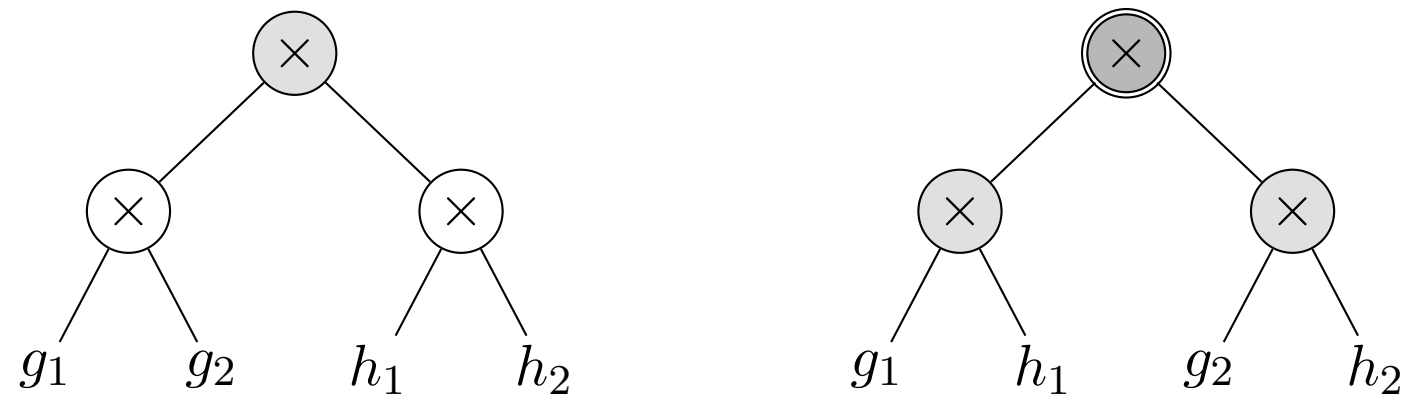
\begin{figure}[ht]
\centering
\begin{tikzpicture}[x=11mm,y=11mm,
  one/.style={circle,draw,inner sep=1.2pt,minimum size=5.5mm},
  mix/.style={circle,draw,fill=black!12,inner sep=1.2pt,minimum size=5.5mm},
  dmix/.style={circle,draw,double,fill=black!28,inner sep=1.2pt,minimum size=5.5mm},
  lf/.style={inner sep=1.5pt}]
\node[mix]  (lr) at (1.5,1.9) {$\times$};
\node[one]  (la) at (0.5,0.95) {$\times$};
\node[one]  (lb) at (2.5,0.95) {$\times$};
\node[lf] (l1) at (0,0) {$g_1$};
\node[lf] (l2) at (1,0) {$g_2$};
\node[lf] (l3) at (2,0) {$h_1$};
\node[lf] (l4) at (3,0) {$h_2$};
\draw (lr)--(la) (lr)--(lb) (la)--(l1) (la)--(l2) (lb)--(l3) (lb)--(l4);
\node at (1.5,-0.75) {one side at a time: $\mu=0$};
\node[dmix] (rr) at (6.5,1.9) {$\times$};
\node[mix]  (ra) at (5.5,0.95) {$\times$};
\node[mix]  (rb) at (7.5,0.95) {$\times$};
\node[lf] (r1) at (5,0) {$g_1$};
\node[lf] (r2) at (6,0) {$h_1$};
\node[lf] (r3) at (7,0) {$g_2$};
\node[lf] (r4) at (8,0) {$h_2$};
\draw (rr)--(ra) (rr)--(rb) (ra)--(r1) (ra)--(r2) (rb)--(r3) (rb)--(r4);
\node at (6.5,-0.75) {interleaved: $\mu=1$};
\end{tikzpicture}
\caption{Two bracketings of one depth-three term $g_1g_2h_1h_2$, the $g_i$ depending on blocks
from $\mathcal X$ and the $h_i$ on blocks from $\mathcal Y$. Shaded gates are mixed, the doubly
drawn gate is doubly mixed. On the left no gate has two mixed children; on the right the root
has two, and with $2k$ factors the interleaved bracketing reaches $\mu=\lceil\log_2
k\rceil$. The mixed-product depth of an unbounded fan-in expression is accordingly read as the
minimum over bracketings.}
\label{fig:bracketing}
\end{figure}

At depth three the inequality $\rkt(f)\geq\rkp(\flt{Y}{Z}{f})$ is immediate: a proof is a
nonnegative tensor decomposition, and each of its terms flattens to a rank-one matrix. A deeper
formula admits no such reading, and its flattening rank can exceed the number of its leaves; the
bound must therefore say by how much. Depth itself is not the governing parameter. Multiplying by a
factor living on one side of the bipartition only permutes and duplicates rows or columns, so
what must be counted is the number of \emph{doubly mixed} product gates on a root-to-leaf path
--- those both of whose children involve blocks from both sides. We call this the
\emph{mixed-product depth} $t$. Gates are binary here, so a depth-three term, whose product has
unbounded fan-in, is bracketed first; multiplying each side separately and only then combining
the two one-sided products makes no gate doubly mixed (Figure~\ref{fig:bracketing}).

\begin{theorem}[Informal; Theorem~\ref{thm:mixed-depth-flattening}]
A monotone formula with $L$ leaves and mixed-product depth $t$ computing a set-multilinear $f$
has $\rkp(\Flat(f))\leq L^{2^{t}}$, so a submatrix of $\Flat(f)$ of nonnegative rank $R$ forces
$L\geq R^{1/2^{t}}$. A depth-three proof has mixed-product depth zero.
\end{theorem}

Mixed-product depth zero is strictly more permissive than depth three. For every
$\epsilon\in(0,1]$ and every bipartition, $\chi(\TS_G)-\epsilon\cdot\chi(1)$ has a monotone
formula of mixed-product depth zero with $O(n^{3})$ leaves (Theorem~\ref{thm:mu0-upper}), while
on a constant-degree expander its tensor rank, the least size of a depth-three proof, is
$2^{\Omega(n)}$ at constant $\epsilon$. Every bipartite flattening of the target therefore has nonnegative rank
$O(n^{3})$, so the flattening bound is
empty for $\TS_G$ at every $t$ (Corollary~\ref{cor:tseitin-flat-easy}) and a DNF to feed it has
to be found elsewhere (Section~\ref{sec:polytopes}).

The restriction to formulas cannot be dropped: an explicit $O(n)$-size set-multilinear monotone
\emph{circuit} with no doubly mixed product gate computes a polynomial of flattening rank $2^{n}$
(Proposition~\ref{prop:circuit-sharing-obstruction}).

Feeding it needs a DNF one of whose flattenings has, after the shift, a submatrix of large
nonnegative rank. Both of our sources of such DNFs use one construction,
Proposition~\ref{prop:local-witness-dnf}. Its input is a matrix $M\geq J$ whose rows and
columns are indexed by Boolean encodings $x$ and $y$, and whose entry $M(x,y)$ counts how many
of some fixed conjunctions $W_1,\dots,W_q$ (the \emph{witnesses}) the pair $(x,y)$ satisfies;
its output is an explicit tautological DNF
whose flattening contains $M$ as the submatrix on the rows and columns that carry valid
encodings. We call this submatrix the \emph{valid slice}. After the shift it becomes
$M-\epsilon\cdot J$, so if $M-\epsilon\cdot J$ has large nonnegative rank, the DNF is hard, with
the leaf bound of Corollary~\ref{cor:local-witness-mixed-depth}. The hypothesis $M\geq J$ is also what makes the
rank hard to certify: it forces $M-\epsilon\cdot J\geq(1-\epsilon)J>0$, so for every
$\epsilon<1$ the slice has no zero entry, its rectangle covering number is one and its fooling
sets are singletons. The entire support-based toolkit is thus unavailable below $\epsilon=1$,
and the rank must come from a method tolerating a strictly positive matrix: hyperplane
separation, or the information-theoretic bounds we quote. Even hyperplane separation survives
only for signed test matrices --- against a nonnegative one the floor $(1-\epsilon)J$ is itself
an admissible rank-one term and the bound collapses, as Section~\ref{sec:flattenings} computes.
The test tensor of Section~\ref{sec:tseitin-witness} and the functional behind
Section~\ref{sec:csp-dnf} both carry negative entries for this reason.

We have two recipes for a count matrix $M\geq J$ as above whose perturbation
$M-\epsilon\cdot J$ has large nonnegative rank. Recipe I
(Proposition~\ref{prop:recipe-polytope}) takes a $0/1$ polytope
$P=\mathsf{conv}\{\chi_H : H\in\mc H\}$ --- say, the perfect matching polytope of $K_{2n}$ ---
together with a family $\mc F$ of covering inequalities $\lta a_f,\ol x\rta\geq b$ with
$a_f\in\{0,1\}^{E}$ valid on it, and sets $M(f,H) := |\supp(a_f)\cap H|$, which is the slack
matrix of $\mc F$ at $\mc H$ shifted by $b$. The nonnegative rank of $M-\epsilon\cdot J$ is
then at least the least extension complexity of a polyhedron between $P$ and the relaxation
that lowers every right-hand side from $b$ to $\epsilon$, and exceeds it by at most one
(Section~\ref{sec:polytopes}); only the lower bound is used. That least extension complexity is an
\emph{approximate extension complexity}. Recipe II (Proposition~\ref{prop:recipe-csp}) takes a Max-$k$-CSP instance with a degree-$d$
Sherali--Adams integrality gap and plants many copies of it in a larger variable set, the rank
coming from that gap; its rows are the planted copies rather than valid inequalities, no
polytope occurs, and the admissible shift is limited only by the quality of the planting
estimate. Beyond the
encoding lemma the two have nothing in common.

\begin{theorem}[Informal; Theorem~\ref{thm:matching-mixed-depth} and
Corollary~\ref{cor:pm-of-lb}]
Recipe I applied to the odd cut inequalities of $K_{2n}$ gives DNFs $A_{\PM,n}$ and $A_{\OF,n}$
--- every odd cut meets every perfect matching, respectively every odd factor --- for which every
monotone formula of mixed-product depth $t$ computing the target has at least
$\rkp(M_{\PM}-\epsilon\cdot J)^{1/2^{t}}$ leaves. By \cite{BraunPokutta15} this is
$2^{\Omega(n/2^{t})}$ for constant $\epsilon$, and still $2^{n^{\Omega(1)}/2^{t}}$ for
$\epsilon \geq c_{0}\log\log n/\log n$ with $c_{0}>8$ fixed.
\end{theorem}

Here $\rkp(M_{\PM}) = O(n^{2})$ while the perturbation alone creates the exponential bound --- the
$\epsilon$-sensitivity Hrube\v{s}' reductions call for, in the form supplied by Rothvo\ss{}
\cite{Rothvoss17} and Braun and Pokutta \cite{BraunPokutta15}.
Appendix~\ref{sec:threshold} makes the constants explicit and proves the sufficient range
$\epsilon\geq c_0\log\log n/\log n$ for every fixed $c_0>8$.
Recipe II reaches further down.

\begin{theorem}[Informal; Theorem~\ref{thm:explicit-csp-dnf-small-shift}]
For every fixed $0<\gamma<1/4$, there is an explicit family of tautological DNFs $A_K$ on $K$
variables with $K^{O(1)}$ terms such that, for all sufficiently large members and every
$K^{-\gamma}\leq\epsilon<1$, the flattening of $\chi(A_K)-\epsilon\cdot\chi(1)$ has nonnegative rank
$K^{\Omega(\log K)}$.
\end{theorem}

It combines a quantitative form of the planted-CSP argument of Chan, Lee, Raghavendra and Steurer
\cite{ChanLeeRaghavendraSteurer16}, in which we track the dependence of the admissible shift on
all parameters, with the linear-degree Sherali--Adams integrality gap for Tseitin on explicit
constant-degree expanders \cite{Grigoriev01}, also used in Section~\ref{sec:selector-lift}.

This is the smallest shift at which we can prove superpolynomial hardness for any explicit
polynomial-size DNF at mixed-product depth above zero, and its mechanism --- a degree-$d$
Sherali--Adams integrality gap at the base instance, planted so as to defeat not that one
relaxation but every linear program of the given size at once --- needs neither the extension
complexity of a combinatorial polytope nor hyperplane separation. It is a statement about a new
DNF family, and does not reach the Tseitin range of Section~\ref{sec:tseitin-lb}, which goes
down to $\log K/K$ but only at depth three.

Tseitin is not the only possible base: the explicit sparse $3$-XOR instances of Hopkins and
Lin \cite{HopkinsLin22}, which have $\mathrm{opt}(I_0)\leq1-\mu_1$ and
$\mathrm{SoS}_{\mu_2m}(I_0)=1$ for constants $\mu_1,\mu_2>0$ with $L=O(m)$ constraints, satisfy
the hypotheses of Proposition~\ref{prop:recipe-csp} equally well and give the same bound, with a
constant gap in place of the $1/n$ available here.

\paragraph{AI disclosure.} The initial version of this paper was written by the author, with minor assistance from
Claude Opus 4.6. It was much shorter and contained the pigeonhole principle upper bound, the depth-three Tseitin lower
bounds, and the lower bounds for the perfect matching and odd factor DNFs. When an early draft
was ready, at the end of summer 2026, the author put a few questions about it to ChatGPT Sol 5.6
for the first time. Several items were retrieved from the literature that the author had not been
aware of: the Sherali--Adams normal form of Fleming, G\"o\"os, Grosser and Robere
\cite{FlemingGoosGrosserRobere22}, the planted-instance argument of Chan, Lee, Raghavendra and
Steurer \cite{ChanLeeRaghavendraSteurer16}, and the moment-matching mixtures of product
distributions of Chen and Moitra \cite{ChenMoitra18}, on which our decomposition of a weakly
biased parity into few nonnegative products is modelled. It then became clear that
several further important results follow and should be included. The mathematics of those new results is
essentially due to ChatGPT Sol 5.6. The writing of the final paper was done in close
collaboration with Claude Opus 4.6--5. The author has revised everything that was generated by AI
and is fully responsible for the content.

\paragraph{Acknowledgements.} The author is grateful to Pavel Hrube\v{s} for introducing him to
$\epsilon$-sensitive complexity and for discussions.

\tableofcontents

\section{Preliminaries}\label{sec:prelim}

\subsection{\texorpdfstring{$\AMC$}{AMC} proof system}\label{sec:AMC-def}

The \emph{algebraic monotone calculus} ($\AMC$) was introduced by
Hrube\v{s} \cite{Hrubes20}. It is a Hilbert-style proof system whose
lines are inequalities $f \amcle g$, where $f, g$ are \emph{monotone}
polynomials over $\mathbb{R}$ (i.e.\ all coefficients nonnegative) in some
variable set $X$. The intended semantics is
\[
   f \amcle g \;\;\Longleftrightarrow\;\; \supp(f)\subseteq
   \supp(g),
\]
where $\supp(p)$ is the set of monomials that occur in $p$ with
nonzero coefficient (with the conventions $\supp(0) = \emptyset$ and
$\supp(a) = \{1\}$ for $a \in \mathbb{R}\setminus\{0\}$).

\begin{definition}[$\AMC$, {\cite[Sec.~2.3]{Hrubes20}}]
\label{def:AMC}
The axioms of $\AMC$ are
\begin{itemize}
\item[(A1)] $f \amcle f$ for any monotone polynomial $f$;
\item[(A2)] $a \amcle b$ for any $a, b \in \mathbb{R}$ with
$a \geq 0$, $b > 0$.
\end{itemize}
The inference rules are
\begin{itemize}
\item[(R1)] (cut/transitivity) from $f_1 \amcle g$ and
$g \amcle f_2$ derive $f_1 \amcle f_2$;
\item[(R2)] (compatibility with $\circ \in \{+, \times\}$) from
$f_1 \amcle g_1$ and $f_2 \amcle g_2$ derive
$f_1 \circ f_2 \amcle g_1 \circ g_2$.
\end{itemize}
An $\AMC$ \emph{proof} of $f \amcle g$ is a sequence of lines
$f_1\amcle g_1, \dots, f_m \amcle g_m$ with $f_m = f$, $g_m = g$
in which every line is an axiom or follows from earlier lines by (R1) or
(R2). The \emph{size} of the proof is the smallest $s \geq m$ such that the
$2m$ polynomials $f_1, g_1, \dots, f_m, g_m$ can be \emph{simultaneously}
computed by a single monotone arithmetic circuit of size $s$.
\end{definition}

\begin{prop}[Soundness and completeness; {\cite[Prop.~2.6]{Hrubes20}}]
\label{prop:AMC-sound-complete}
For monotone polynomials $f, g$ the following are equivalent:
\begin{enumerate}
\item $\supp(f) \subseteq \supp(g)$;
\item there exists $\epsilon > 0$ such that $g - \epsilon\cdot f$ is monotone;
\item there exists an $\AMC$ proof of $f \amcle g$.
\end{enumerate}
\end{prop}

\begin{theorem}[Size characterization; {\cite[Thm.~2.7]{Hrubes20}}]
\label{thm:AMC-size}
Let $f, g$ be monotone polynomials.
\begin{enumerate}
\item[(a)] If $f \amcle g$ has an $\AMC$ proof of size $s$,
then there exists $\epsilon > 0$ such that $f$, $g$, and
$g - \epsilon\cdot f$ are \emph{simultaneously} computable by a single
monotone arithmetic circuit of size $O(s)$.
\item[(b)] Conversely, if $\epsilon > 0$ and $f$, $g$, $g - \epsilon\cdot f$
are simultaneously computable by a single monotone arithmetic circuit of
size $s$, then $f \amcle g$ has an $\AMC$ proof of size $O(s)$.
\end{enumerate}
\end{theorem}

We use the certificate presentation supplied by Theorem~\ref{thm:AMC-size}: a certificate for
$f\amcle g$ is a monotone circuit for $g-\epsilon\cdot f$ for some $\epsilon>0$. The
conversion back to a derivation also includes the circuit costs of $f$ and $g$. For the DNF
encodings below these costs are $O(mn)$, so the two presentations are polynomially equivalent.
Depth and formula restrictions in this paper are imposed directly on the certificate. In
particular, a depth-$3$ certificate is a $\Sigma\Pi\Sigma$ expression
$g-\epsilon\cdot f=\sum_{i=1}^{r}\prod_j L_{i,j}$, with nonnegative-coefficient affine forms
$L_{i,j}$. We use the term ``proof'' for these certificates as well.

\begin{definition}\label{def:SHAMC}
$\SHAMC$ is the certificate proof system whose proofs are depth-$3$ certificates, taken as
$\Sigma\Pi\Sigma$ formulas and measured by the number $r$ of product terms. As noted in
Section~\ref{sec:contributions}, at depth three the minimum circuit and formula sizes differ by
at most a factor $O(n)$, so nothing below turns on this choice.
\end{definition}

\subsection{\texorpdfstring{$\AMC$}{AMC} as a propositional proof system}\label{sec:encoding}

Following \cite{Hrubes20}, the statement ``the DNF $A$ is a tautology'' is
encoded as an inequality $\chi(1) \amcle \chi(A)$ as follows. For a DNF
$A = A_1 \vee \cdots \vee A_m$ in $n$ Boolean variables one works in $2n$
variables $x_1^0, x_1^1, \dots, x_n^0, x_n^1$, one per literal, grouped into
blocks $X_i = \{x_i^0, x_i^1\}$; to an assignment $\sigma \in \{0,1\}^n$
corresponds the set-multilinear monomial
$\mathbf{x}^{\sigma} = \prod_{i=1}^{n} x_i^{\sigma_i}$, and
\[
   \chi(A) := \sum_{\sigma \in \{0,1\}^n} c_A(\sigma)\, \mathbf{x}^{\sigma},
   \qquad
   c_A(\sigma) := \bigl|\{\, j : \sigma \models A_j \,\}\bigr| .
\]
Equivalently $\chi$ is additive over the terms of $A$, and the characteristic
polynomial of a single term
$A_j = \bigwedge_{i \in S_j} x_i \wedge \bigwedge_{i \in T_j} \neg x_i$
(with $S_j, T_j \subseteq [n]$ disjoint) is
\[
   \chi(A_j) \;=\; \prod_{i \in S_j} x_i^1 \;\prod_{i \in T_j} x_i^0
   \;\prod_{i \notin S_j \sqcup T_j}\!\bigl(x_i^0 + x_i^1\bigr) .
\]
Each $\chi(A_j)$ is thus a single shallow monotone product, which is why
$\chi(A)$ and $\chi(1) = \prod_{i}(x_i^0+x_i^1)$ admit small, shallow monotone
circuits. Since $\chi(1)$ collects \emph{all} set-multilinear monomials with
coefficient $1$, the DNF $A$ is a tautology precisely when
$\supp(\chi(1)) \subseteq \supp(\chi(A))$.

We write $\AMC^{\epsilon}$, $\SHAMC^{\epsilon}$, $\AMCb^{\epsilon}$ for the
system in which the parameter is required to be \emph{at least} $\epsilon$, and drop the
superscript when it is existentially quantified as in
Proposition~\ref{prop:AMC-sound-complete}. Demanding a lower bound rather than equality costs
nothing: since $\chi(1)$ is itself a single rank-one term, adding
$(\epsilon'-\epsilon)\chi(1)$ turns a proof with parameter $\epsilon'$ into one with parameter
$\epsilon<\epsilon'$ at the cost of one further term. 

Keeping track of both the parameter and the fragment is essential. With the exception of
Corollary~\ref{cor:all-eps-amcb}, our lower bounds apply over specified ranges of $\epsilon$
to depth-three certificates or formulas of bounded mixed-product depth. A lower bound for an
existentially quantified fragment must hold uniformly for every $\epsilon>0$; a lower bound
for unrestricted $\AMC$ must additionally allow arbitrary monotone circuits.

\subsection{\texorpdfstring{$\SHAMC$, $\AMCb$}{Sigma Pi Sigma-AMC, Sigma Pi Sigma-AMC-b} and conical juntas}\label{sec:juntas}

It is convenient to record the form the two fragments take as identities of functions on the
Boolean cube.
For a DNF $A = A_1 \vee \dots \vee A_m$ in variables $x_1, \dots, x_n$ the
coefficient $c_A(\sigma)$ defined above is also the
number of clauses of the CNF $C := \neg A$ falsified by $\sigma$, the quantity written
$\viol_{C}(\sigma)$ in the semialgebraic proof complexity literature. Since
$[\mathbf{x}^\sigma]\chi(1) = 1$ for every $\sigma$, we have
$[\mathbf{x}^{\sigma}]\bigl(\chi(A)-\epsilon\cdot\chi(1)\bigr) = c_A(\sigma)-\epsilon$. Sending a
set-multilinear tensor $f$ to its coefficient function $\sigma\mapsto[\mathbf x^{\sigma}]f$ is a
linear bijection onto the real functions on $\{0,1\}^{n}$, the set-multilinear monomials being
indexed by $\sigma$; under it the target $\chi(A)-\epsilon\cdot\chi(1)$ becomes
$c_A-\epsilon = \viol_{C}-\epsilon$. An $\AMC$ proof is therefore an identity of functions on
the cube with $\viol_{C}-\epsilon$ on the left; the two fragments differ in what may stand on the
right, and that is what the rest of this subsection records.

\begin{definition}[$\AMCb$; the bounded-bias fragment]\label{def:AMCb}
An \emph{$\AMCb$ proof} of a DNF tautology $A$ with parameter $\epsilon > 0$ is a family
$\{(\eta_j, S_j, T_j)\}_{j=1}^{s}$ with $\eta_j \in \mathbb{R}_{>0}$ and
$S_j, T_j \subseteq [n]$ disjoint such that
\[
   \chi(A) - \epsilon\cdot\chi(1)
   \;=\; \sum_{j=1}^{s} \eta_j \prod_{i \in S_j} x_i^1 \prod_{i \in T_j} x_i^0
         \prod_{i \notin S_j \sqcup T_j}\bigl(x_i^0 + x_i^1\bigr).
\]
Its \emph{size} is $s$ and its \emph{width} is $\max_j |S_j \sqcup T_j|$, the largest
number of coordinates a term fixes. Polynomial degree is not a useful measure here: every term
has degree exactly $n$.
\end{definition}

Reading off coefficients, the $j$-th summand contributes $\eta_j$ to exactly those $\sigma$
lying in the subcube
$C_j := \{\sigma : \sigma_i = 1 \ (i \in S_j),\ \sigma_i = 0 \ (i \in T_j)\}$, so
Definition~\ref{def:AMCb} says precisely that
\begin{equation}\label{eq:junta}
   \viol_{C}(\sigma) - \epsilon \;=\; \sum_{j=1}^{s}\eta_j\,\mathbf{1}_{C_j}(\sigma)
   \qquad\text{for all }\sigma \in \{0,1\}^n,
\end{equation}
that is, that $\viol_{C} - \epsilon$ is a \emph{conical junta} of size $s$ and width
$\max_j|S_j\sqcup T_j|$.

The fragment $\SHAMC$ relaxes~\eqref{eq:junta} in one step. A rank-$1$ monotone
set-multilinear tensor is
\[
   \eta\cdot\prod_{i=1}^{n}\bigl(p_i\,x_i^1 + (1-p_i)\,x_i^0\bigr),
   \qquad \eta \geq 0,\ p_i \in [0,1],
\]
whose coefficient at $\mathbf{x}^\sigma$ is
$\eta\cdot\prod_i p_i^{\sigma_i}(1-p_i)^{1-\sigma_i}$, i.e.\ $\eta$ times the value at
$\sigma$ of a \emph{product distribution} on $\{0,1\}^n$. Hence a $\SHAMC$ proof of size
$r$ is exactly a representation
\begin{equation}\label{eq:prodmix}
   \viol_{C}(\sigma) - \epsilon \;=\; \sum_{i=1}^{r}\eta_i\,\mu_i(\sigma),
   \qquad \eta_i \geq 0,\ \mu_i \text{ a product distribution},
\end{equation}
and $\AMCb$ is the special case $p_i \in \{0, \tfrac12, 1\}$. We call the numbers $p_i$ the
\emph{biases} of the term and accordingly refer to $\SHAMC$ as the arbitrary-bias and to
$\AMCb$ as the bounded-bias fragment. A single term of~\eqref{eq:prodmix} can fail to be a
conical junta of subexponential size --- not, however, merely because its biases lie in the open
interval, since $p_i = \tfrac12$ for every $i$ is one subcube term; what is needed is that the
biases be generic. Taking odds ratios $p_i/(1-p_i) = e^{2^{i-1}}$ makes the coefficient at
$\mathbf x^{\sigma}$ proportional to $e^{m(\sigma)}$ with $m(\sigma) = \sum_i\sigma_i2^{i-1}$, so
the $2^{n}$ coefficients of the term are proportional to $1,e,e^{2},\dots,e^{2^{n}-1}$. Let
$\chi_{C_1},\dots,\chi_{C_r}$ be subcube indicators with $r<2^{n}$ and consider the homogeneous
system $\lta\chi_{C_j},\varphi\rta = 0$, $j\leq r$, in the unknown $\varphi$: it has $0/1$
coefficients and fewer equations than unknowns, so it has a nonzero \emph{rational} solution
$\varphi$, which by linearity annihilates the whole real span of the $\chi_{C_j}$. Pairing
$\varphi$ with the coefficient vector of the term would then give a nontrivial rational linear
relation among $1,e,\dots,e^{2^{n}-1}$, which transcendence of $e$ forbids. So the term lies
outside the real span of any $r<2^{n}$ subcube indicators.
Hence~\eqref{eq:prodmix} is a genuine relaxation of~\eqref{eq:junta} at the level of
representation. Whether it is one at the level of \emph{proof} size is a different question,
since a proof target must have the special form $\viol_C - \epsilon$ with $\viol_C$
integer-valued; Corollary~\ref{cor:tiny-shift-separation} answers it at exponentially small
shift, where the two fragments separate exponentially on Tseitin. We write $\rkt(\,\cdot\,)$ for the least $r$ in~\eqref{eq:prodmix} and
$\rktu(\,\cdot\,)$ for the least $s$ in~\eqref{eq:junta}, so that
$\rkt \leq \rktu$ always.

The two subsections on Tseitin work in $\pm1$ coordinates, and we record once how the terms of
\eqref{eq:prodmix} look there.

\begin{lem}[Rank-one terms in $\pm1$ coordinates]\label{lem:pm-rank-one}
For $\sigma\in\{0,1\}^{n}$ put $z_i := (-1)^{\sigma_i}$, and read a set-multilinear tensor as
the function $\sigma\mapsto$ its coefficient at $\mathbf x^{\sigma}$. Call $\prod_{i}g_i(z_i)$
with every $g_i:\{-1,1\}\to\mathbb{R}_{\geq0}$ a \emph{nonnegative product}. Since $z_i = -1$
exactly when $\sigma_i = 1$, each factor $p_i^{\sigma_i}(1-p_i)^{1-\sigma_i}$ equals
$\bigl(1+\beta_iz_i\bigr)/2$ for $\beta_i := 1-2p_i$, so the rank-one
term of \eqref{eq:prodmix} with weight $\eta\geq0$ and biases $p_i\in[0,1]$ reads, in these
coordinates,
\begin{equation}\label{eq:pm-rank-one}
   \eta\prod_{i=1}^{n}p_i^{\sigma_i}(1-p_i)^{1-\sigma_i}
   \;=\;
   \frac{\eta}{2^{n}}\prod_{i=1}^{n}\bigl(1+\beta_iz_i\bigr),
   \qquad \beta_i \;:=\; 1-2p_i \;\in\; [-1,1] ,
\end{equation}
and conversely $\eta'\prod_i(1+\beta_iz_i)$ with $\eta'\geq0$ and $\beta_i\in[-1,1]$ is the
rank-one term with $\eta = 2^{n}\eta'$ and $p_i = (1-\beta_i)/2$. The rank-one terms are
therefore exactly the nonnegative products: a nonnegative function on $\{-1,1\}$ is
$\bar g\,(1+\beta z)$ with $\bar g := \tfrac12(g(1)+g(-1))$, so one with no $g_i$ identically zero
equals the right hand side of \eqref{eq:pm-rank-one} for
\[
   \eta' \;=\; \prod_{i}\tfrac12\bigl(g_i(1)+g_i(-1)\bigr),
   \qquad
   \beta_i \;=\; \frac{g_i(1)-g_i(-1)}{g_i(1)+g_i(-1)} \;\in\; [-1,1] ,
\]
and one with some $g_i$ identically zero, that is with $\bar g_i = 0$, is the zero tensor. In particular $\rkt(f)$ is the
least number of nonnegative products summing to $f$, and $\AMCb$ is the case
$\beta_i\in\{-1,0,1\}$.
\end{lem}

The bounds below are proved by induction on a formula, gate by gate, and what they need is
set-multilinearity as a condition on the formula rather than on the polynomial it computes. For
monotone formulas the two agree after zero gates are pruned, as we record below. Gates have fan-in two throughout; an
expression whose sums and products have larger fan-in, such as the depth-three expressions of
Definition~\ref{def:SHAMC}, is read as a formula by bracketing them, which changes neither the
polynomial computed nor the number of leaves. It can change the mixed-product depth of
Definition~\ref{def:mixed-depth}, and the convention there says how we read it.

\begin{definition}[Syntactic set-multilinearity]\label{def:syntactic-sml}
The \emph{block support} $S_v$ of a gate $v$ is the set of blocks occurring below it, so
$S_v$ is a single block at a variable leaf and empty at a constant leaf. A formula
$\Phi$ is \emph{syntactically set-multilinear} if, at
every internal gate $v$ with children $u,w$,
\[
   S_u = S_w = S_v \quad\text{at a sum gate},
   \qquad
   S_u\cap S_w=\emptyset \ \text{ and } \ S_v = S_u\sqcup S_w \quad\text{at a product gate}.
\]
An induction up the formula then gives that $F_v$ is set-multilinear with block support $S_v$ at
every gate, which is the form in which the condition is used.
\end{definition}

For a general formula this is a genuine restriction: cancellation lets a formula compute a
set-multilinear polynomial through gates that are not set-multilinear. In the monotone case it
costs nothing.

\begin{lem}[Automatic syntactic set-multilinearity]\label{lem:autosml-depth}
Let $\Phi$ be a monotone formula, none of whose gates computes the zero polynomial, whose output
$F$ is set-multilinear with block support $S$. Then $\Phi$ is syntactically set-multilinear, and
$S_v$ is the common block support of the monomials of $F_v$ at every gate $v$.
\end{lem}

\begin{proof}
Since $\Phi$ is a tree, each gate $v$ reaches the output through a single multiplicative
context: writing $M_v$ for the product of the polynomials computed at the siblings of the gates
on the path from $v$ to the root that are children of a product gate, and $R$ for the remaining
contributions, $F = M_v\cdot F_v + R$. All coefficients are nonnegative, so nothing cancels and
every monomial of $M_v F_v$ is a monomial of $F$.

Fix a monomial $\mu$ of $M_v$, which exists as $M_v\neq0$, and let $m_1,m_2$ be monomials of
$F_v$. Then $\mu m_1$ and $\mu m_2$ are monomials of $F$, hence set-multilinear over $S$, so
$\mathrm{blocks}(\mu)\sqcup\mathrm{blocks}(m_1)=S=\mathrm{blocks}(\mu)\sqcup\mathrm{blocks}(m_2)$
and $\mathrm{blocks}(m_1)=\mathrm{blocks}(m_2)$. Every monomial of $F_v$ therefore has the same
block support, which we call $S_v$, and uses each of its blocks once; that is, $F_v$ is
set-multilinear with block support $S_v$, and $S_v$ is the set of blocks occurring below $v$.

The two clauses follow. At a sum gate, no cancellation makes every monomial of $F_u$ and of
$F_w$ a monomial of $F_v$, so $S_u=S_w=S_v$. At a product gate, a monomial of $F_v$ is $m_um_w$
with $\mathrm{blocks}(m_um_w)=S_u\cup S_w$, and it meets each of its blocks once, so
$S_u\cap S_w=\emptyset$ and $S_v=S_u\sqcup S_w$.
\end{proof}

The targets $\chi(A)-\epsilon\cdot\chi(1)$ of Section~\ref{sec:encoding} are set-multilinear.
A formula computing a nonzero target can first be pruned of gates computing zero, without
increasing its leaf count, ordinary depth or mixed-product depth. The resulting formula is
syntactically set-multilinear by the lemma, so we omit this restriction in the lower bounds
below. A zero target has the constant-zero certificate and requires no such reduction.

This is what the next corollary is for. By Definition~\ref{def:SHAMC} a $\SHAMC$ proof is a
depth-three monotone expression $\sum_{i}\prod_{j}L_{i,j}$ in which each $L_{i,j}$ is an
\emph{arbitrary} affine form with nonnegative coefficients in all $2n$ variables at once, with no
set-multilinearity imposed. The corollary below, which is Lemma~\ref{lem:autosml-depth} read on
such an expression, says that the freedom buys nothing --- the claim made informally in
Section~\ref{sec:contributions}.

\begin{corollary}[Automatic set-multilinearity at depth three]\label{lem:autosml}
Let $F$ be a nonzero monotone polynomial in the variables
$\{x_i^0, x_i^1\}_{i\in[n]}$ that is homogeneous of degree $n$ and set-multilinear with
respect to the blocks $X_i = \{x_i^0, x_i^1\}$, and suppose
$F = \sum_{i=1}^{r}\prod_{j=1}^{k_i}L_{i,j}$ with every $L_{i,j}$ a nonzero affine form
with nonnegative coefficients. Then each product term equals
$\eta_i\prod_{b=1}^{n}\ell_{i,b}$ with $\eta_i > 0$ and $\ell_{i,b}$ a linear form
supported in the block $X_b$. In particular every $\SHAMC$ proof of a DNF tautology $A$
is a nonnegative tensor decomposition of $\chi(A) - \epsilon\cdot\chi(1)$ with at most as
many rank-one terms as the proof has size.
\end{corollary}

\begin{proof}
Read the expression as a monotone formula: bracket its sums and products into gates of fan-in
two, each $L_{i,j}$ becoming a sum over its constant leaf and its variable leaves, and discard
the gates computing zero, which changes nothing as $F\neq0$.
The output is set-multilinear with block support $[n]$, so by Lemma~\ref{lem:autosml-depth} the
monomials of every gate share a block support. At the gate $L_{i,j}$ those monomials are the
constant term, of support $\emptyset$, and the variables occurring in it, each of support its own
block; so $L_{i,j}$ is either a positive constant, which we absorb into $\eta_i$, or a linear
form $\ell_{i,b}$ inside a single block $X_b$. Within a term the product clause makes these
blocks disjoint and the sum clause makes their union $[n]$, so its linear factors are exactly one
per block.
\end{proof}

\subsection{Flattenings and the flattening bound}\label{sec:flattenings}\label{sec:mixed-depth}

Let $A$ be a DNF in Boolean variables $x_1, \dots, x_n$, encoded as in
Section~\ref{sec:encoding} via the blocks $X_i := \{x_i^0, x_i^1\}$. For a subset
$I \subseteq [n]$ write $X_I := \bigcup_{i\in I} X_i$. Let $f$ be a monotone polynomial that is
set-multilinear with block support $S\subseteq[n]$, and let $I \sqcup \ol I = [n]$ be a
bipartition of the blocks. The \emph{flattening} of $f$ across it is the nonnegative
matrix $\flt{X_I}{X_{\ol I}}{f}$ whose rows are indexed by the set-multilinear monomials in
$X_{S\cap I}$, whose columns are indexed by those in $X_{S\cap\ol I}$, and whose entry at a pair
of monomials is the coefficient in $f$ of their product; the monomials in $X_\emptyset$ are the
single empty one. The targets $\chi(A)-\epsilon\cdot\chi(1)$ have $S = [n]$, where this is the
full coefficient matrix of $f$; the gates of a formula compute polynomials of smaller support,
where the index sets shrink accordingly, a variable giving a single column or a single row and a
constant a $1\times1$ matrix. When the bipartition is clear from the
context we abbreviate this to $\Flat(f)$.

Fix such a bipartition. Since $\chi(1) = \prod_{i=1}^{n}(x_i^0 + x_i^1)$ flattens
to the all-ones matrix, which we denote $J_{X_I \mid X_{\ol I}}$,
\begin{equation}\label{eq:flatten-perturbed}
   \flt{X_I}{X_{\ol I}}{\chi(A) - \epsilon\cdot\chi(1)}
   \;=\;
   \flt{X_I}{X_{\ol I}}{\chi(A)} \;-\; \epsilon\cdot J_{X_I \mid X_{\ol I}}.
\end{equation}
Because each rank-one tensor of~\eqref{eq:prodmix} flattens to a nonnegative rank-one
matrix, a $\SHAMC$ derivation of $A$ with parameter $\epsilon$ has at least
\begin{equation}\label{eq:flat-lb}
   \rkp\!\bigl(\flt{X_I}{X_{\ol I}}{\chi(A)} \;-\; \epsilon\cdot J_{X_I \mid X_{\ol I}}\bigr)
\end{equation}
terms, and, $\rkp$ being monotone under taking submatrices, also at least $\rkp(N)$ for every
submatrix $N$ of the matrix in~\eqref{eq:flatten-perturbed}. Section~\ref{sec:polytopes}
proceeds throughout by exhibiting such a submatrix. A derivation in $\SHAMC^{\epsilon}$ carries
some parameter $\epsilon'\geq\epsilon$, so what bounds its size is the minimum of
\eqref{eq:flat-lb} over $\epsilon'\geq\epsilon$; the lower bounds below hold throughout a range
of shifts and so bound that minimum as well.

The rest of this subsection is not standard material: it states this bound for monotone
formulas of arbitrary \emph{ordinary} depth, the parameter being instead the mixed-product depth
$t$ of Definition~\ref{def:mixed-depth}. That is the form used throughout the paper, and the
display above is its case $t = 0$ read at depth three.

The governing parameter is not depth. Multiplying by a factor that lives on one side of the
bipartition permutes and duplicates rows, or columns, and cannot raise the rank, so what has to
be counted is only the product gates that genuinely mix the two sides.

\begin{definition}[Mixed-product depth]\label{def:mixed-depth}
Fix a partition of the variable blocks into $\mathcal X \sqcup \mathcal Y$ and let $\Phi$ be a
syntactically set-multilinear monotone arithmetic formula. A gate is \emph{mixed} if its block
support meets both $\mathcal X$ and $\mathcal Y$. A product gate is \emph{doubly mixed} if both
of its children are mixed. For a gate $v$ let $\mu(v)$ be the maximum number of doubly mixed
product gates on a path from $v$ to a leaf and $L(v)$ the number of leaves below $v$. The
\emph{mixed-product depth} $\mu(\Phi)$ of $\Phi$ is $\mu$ of its output gate. A sum or product of larger
fan-in has first to be bracketed into gates of fan-in two, and $\mu$ depends on the choice;
since Theorem~\ref{thm:mixed-depth-flattening} applies to every bracketing, we read $\mu$ of
such an expression as the minimum over bracketings.
\end{definition}

\begin{theorem}[Flattening bound at bounded mixed-product depth]
\label{thm:mixed-depth-flattening}
Let $\Phi$ be a syntactically set-multilinear monotone arithmetic formula with respect to the
block partition $\mathcal X\sqcup\mathcal Y$. For every gate $v$, with $F_v$ the polynomial
computed at $v$ and $\Flat(F_v)$ its flattening across the induced
$\mathcal X\mid\mathcal Y$ bipartition,
\[
   \rkp\bigl(\Flat(F_v)\bigr)\;\leq\; L(v)^{2^{\mu(v)}} .
\]
Consequently, if $\Phi$ has $L$ leaves and mixed-product depth at most $t$, then
\begin{equation}\label{eq:method}
   L \;\geq\; \rkp(N)^{1/2^{t}}
   \qquad\text{for every submatrix $N$ of $\Flat(F_\Phi)$.}
\end{equation}
\end{theorem}

\begin{proof}
Write $r(v) := \rkp(\Flat(F_v))$ and induct. A leaf computes a variable, of block support a
single block $X_i$, or a constant, of empty block support, so by the definition above its
flattening is $|X_i|\times1$ or $1\times|X_i|$ according to the side of $X_i$, or $1\times1$; in
each case it has one nonzero entry and $r\leq1$.

At a sum gate $v$ with children $u, w$, Definition~\ref{def:syntactic-sml} gives
$S_u = S_w = S_v$. The rows of the flattening of a set-multilinear polynomial with block support
$S$ are indexed by the set-multilinear monomials in the blocks of $S\cap\mathcal X$ and its
columns by those in $S\cap\mathcal Y$, index sets determined by $S$ alone; so
$\Flat(F_u)$ and $\Flat(F_w)$ have the same rows and the same columns, and flattening being a
re-indexing of the coefficient vector, $\Flat(F_v)=\Flat(F_u)+\Flat(F_w)$ entrywise. All three
matrices are nonnegative because $\Phi$ is monotone, so concatenating nonnegative rank-one
decompositions of the two summands gives one of the sum: $r(v) \leq r(u)+r(w)$. Put
$q := 2^{\mu(v)} \geq 1$; since
$\mu(u),\mu(w)\leq\mu(v)$, induction and $a^{q}+b^{q}\leq(a+b)^{q}$ for $q \geq 1$,
$a,b\geq 0$ give $r(v)\leq L(u)^{q}+L(w)^{q}\leq L(v)^{q}$.

At a product gate $S_u\cap S_w=\emptyset$ by Definition~\ref{def:syntactic-sml}, so up to
permutations $\Flat(F_v) = \Flat(F_u)\otimes\Flat(F_w)$ and $r(v)\leq r(u)r(w)$. If $v$ is not
doubly
mixed, one child is one-sided, its flattening has a single row or column, and
$r(v)\leq r(u)\leq L(v)^{2^{\mu(v)}}$. If $v$ is doubly mixed then
$\mu(v)=1+\max\{\mu(u),\mu(w)\}$, so with $q := 2^{\mu(v)}$ both child exponents are at
most $q/2$ and
\[
   r(v)\;\leq\; L(u)^{q/2}L(w)^{q/2}\;\leq\;(L(u)+L(w))^{q}\;=\;L(v)^{q}.
\]
Finally, nonnegative rank is monotone under taking submatrices.
\end{proof}

\begin{remark}[Depth three gives mixed-product depth zero]\label{rem:t0}
A depth-three monotone $\Sigma\Pi\Sigma$ expression for a set-multilinear polynomial is, by
Corollary~\ref{lem:autosml}, a sum of rank-one set-multilinear tensors. Its product gates have
unbounded fan-in while $\mu$ counts binary gates, so a bracketing has to be chosen, and the
choice matters: interleaving the two sides makes a gate doubly mixed at every level. Bracketing
each term by multiplying all $\mathcal X$-factors, then all $\mathcal Y$-factors, and finally
the two one-sided products leaves no product gate with two mixed children
(Figure~\ref{fig:bracketing}), so the mixed-product depth is zero and \eqref{eq:method} reads
\[
   \rkt(f) \;\geq\; \rkp\bigl(\Flat(f)\bigr) \;\geq\; \rkp(N),
\]
the bound on nonnegative tensor rank quoted in Section~\ref{sec:flattenings}. Sharing the
bottom linear forms in a depth-three \emph{circuit} also does not help: its flattening rank is
at most the number of top product gates. The converse implication fails, and fails
exponentially: Remark~\ref{rem:mu0-vs-depth3} exhibits formulas of mixed-product depth zero,
of polynomial size and logarithmic depth, for targets of exponential tensor rank. Mixed-product
depth is thus not a measure of ordinary depth.
\end{remark}

The restriction to formulas is not cosmetic. Reuse of an already mixed partial polynomial can
compress a flattening of exponential rank into a linear-size circuit, even one with no doubly
mixed product gate at all.

\begin{prop}[Sharing defeats the bound for circuits]
\label{prop:circuit-sharing-obstruction}
For every $n$ there is a set-multilinear monotone polynomial $P_n$ whose
flattening has nonnegative rank $2^{n}$, but which has a syntactically
set-multilinear monotone circuit of size $O(n)$ containing no doubly mixed product gate.
\end{prop}

\begin{proof}
Use the $2n$ blocks $X_i = \{x_i^0,x_i^1\}$ and $Y_i = \{y_i^0,y_i^1\}$, partitioned into all
$X_i$ versus all $Y_i$, and set $P_n = \prod_{i=1}^{n}(x_i^0y_i^0+x_i^1y_i^1)$. The
coefficient of $\prod_i x_i^{a_i}\prod_i y_i^{b_i}$ is $1$ if $a=b$ and $0$ otherwise, so the
flattening is the identity matrix of order $2^{n}$, of nonnegative rank $2^{n}$. On the other
hand $P_i = x_i^0(y_i^0P_{i-1})+x_i^1(y_i^1P_{i-1})$, reusing one gate for $P_{i-1}$ in both
summands, adds $O(1)$ gates per $i$; every gate is syntactically set-multilinear, and at each
product gate one child is a single variable, hence one-sided.
\end{proof}

The polynomial itself is not hard for formulas: the displayed product is one, with $4n$ leaves.
But its product gates are doubly mixed, and bracketed in a balanced tree it has mixed-product
depth $\lceil\log_2n\rceil$, where \eqref{eq:method} asks only for $L\geq2$. At mixed-product
depth zero the same bound forces $2^{n}$ leaves, and that is what the circuit above evades by
reusing $P_{i-1}$.

So the method allows arbitrary ordinary depth and arbitrary one-sided multiplication, but a
bounded number of genuinely two-sided multiplications per path and no reuse of a mixed partial
result.
Writing $R=\rkp(N)$, the bound is $\log L\geq (\log R)/2^t$. It diverges when
$2^t=o(\log R)$, and is superpolynomial in an input parameter $n$ when
$(\log R)/2^t=\omega(\log n)$.

The following elementary lemma is the standard way of producing a DNF whose flattening
contains a prescribed nonnegative integer matrix. We use it in
Section~\ref{sec:csp-dnf}.

\begin{prop}[Local-witness DNF encoding]\label{prop:local-witness-dnf}
Let $R\subseteq\{0,1\}^{X}$ and $C\subseteq\{0,1\}^{Y}$ be sets of \emph{valid} row and
column encodings over disjoint variable sets $X$ and $Y$. Suppose invalid encodings are
recognised by DNFs $I_R(X)$ and $I_C(Y)$, that is, $I_R(x) = 1$ iff $x \notin R$ and
$I_C(y) = 1$ iff $y \notin C$, and suppose $W_1,\ldots,W_q$ are conjunctions of literals
such that for every $(x,y)\in R\times C$
\[
 M(x,y):=|\{i:W_i(x,y)=1\}|\geq1 .
\]
Then
\[
 A(X,Y):=I_R(X)\vee I_C(Y)\vee\bigvee_{i=1}^{q}W_i(X,Y)
\]
is a tautological DNF with $|I_R|+|I_C|+q$ terms whose \emph{valid slice} --- the submatrix
of the $X \mid Y$ flattening of $\chi(A)$ indexed by $R\times C$, that is, the rows and columns
carrying valid encodings --- is exactly $M$. Since $\chi(1)$ flattens to the all-ones matrix, the
submatrix of the flattening of $\chi(A)-\epsilon\cdot\chi(1)$ indexed by $R\times C$ is then
$M-\epsilon\cdot J$; we call it the \emph{perturbed valid slice}.
\end{prop}

\begin{corollary}[Leaf bound from the perturbed valid slice]\label{cor:local-witness-mixed-depth}
In the setting of Proposition~\ref{prop:local-witness-dnf}, every monotone formula of
mixed-product depth at most $t$ computing $\chi(A)-\epsilon\cdot\chi(1)$ has at least
$\rkp(M-\epsilon\cdot J)^{1/2^{t}}$ leaves.
\end{corollary}

\begin{proof}
By Proposition~\ref{prop:local-witness-dnf} the $X\mid Y$ flattening of
$\chi(A)-\epsilon\cdot\chi(1)$ contains $M-\epsilon\cdot J$ as the submatrix indexed by the
valid rows and columns. The target is set-multilinear, so the formula is syntactically
set-multilinear by Lemma~\ref{lem:autosml-depth} and \eqref{eq:method} applies.
\end{proof}

The hypothesis $M \geq J$ is exactly what makes $A$ a tautology, and its cost, announced in
Section~\ref{sec:contributions}, is this. Since $M-\epsilon\cdot J\geq(1-\epsilon)J>0$ for $\epsilon<1$, the matrix has no zero entry, so
its rectangle covering number is one and its largest fooling set is a single position. The
hyperplane separation bound survives, but only for signed test matrices. Writing
$\alpha_W:=\max\{\langle W,R\rangle: R$ nonnegative of rank one, $R\leq M-\epsilon\cdot J\}$, the
$r$ summands of a nonnegative factorization each lie entrywise below $M-\epsilon\cdot J$ and so
score at most $\alpha_W$ apiece, giving
$\rkp(M-\epsilon\cdot J)\geq\langle W,M-\epsilon\cdot J\rangle/\alpha_W$ whenever $\alpha_W>0$.
For $W\geq0$ this is vacuous: the floor $(1-\epsilon)J$ is itself an admissible $R$, so
$\alpha_W\geq(1-\epsilon)\|W\|_1$, while
$\langle W,M-\epsilon\cdot J\rangle\leq\langle W,M\rangle\leq\|W\|_1\max M$, and the quotient
never exceeds $\max_{x,y}M(x,y)/(1-\epsilon)$.

A witness must therefore have negative entries, and the achievable bound is governed by how small
its total mass is relative to its correlation with $M$.

\section{The depth-three fragment}\label{sec:SA}

Representation~\eqref{eq:junta} is exactly the normal form for Sherali--Adams refutations
of Fleming, G\"o\"os, Grosser and Robere \cite[Thm.~5]{FlemingGoosGrosserRobere22}: an
unsatisfiable CNF $C$ with a Sherali--Adams refutation of degree $d$ and size $s$, measured
in expanded ordinary monomials, has an $\epsilon > 0$ and a conical junta $J$ of degree $d$
and size $s2^{d}$ with $\viol_{C} - \epsilon = J$, and conversely; read as widths, the two
sides agree once each is maximised with the clause width $w(C)$
\cite[Lem.~10]{FlemingGoosGrosserRobere22}. In particular the existential quantifier
$\exists\,\epsilon > 0$ built into $\AMC$ by Proposition~\ref{prop:AMC-sound-complete} is
precisely the one appearing on the Sherali--Adams side, and $\AMCb$ and Sherali--Adams
coincide as \emph{width} measures, each read as its maximum with the clause width $w(C)$:
\begin{equation}\label{eq:degree-equality}
   \max\bigl\{\SA(C),\,w(C)\bigr\}
   \;=\;
   \max\Bigl\{\min_{\epsilon > 0}
   \mathrm{width}\bigl(\AMCb^{\epsilon}\text{-proof of } A\bigr),\;w(C)\Bigr\} ,
   \qquad C := \neg A .
\end{equation}
That maximum cannot be dropped on either side.

In size the two directions are not symmetric. In a Sherali--Adams refutation
$-\sum_i B_iJ_i + J_0 = -1$ the multipliers $J_i$ may be arbitrary conical juntas, whereas
\eqref{eq:junta} rearranges to
\begin{equation}\label{eq:amcb-as-sa}
   -\sum_{i=1}^{m}\frac{B_i}{\epsilon} \;+\; \frac{J}{\epsilon} \;=\; -1 ,
\end{equation}
the special case of constant multipliers. An $r$-term $\AMCb$ proof is therefore a
Sherali--Adams refutation of size $r+m$: the $r$ conjunctions of $J/\epsilon$, and the one
constant multiplier $1/\epsilon$ attached to each of the $m$ clauses. The converse is \cite[Thm.~5]{FlemingGoosGrosserRobere22}, which reduces arbitrary
multipliers to constants at a cost $2^{d}$ --- incurred by writing the complement of an $r$-literal
conjunction as the sum of all $2^{r}-1$ complete cells. We replace those by $r$ cells, which
brings the cost down to $d$.

\subsection{Simulation of Sherali--Adams and the transfer of size lower bounds}
\label{sec:size-conventions}\label{sec:no-transfer}

We work in the twin variables $x_i$ and $\bar x_i$, taken as syntactically distinct; they are
the variables $x_i^{1}$ and $x_i^{0}$ of Section~\ref{sec:encoding} under the names standard in
the Sherali--Adams literature. They live in the quotient ring
\[
   \mathcal{B}_n \;:=\; \mathbb{R}[x_1,\bar x_1,\dots,x_n,\bar x_n]\big/\mathcal{I}_n ,
   \qquad
   \mathcal{I}_n \;:=\; \bigl\langle\; x_i^{2}-x_i,\;\; \bar x_i^{2}-\bar x_i,\;\;
   x_i+\bar x_i-1 \;:\; i \in [n] \;\bigr\rangle .
\]
Evaluation on $\{0,1\}^n$ identifies $\mathcal{B}_n$ with the ring of real functions on the
cube, so an identity of Boolean functions and a congruence modulo $\mathcal{I}_n$ are the same
statement; we phrase all manipulations symbolically, in $\mathcal{B}_n$. A \emph{conjunction} is
a product of consistent literals and a \emph{conical junta} a nonnegative combination of
conjunctions. For each clause
$D_i = \ell_{i1}\vee\cdots\vee\ell_{iw_i}$ of $C = D_1\wedge\cdots\wedge D_m$ let
$B_i := \bar\ell_{i1}\cdots\bar\ell_{iw_i}$ be its falsifying conjunction. Under the
identification just made, the count $\viol_C$ of Section~\ref{sec:juntas} is the element
$\sum_i B_i$ of $\mathcal{B}_n$. A Sherali--Adams
refutation of $C$ is an identity
\begin{equation}\label{eq:SA-refutation}
   -\sum_{i=1}^{m} B_i J_i \;+\; J_0 \;=\; -1,
   \qquad
   J_i = \sum_{a=1}^{t_i}\lambda_{ia}Q_{ia},
   \qquad
   J_0 = \sum_{b=1}^{t_0}\mu_b R_b,
\end{equation}
where the $Q_{ia}$ and $R_b$ are conjunctions and the coefficients $\lambda_{ia},\mu_b$ are
positive. Its \emph{degree} $d$ is the largest degree of a nonzero
$B_iQ_{ia}$ or $R_b$ before cancellation, and its \emph{size} is the number of conjunctions
displayed,
\begin{equation}\label{eq:sa-size}
   s \;:=\; t_0 + \sum_{i=1}^{m}t_i ,
\end{equation}
repetitions counted, coefficients free. In the twin variables a conjunction is a single
monomial, so this is monomial size. \cite{FlemingGoosGrosserRobere22} work instead in the
single-variable multilinear quotient $\mathbb{R}[x]/\langle x_i^{2}-x_i\rangle$, where a
conjunction is $\prod_{i\in S}x_i\prod_{j\in T}(1-x_j)$ and there is no $\bar x_j$ to begin
with, and count the distinct monomials of the expansion; one conjunction is charged up to
$2^{d}$ times.

The normal form \eqref{eq:junta} is the case of constant multipliers. To reach it, an occurrence
$-\lambda B_iQ$ in \eqref{eq:SA-refutation} is rewritten as
$-\lambda B_i + \lambda B_i(1-Q)$, which leaves the clause $B_i$ with a constant multiplier and
moves the remainder to the positive side --- provided the complement $1-Q$ is itself a conical
junta. Writing it as the sum of all $2^{r}-1$ complete cells on the variables of $Q$ does that;
the following partition uses $r$ of them.

\begin{lem}[First mismatch]\label{lem:first-mismatch}
Let $Q = \ell_1\ell_2\cdots\ell_r$ be a conjunction of consistent literals and write
$\bar\ell_j$ for the complementary literal on the same variable. Then
\begin{equation}\label{eq:first-mismatch}
   1 - Q \;=\; \sum_{j=1}^{r}\Bigl(\prod_{h<j}\ell_h\Bigr)\bar\ell_j
\end{equation}
is an identity of polynomials in the twin variables, a consequence of the linear relations
$\ell_j + \bar\ell_j = 1$ alone; the Boolean relations $x_i^{2} = x_i$ are not used. The $r$
conjunctions on the right-hand side are moreover pairwise disjoint: the product of any two of
them is divisible by some $\ell_j\bar\ell_j$, hence vanishes modulo $x_i^{2}-x_i$.
\end{lem}

\begin{proof}
Substituting $\bar\ell_j = 1 - \ell_j$, the $j$-th summand of the right-hand side is
$\prod_{h<j}\ell_h - \prod_{h\leq j}\ell_h$, so the sum telescopes to
$1 - \prod_{h\leq r}\ell_h = 1 - Q$. For the disjointness let $j < k$: the $k$-th summand
contains $\ell_j$ in its prefix, while the $j$-th contains $\bar\ell_j$.
\end{proof}

The identity coarsens the $2^{r}-1$ complete cells by the location of the first mismatch, and
is exact. It also separates two factors of $2^{d}$ that are easily conflated. The one in
\cite[Thm.~5]{FlemingGoosGrosserRobere22} is a cost of replacing the
multipliers by constants: their argument replaces a multiplier monomial $\ell_1\cdots\ell_r$
by the complement junta obtained
from $\sum_{S\subseteq[r]}\prod_{i\in S}\ell_i\prod_{j\notin S}\bar\ell_j = 1$, that is, by a
sum of $2^{r}-1$ complete cells; this is the factor the lemma removes, replacing those cells by
$r$. The second is a cost of the size convention alone, independent of how the multipliers are
handled: under expanded ordinary-monomial counting a width-$d$ conjunction is worth up to
$2^{d}$ monomials however it was produced. Our accounting avoids the first outright; the second
it avoids only in the twin-variable measure.

\begin{theorem}[Sherali--Adams to conical junta normal form]\label{thm:sa-to-amcb}
Let $C$ be an unsatisfiable CNF with $m$ clauses and maximum clause width $w(C)$, and let
\eqref{eq:SA-refutation} be a Sherali--Adams refutation of $C$ of degree $d \geq 1$ and size
$s$, the size \eqref{eq:sa-size} being counted in the twin variables, where a conjunction is a
single monomial. Put $N := \sum_{i=1}^{m} t_i$. Then there are an $\epsilon > 0$ and a
conical junta $K$ with
\begin{equation}\label{eq:normal-form}
   \viol_{C} - \epsilon \;=\; K ,
\end{equation}
where the displayed representation of $K$ uses at most
\[
   t_0 + dN + m \;\leq\; d\,s + m
\]
conjunctions and has width at most $\max\{d, w(C)\}$.
\end{theorem}

\begin{proof}
Expand the chosen conical-junta representations of the multipliers, but do \emph{not} expand
literal conjunctions into ordinary monomials. The negative part of \eqref{eq:SA-refutation}
becomes $-\sum_{i=1}^{m}\sum_{a=1}^{t_i}\lambda_{ia}B_iQ_{ia}$. If $B_iQ_{ia}$ contains two
complementary literals it is divisible by some $x_i\bar x_i$, hence zero in $\mathcal{B}_n$, and
may be deleted; we do so once and for all.

Consider one surviving occurrence and abbreviate $\lambda = \lambda_{ia}$, $B = B_i$ and
$Q = Q_{ia} = \ell_1\cdots\ell_r$. Lemma~\ref{lem:first-mismatch} gives
\begin{equation}\label{eq:replacement}
   -\lambda BQ \;=\; -\lambda B \;+\; \lambda B(1-Q)
   \;=\; -\lambda B \;+\; \lambda\sum_{j=1}^{r}B\Bigl(\prod_{h<j}\ell_h\Bigr)\bar\ell_j .
\end{equation}
Since $r \leq \deg(BQ) \leq d$, replacing this occurrence creates at most $d$ positive
conjunctions, each of degree at most $\deg(BQ)$, and so does not increase the degree.

Apply \eqref{eq:replacement} to every surviving occurrence and gather all the positive terms,
together with $J_0$, into a conical junta $J'$. This yields
\begin{equation}\label{eq:after-first-mismatch}
   -\sum_{i=1}^{m}a_iB_i \;+\; J' \;=\; -1,
\end{equation}
where $a_i \geq 0$ is the sum of the coefficients $\lambda_{ia}$ of the surviving multiplier
occurrences attached to clause $i$. The junta $J'$ has at most $t_0 + dN$ displayed
conjunctions and width at most $d$.

Not all $a_i$ vanish: otherwise \eqref{eq:after-first-mismatch} would assert that the
nonnegative function $J'$ equals $-1$. Let $a := \max_i a_i > 0$ and \emph{equalise} the
clause coefficients by setting
\[
   J'' \;:=\; J' \;+\; \sum_{i=1}^{m}(a - a_i)B_i .
\]
This is again a conical junta; it uses at most $m$ additional conjunctions and has width at
most $\max\{d, w(C)\}$, the second alternative arising only when a clause with $a_i = 0$ must
be added and contributes its full width. Now \eqref{eq:after-first-mismatch} reads
$-a\sum_{i}B_i + J'' = -1$. Dividing by $a$ and setting $\epsilon := 1/a$, and using
$\sum_i B_i = \viol_C$, gives $\viol_C - \epsilon = J''/a$, which is
\eqref{eq:normal-form} with the claimed term count and degree. Finally
$t_0 + dN + m \leq d(t_0 + N) + m = ds + m$ because $d \geq 1$.
\end{proof}

It remains to translate \eqref{eq:normal-form} into the $\AMC$ encoding of
Section~\ref{sec:encoding}. For $f \in \mathcal{B}_n$, written as the function it represents,
define its \emph{coefficient lift}
\[
   \Lift(f) \;:=\; \sum_{\sigma \in \{0,1\}^n} f(\sigma)\prod_{i=1}^{n}x_i^{\sigma_i} ,
\]
a linear bijection onto the set-multilinear polynomials in the paired variables
$x_i^0, x_i^1$.

\begin{lem}\label{lem:lift}
If $Q = \prod_{i\in S}x_i\prod_{j\in T}\bar x_j$ is a conjunction, then
\begin{equation}\label{eq:lift-conj}
   \Lift(Q) \;=\; \prod_{i\in S}x_i^1\prod_{j\in T}x_j^0
   \prod_{k \notin S\cup T}\bigl(x_k^0+x_k^1\bigr).
\end{equation}
Moreover $\chi(A) = \Lift(\viol_C)$ and $\chi(1) = \Lift(1)$ for $A = \neg C$: both sides of
\eqref{eq:lift-conj} are the sum of the monomials $\mathbf x^{\sigma}$ over the subcube
$Q = 1$, and the last two identities are those of Section~\ref{sec:encoding} read through
$c_A = \viol_C$.
\end{lem}

A conjunction thus becomes exactly one $\AMCb$ subcube term, with no expansion over its free
coordinates, and applying $\Lift$ to \eqref{eq:normal-form} termwise gives
\begin{equation}\label{eq:lifted-normal-form}
   \chi(A) - \epsilon\cdot\chi(1)
   \;=\; \sum_{u=1}^{T}\eta_u\prod_{i\in S_u}x_i^1\prod_{j\in T_u}x_j^0
         \prod_{k\notin S_u\cup T_u}\bigl(x_k^0+x_k^1\bigr),
   \qquad \eta_u > 0,
\end{equation}
with $T \leq t_0 + dN + m$. By Definition~\ref{def:AMCb} this is precisely an
$\AMCb$ proof of $A$.

\begin{corollary}[$\AMCb$ and Sherali--Adams are polynomially equivalent]
\label{cor:amcb-sa-equivalent}
Let $C$ be an unsatisfiable CNF with $m$ clauses, let $A = \neg C$, and measure
Sherali--Adams by the size \eqref{eq:sa-size}.
\begin{enumerate}
\item A degree-$d$, size-$s$ Sherali--Adams refutation of $C$ yields a depth-three
  $\AMCb$ proof of $A$, for some $\epsilon > 0$, with $O(ds + m)$ product terms and width at most
  $\max\{d, w(C)\}$. As a monotone $\Sigma\Pi\Sigma$ circuit it has top fan-in $O(ds+m)$,
  product gates of fan-in $n$ and bottom linear forms of fan-in at most $2$, hence
  $O\bigl(n(ds + m)\bigr)$ wires.
\item Conversely, an $\AMCb$ proof of $A$ with $r$ terms is, by \eqref{eq:amcb-as-sa}, a
  Sherali--Adams refutation of $C$ with $s \leq r+m$: one constant multiplier for each
  clause, and the $r$ terms of the junta $J/\epsilon$.
\end{enumerate}
Since $d \leq n$, the two systems polynomially simulate one another as existential-$\epsilon$
proof systems; by \eqref{eq:degree-equality} the minimum Sherali--Adams degree and the minimum
$\AMCb$ width agree above the clause width $w(C)$, and either may fall below it.
\end{corollary}

\begin{proof}
In (1), Theorem~\ref{thm:sa-to-amcb} displays $T \leq ds+m$ conjunctions of width at most
$\max\{d,w(C)\}$, each of which lifts by Lemma~\ref{lem:lift} to one product term of
\eqref{eq:lifted-normal-form}. Its three layers are an output addition of fan-in $T$, product
gates of fan-in $n$, and bottom linear forms $x_i^1$, $x_i^0$ or $x_k^0+x_k^1$, so there are at
most $2nT+T$ wires. Part (2) is \eqref{eq:amcb-as-sa} read from right to left.
\end{proof}

The simulation also stays inside the bounded-bias
fragment: its output \eqref{eq:lifted-normal-form} uses only subcube terms, so the arbitrary
biases available in general $\AMC$ are never needed.

Three qualifications belong with the statement. It is an equivalence of
\emph{existential}-$\epsilon$ systems: neither direction normalises a proof to a prescribed
$\epsilon$, the value produced by Theorem~\ref{thm:sa-to-amcb} being $1/\max_i a_i$. It counts product terms and
displayed conjunction occurrences, so a wire count carries the additional
factor $O(n)$ of part~(1), and real coefficients have unit cost on
both sides, so it is not a bit-complexity statement. Above all it is an equivalence in the
size measure of \eqref{eq:sa-size} and not in the expanded ordinary-monomial measure of
\cite{FlemingGoosGrosserRobere22}.

\begin{corollary}[Resolution]\label{cor:resolution}
Let $C$ be an unsatisfiable CNF with $m$ clauses in $n$ variables and $A = \neg C$. A
Resolution refutation of $C$ with $k$ lines yields, for some $\epsilon>0$, an $\AMCb$ proof of
$A$ with at most $2nk+m$ terms. In particular $\AMCb$ polynomially simulates Resolution.
\end{corollary}

\begin{proof}
Work in $\mathcal B_n$ and write $B_D$ for the falsifying conjunction of a clause $D$, so that
$B_{\emptyset} = 1$, and $B_D = 0$ when $D$ contains complementary literals. Let a line
$D\vee D'$ be derived by resolving $D\vee x$ with $D'\vee\neg x$. Then
$B_{D\vee x} = B_D\bar x$, $B_{D'\vee\neg x} = B_{D'}x$ and $B_{D\vee D'} = B_DB_{D'}$, and
splitting $B_DB_{D'} = B_DB_{D'}\bar x + B_DB_{D'}x$ gives
\[
   B_{D\vee x} + B_{D'\vee\neg x}
   \;=\; B_{D\vee D'} \;+\; B_D\bar x\,(1-B_{D'}) \;+\; B_{D'}x\,(1-B_D) .
\]
By Lemma~\ref{lem:first-mismatch} the last two terms form a conical junta $J_\ell$ with at
most $|D'|+|D|\leq2n$ conjunctions, products containing a complementary pair being zero and
dropped.

Weight the lines from the last to the first: the final line, the empty clause, gets weight
$1$, and each derived line passes its weight $w_\ell$ to both of its premises, a line used
several times collecting the sum. Multiply the identity of each derived line $\ell$ by $w_\ell$
and add. A derived line other than the last occurs once as a conclusion, with coefficient
$-w_\ell$, and as a premise with total coefficient $w_\ell$, so it cancels, and what remains is
\[
   \sum_{i=1}^{m}W_iB_i \;=\; 1+\sum_{\ell}w_\ell J_\ell ,
\]
where $W_i\geq0$ is the weight collected by the $i$-th clause of $C$. Not all $W_i$ vanish,
since the right-hand side is at least $1$. With $a := \max_iW_i$, equalising as in the proof
of Theorem~\ref{thm:sa-to-amcb} gives
$a\viol_C - 1 = \sum_\ell w_\ell J_\ell + \sum_i(a-W_i)B_i$, a conical junta with at most
$2nk+m$ conjunctions. Dividing by $a$ and applying Lemma~\ref{lem:lift} gives an $\AMCb$ proof
with $\epsilon = 1/a$.
\end{proof}

Hrube\v{s}' simulation \cite[Prop.~5.8]{Hrubes20} bounds $\AMC$ proof size, a circuit
measure, by $O((m+k)n)$; the corollary counts product terms instead, $O(nk+m)$ of them and so
$O(n^{2}k+nm)$ wires as a $\Sigma\Pi\Sigma$ circuit. Both simulations have polynomial size, and
we record the corollary to fix the position of $\AMCb$ among the standard systems.

\paragraph{Size lower bounds transfer.}
Atserias and Hakoniemi \cite{AtseriasHakoniemi19} work in the twin variables modulo
$\mathcal I_n$ and measure a sums-of-squares refutation by the number of twin-variable
monomials in the polynomials under the squares and in the multipliers of the constraint
equations, multipliers of the Boolean axioms being uncharged. In this measure an $r$-term
$\AMCb$ proof is a sums-of-squares refutation of $\{B_i = 0\}$ with at most $r+m$ monomials:
dividing the identity $\viol_C-\epsilon = \sum_j\eta_jQ_j$ of \eqref{eq:junta} by $\epsilon$ and
using $Q_j^{2}\equiv Q_j$ gives
\[
   -1 \;\equiv\; \sum_{j=1}^{r}\Bigl(\sqrt{\eta_j/\epsilon}\;Q_j\Bigr)^{2}
   \;+\;\sum_{i=1}^{m}\Bigl(-\frac1\epsilon\Bigr)B_i \pmod{\mathcal I_n},
\]
with one monomial under each square and one constant multiplier per clause; a conjunction
counts as a single monomial, and there is no expansion over $\bar x_i = 1-x_i$. Lower bounds come
from their size--degree trade-off, Theorem~\ref{thm:ah-size-degree}.

\begin{theorem}[\cite{AtseriasHakoniemi19}]\label{thm:ah-size-degree}
If a system of degree-$k$ polynomial constraints in $n$ Boolean variables has a sums-of-squares
refutation of monomial size at most $s$, then it has one of degree
$O\bigl(\sqrt{n\log s}\,\bigr)+k$.
\end{theorem}

A degree lower bound may be invoked only for the system of equations it was proved for. For
random $3$-SAT their system is the clause system itself: their equation $p_D = 1$ for a clause
$D$ is $B_D = 1-p_D = 0$. For Tseitin it is not: they use the parity equations $P_v = 0$,
$P_v := \prod_{e\ni v}(1-2x_e)-(-1)^{c(v)}$, for which \cite{Grigoriev01} supplies the degree
lower bound. The clause system converts at no cost. If $B_{v,a}$ is the falsifying conjunction
of an assignment $a$ to $\delta(v)$ violating the constraint at $v$, then
$\prod_{e\ni v}(1-2x_e) = -(-1)^{c(v)}$ on its support, so
\[
   B_{v,a}P_v \;\equiv\; \kappa_vB_{v,a} \pmod{\mathcal I_n},
   \qquad \kappa_v := -2(-1)^{c(v)} \neq 0,
\]
so in the refutation above the constant multiplier of $B_{v,a}$ can be traded for the
one-monomial multiplier $-B_{v,a}/(\epsilon\kappa_v)$ of $P_v$. An $r$-term $\AMCb$ proof of $\TS_G$ thus
gives a sums-of-squares refutation of the parity system with at most $r+m$ monomials. With these
inputs the trade-off gives monomial size $2^{\Omega(n)}$ in both cases
\cite[Cor.~11 and Cor.~14]{AtseriasHakoniemi19}. Hence:

\begin{corollary}[$\AMCb$ size lower bounds, uniformly in $\epsilon$]\label{cor:all-eps-amcb}
Let $G$ be a constant-degree expander on $n$ vertices. Every $\AMCb$ proof of $\TS_G$ has size
$2^{\Omega(n)}$, and so does every $\AMCb$ proof of the negation of a random $3$-SAT instance as
in \cite[Cor.~14]{AtseriasHakoniemi19}, in both cases simultaneously for every $\epsilon>0$.
\end{corollary}

What has been studied in this normal form is its \emph{degree}: the least width of a conical
junta representing $\viol_C-1$, that is the case $\epsilon = 1$ of \eqref{eq:junta}, which
G\"o\"os, Jain and Watson treat under the name of the ``$\exists$-$1$ game'' and for which they
prove degree lower bounds for Tseitin formulas \cite{GoosJainWatson18}, and whose restriction to
integer coefficients Filmus, Mahajan, Sood and Vinyals relate to the complexity of MaxSAT
resolution \cite{FilmusMahajanSoodVinyals20}; see
\cite[\S1.2]{FlemingGoosGrosserRobere22} for the comparison. The \emph{size} of a conical junta
of unrestricted width, which is the measure of this paper, seems to have received little
attention.

\subsection{An upper bound for the pigeonhole principle}\label{sec:tensor-CC-UB}

The fragment is thus at least as strong as Sherali--Adams, and the polynomial-size
Sherali--Adams refutation of the unary pigeonhole principle, noted by Rhodes \cite{Rhodes07}
(see also \cite{DantchevGhaniMartin20}), already gives a polynomial $\AMCb$ proof by
Corollary~\ref{cor:amcb-sa-equivalent}. We include the construction below, obtained
independently of the simulation, because it is explicit: $O(n^{4})$ subcube terms with
coefficients $\eta\in\{0,1\}$ and no dependence on the occurrence convention.

The two arguments share their core. One of the three families of terms below consists of the
juntas $\prod_{i}x_{i,k}^{0}$, that is $\prod_i(1-x_{i,k})$, saying that hole $k$ is empty, and
these are also the multipliers carrying the Sherali--Adams refutation. The other two families
are the bookkeeping demanded by the exact pointwise identity \eqref{eq:junta}, which an
inequality-based refutation may discard.

We consider the DNF encoding of $\mathsf{PHP}^{n+1}_n$ in the variables
$\{x_{i,k}^0, x_{i,k}^1\}_{i\in[n+1],\,k\in[n]}$, so that
$\chi(1) = \prod_{i,k}(x_{i,k}^0+x_{i,k}^1)$. For a Boolean assignment $\sigma$ set
\[
   c_k(\sigma) := |\{i : \sigma(i,k) = 1\}|,\qquad
   d_i(\sigma) := |\{k : \sigma(i,k) = 1\}|,\qquad
   E^{\text{pig}}_i(\sigma) := \mathbf{1}[d_i(\sigma) = 0].
\]
The tautology is
\[
   \mathsf{PHP}^{n+1}_n \;:=\;
   \bigvee_{i\in[n+1]}\ \bigwedge_{k\in[n]}\neg x_{i,k}
   \;\vee\;
   \bigvee_{k\in[n]}\ \bigvee_{i<i'}\bigl(x_{i,k}\wedge x_{i',k}\bigr) ,
\]
one term per empty pigeon and one per collision in a hole, so that
$\neg\mathsf{PHP}^{n+1}_n$ is the usual CNF with the pigeon clauses $\bigvee_k x_{i,k}$ and the
hole clauses $\neg x_{i,k}\vee\neg x_{i',k}$. Counting satisfied terms,
\[
   [\mathbf{x}^\sigma]\,\chi(\mathsf{PHP}^{n+1}_n)
   \;=\; \sum_{i=1}^{n+1} E^{\text{pig}}_i(\sigma)
        \;+\; \sum_{k=1}^{n} \binom{c_k(\sigma)}{2}.
\]

\begin{theorem}\label{thm:tensor-PHP-UB}
$\mathsf{PHP}^{n+1}_n$ admits an $\AMCb^{\epsilon}$ proof of size $O(n^4)$, for every
$\epsilon \in (0,1]$, using only coefficients $\eta \in \{0,1\}$ together with a single
coefficient $1-\epsilon$.
\end{theorem}

\begin{proof}
Write $t_+ := \max\{t,0\}$. The combinatorial identity
$\binom{c}{2} = (c-1)_+ + \binom{(c-1)_+}{2}$, valid for every integer $c \geq 0$, applied to
each hole gives
\[
   \sum_{k}\binom{c_k}{2}
   \;=\; \sum_{k:\,c_k\geq 1}(c_k - 1) + \sum_{k}\binom{(c_k - 1)_+}{2}
   \;=\; \sum_{i} d_i - (n - E^{\text{hol}}) + \sum_{k}\binom{(c_k - 1)_+}{2},
\]
where $E^{\text{hol}} := |\{k: c_k = 0\}|$ and we used $\sum_k c_k = \sum_i d_i$ (edge
conservation); here and in what follows the fixed assignment $\sigma$ is suppressed from
$c_k$, $d_i$, $E^{\text{pig}}_i$ and $E^{\text{hol}}$. Substituting,
\begin{align*}
   [\mathbf{x}^\sigma]\bigl(\chi(\mathsf{PHP}^{n+1}_n) - \chi(1)\bigr)
   &= \sum_i E^{\text{pig}}_i + \sum_i d_i + E^{\text{hol}} - n
      + \sum_k\binom{(c_k - 1)_+}{2} - 1 \\
   &= \underbrace{\sum_i\bigl(E^{\text{pig}}_i + d_i - 1\bigr)}_{=:\,F_{\text{pig}}}
      \;+\; E^{\text{hol}}
      \;+\; \sum_k \binom{(c_k - 1)_+}{2},
\end{align*}
using $\sum_i 1 = n+1$ to fold the $-n-1$ into the per-pigeon sum. Checking cases,
$E^{\text{pig}}_i + d_i - 1$ is $0$ if $d_i \in \{0,1\}$ and equals $d_i - 1$ if
$d_i \geq 1$, so $F_{\text{pig}}$ counts excess hole assignments per pigeon. All three terms are
nonnegative, so, writing
\[
   F_{\text{pig}} := \sum_\sigma\Bigl(\sum_i (d_i-1)_+\Bigr)\mathbf{x}^{\sigma},
   \qquad
   E_{\text{holes}} := \sum_\sigma E^{\text{hol}}\,\mathbf{x}^{\sigma},
   \qquad
   T_{\text{holes}} := \sum_\sigma \sum_k\binom{(c_k - 1)_+}{2}\mathbf{x}^{\sigma}
\]
(each coefficient evaluated at the indexing $\sigma$), we obtain the coefficientwise identity
$\chi(\mathsf{PHP}^{n+1}_n) - \chi(1) = F_{\text{pig}} + E_{\text{holes}} + T_{\text{holes}}$
and hence
\begin{equation}\label{eq:DNFphpdecomp}
   \chi(\mathsf{PHP}^{n+1}_n) - \epsilon\cdot\chi(1)
   \;=\; (1-\epsilon)\cdot\chi(1)
         \;+\; F_{\text{pig}} \;+\; E_{\text{holes}} \;+\; T_{\text{holes}}.
\end{equation}

Each family decomposes into bounded-bias rank-$1$ tensors as follows. Below,
``free'' means that the variable $v$ contributes the factor $(x_v^0 + x_v^1)$.

\emph{Floor.} A single rank-$1$ tensor $(1-\epsilon)\cdot\chi(1)$. One term.

\emph{$F_{\text{pig}}$ (excess pigeons).} For each $i\in[n+1]$ and each
pair of holes $k < k'$, the indicator ``$k$ is the first hole of pigeon $i$
and $k'$ is also a hole of $i$'' is
\[
   F_{i,k,k'}(X)
   \;:=\; \Bigl(\prod_{p<k} x_{i,p}^0\Bigr)\cdot x_{i,k}^1\cdot x_{i,k'}^1
         \cdot \prod_{v\text{ free}}(x_v^0 + x_v^1).
\]
Summing over $(i,k,k')$ recovers $F_{\text{pig}}$, whose coefficient at $\sigma$ is
$\sum_i (d_i - 1)_+$. Number of terms: $(n+1)\binom{n}{2} = O(n^3)$.

\emph{$E_{\text{holes}}$ (empty holes).} For each $k\in[n]$,
\[
   E_k(X) \;:=\; \Bigl(\prod_{i\in[n+1]} x_{i,k}^0\Bigr)
                 \cdot \prod_{v\text{ free}}(x_v^0 + x_v^1),
\]
and summing over $k$ recovers $E_{\text{holes}}$. Number of terms: $n$.

\emph{$T_{\text{holes}}$ (excess pairs).} For each hole $k$, each first
pigeon $i$ in $k$, and each pair of further pigeons $i < l < l'$:
\[
   T_{k,i,l,l'}(X)
   \;:=\; \Bigl(\prod_{q<i} x_{q,k}^0\Bigr)
          \cdot x_{i,k}^1\cdot x_{l,k}^1\cdot x_{l',k}^1
          \cdot \prod_{v\text{ free}}(x_v^0 + x_v^1),
\]
and summing recovers $T_{\text{holes}}$, whose coefficient at $\sigma$ is
$\sum_k\binom{(c_k - 1)_+}{2}$.
Number of terms: $n\cdot\binom{n+1}{3} = O(n^4)$.

All the tensors listed are subcube terms in the sense of Definition~\ref{def:AMCb}, and their
total number is $O(n^4)$.
\end{proof}

\subsection{Tseitin: a direct tensor rank lower bound}
\label{sec:tseitin-lb}

Our principal direct lower bound is proved here. It is not a reduction to the extension
complexity of a polytope, and it cannot pass through a bipartite flattening: by
Corollary~\ref{cor:tseitin-flat-easy} every bipartite flattening of the perturbed Tseitin target
has nonnegative rank $O(n^{3})$, at every $\epsilon\in(0,1]$, so the route of
Section~\ref{sec:polytopes} is empty here whatever the mixed-product depth. We adapt instead
Rothvo\ss{}' hyperplane separation method \cite{Rothvoss17} directly to monotone tensor rank and
to the regime $\epsilon < 1$.

The argument runs in two stages. Sections~\ref{sec:tseitin-setup}--\ref{sec:tseitin-main}
carry it out in the bounded-bias fragment $\AMCb$, where a rank-one term is a subcube and the
defect-stratum bookkeeping is available; Section~\ref{sec:tseitin-refinement} then removes the
bias restriction, by refining an arbitrary product distribution exactly into subcubes and
summing the refinements with their actual weights. The outcome,
Theorem~\ref{thm:tseitin-lb-arbitrary}, is the same $2^{\Omega(\epsilon\bar hn)}$ bound, with
$\bar h := \min\{h,1\}$ for the edge expansion $h$ defined below, for
all of $\SHAMC$, that is, for every general depth-three $\AMC$ proof.
Section~\ref{sec:tiny-shift} then shows why it cannot reach all the way: Tseitin has a
polynomial-size arbitrary-bias proof once $\epsilon$ is exponentially small.

One caveat belongs at the front. At bounded bias the transfer of
Corollary~\ref{cor:all-eps-amcb} already gives $2^{\Omega(n)}$ on these graphs for
\emph{every} $\epsilon>0$, so Theorem~\ref{thm:tseitin-lb-eps} is not the strongest
$\AMCb$ statement available for Tseitin, and Theorem~\ref{thm:tseitin-lb-arbitrary} implies it
up to the polynomial prefactor, since $\rktu\geq\rkt$. The contribution here is the
arbitrary-bias bound, which no transfer currently reaches; the bounded-bias argument is kept
because the refinement of Section~\ref{sec:tseitin-refinement} reuses it verbatim.

\subsubsection{Setting}\label{sec:tseitin-setup}

Let $G = (V,E)$ be a graph on $n = |V|$ vertices with a charge function
$c : V \to \mathbb{F}_2$ of odd total charge $\sum_v c(v) \equiv 1 \pmod 2$. The Tseitin
formula asserts $\bigoplus_{e \ni v}x_e = c(v)$ for every $v$; it is unsatisfiable, and
$\TS_G$ denotes the DNF, in the variables $\{x_e\}_{e\in E}$, whose terms are the local
assignments violating one of these constraints. For a total assignment
$x \in \{0,1\}^{E}$ its \emph{defect set} is
\[
   \partial x + c \;\in\; \mathbb{F}_2^{V},
   \qquad (\partial x)_v := \bigoplus_{e \ni v}x_e ,
\]
the set of violated constraints, and its \emph{defect count} $k(x)$ is the size of that set.
Since $\chi(\TS_G)$ has coefficient $k(x)$ at $\mathbf{x}^{x}$, we study
\[
   T^{\epsilon} \;:=\; \chi(\TS_G) - \epsilon\cdot\chi(1),
   \qquad
   [\mathbf{x}^{x}]\,T^{\epsilon} \;=\; k(x) - \epsilon .
\]
We write $Q_k$ for the stratum of assignments with exactly $k$ defects, and
$Z(G) := \ker\partial$ for the cycle space of $G$, whose dimension is $|E|-n+1$ when $G$ is
connected and $|E|-n+p$ when $G$ has $p$ components.

Throughout Section~\ref{sec:tseitin-lb} the graph $G$ is \emph{$d$-regular and connected} with $d\geq3$ and edge
expansion
\[
   h \;:=\; \min_{0<|S|\leq n/2}\frac{|E(S, V\setminus S)|}{|S|} \;>\; 0;
\]
$h>0$ for every connected $G$, and what makes a family of such graphs an \emph{expander family}
is a bound $h = \Omega(1)$ holding uniformly as $n$ grows; that is the regime in which the bounds
below are read. Note that $h\leq d$, since $|E(S,V\setminus S)|\leq d|S|$ for every $S$. We also assume $G$ is \emph{bridgeless}, which
holds whenever $h > \tfrac12$.

\begin{lem}[Defect strata]\label{lem:strata}
Let $G$ carry a charge $c$.
\begin{enumerate}
\item For every $x\in\{0,1\}^{E}$ the defect count $k(x)$ has the same parity as
  $\sum_v c(v)$; so $Q_k = \emptyset$ unless $k \equiv \sum_v c(v) \pmod 2$, and when the
  formula is unsatisfiable every total assignment has an odd number of defects.
\item If moreover $G$ is connected, on $n$ vertices, then
  $|Q_k| = \binom{n}{k}\,2^{|E|-n+1}$ for every $k \equiv \sum_v c(v) \pmod 2$.
\end{enumerate}
\end{lem}

\begin{proof}
Working in $\mathbb{F}_2$, $\sum_{v}(\partial x)_v = \sum_{v}\bigoplus_{e\ni v}x_e
= \sum_{e}2x_e = 0$, because each edge is counted at its two endpoints; hence
$\sum_v(\partial x + c)_v = \sum_v c(v)$, which is (1). In particular
$\mathrm{im}\,\partial$ lies in the even-weight subspace of $\mathbb{F}_2^{V}$, and for
connected $G$ it is all of it: the edge set of a path from $u$ to $v$ has boundary $e_u+e_v$,
and such vectors span that subspace, of dimension $n-1$. By rank--nullity
$\dim\ker\partial = |E|-n+1$, so the defect vector $\partial x + c$ takes each value of the
parity of $c$ at exactly $2^{|E|-n+1}$ assignments, and $\binom{n}{k}$ of those values have
weight $k$.
\end{proof}

\subsubsection{Bounded-bias rank-one tensors as partial assignments}\label{sec:tseitin-rank1}

A rank-one monotone tensor in $\bigotimes_{e\in E}\langle x_e^0, x_e^1\rangle_{+}$ is a product
\[
  R \;=\; \eta\cdot\prod_{e\in E}\bigl(p_e\, x_e^1 + (1-p_e)\, x_e^0\bigr),
  \qquad p_e\in[0,1],\;\eta\geq 0,
\]
and it defines a probability distribution $\mathcal{D}_R$ on total assignments, edge $e$ being
set to $1$ independently with probability $p_e$. We write
\[
   \omega_k(R) \;:=\; \eta\cdot\Pr_{\mathcal D_R}[\,x \in Q_k\,]
\]
for the mass that $R$ places on the $k$-th defect stratum, and extend $\omega_k$ additively to
sums of rank-one tensors. Until Section~\ref{sec:tseitin-refinement} all biases are
\emph{bounded}, i.e.\ $p_e\in\{0,1,\tfrac12\}$, which is Definition~\ref{def:AMCb}; we call
such a term \emph{uniform}. Its \emph{residual graph} $G'(R)$ has edge set
$\{e : p_e = \tfrac12\}$ and residual charges
$c'(v) = c(v)\oplus\bigoplus_{e\ni v,\,p_e\in\{0,1\}}p_e$; its components (isolated vertices
included) are called \emph{odd} or \emph{even} according to the parity of their total residual
charge. For a uniform term, every assignment in the support of $R$ has the same value
$\eta\,2^{-|E(G'(R))|}$, so
\begin{equation}\label{eq:omega-uniform}
   \omega_k(R) \;=\; \eta\,2^{-|E(G'(R))|}\,|Q_k(R)|,
   \qquad Q_k(R) := Q_k\cap\mathrm{supp}\,R .
\end{equation}

\begin{prop}[Odd-component characterisation]\label{prop:oddcomp}
Let $R$ be a uniform rank-one tensor and let $\ell = \ell(R)$ be the number of odd components
of $G'(R)$. Then the minimum defect count over the support of $R$ equals $\ell$, and $\ell$ is
odd. In particular
\[
  \omega_1(R) = 0 \;\iff\; \ell \geq 3 .
\]
\end{prop}

\begin{proof}
Fix the deterministic edges. By Lemma~\ref{lem:strata}(2) applied to each component $X$ of
$G'(R)$, the defect set of an assignment in the support of $R$ meets $X$ in an arbitrary
subset of $X$ whose parity is that of the residual charge of $X$, independently across
components. The minimum total is therefore one defect in each odd component and none in the
even ones, that is $\ell$. Lemma~\ref{lem:strata}(1) makes every realisable defect
count odd, so $\ell$ is odd. Finally $\omega_1(R) > 0$ iff the minimum is $1$ iff $\ell = 1$.
\end{proof}

\subsubsection{The witness and the plan of the proof}\label{sec:tseitin-witness}

Fix an odd parameter $k^{*}\geq5$ and define the symmetric functional
$W^{\epsilon}$ on assignments by $\langle W^{\epsilon}, \mathbf{x}^{x}\rangle := W^{\epsilon}_{k(x)}$ with
\begin{equation}\label{eq:witness-eps}
  W^{\epsilon}_1 \;:=\; -\infty,\qquad
  W^{\epsilon}_3 \;:=\; \frac{2}{\epsilon|Q_3|},\qquad
  W^{\epsilon}_{k^*} \;:=\; -\frac{1}{(k^*-1)|Q_{k^*}|},\qquad
  W^{\epsilon}_k \;:=\; 0\;\text{ for }k\notin\{1,3,k^*\}.
\end{equation}
The value $-\infty$ is a convention recording that $\langle W^{\epsilon},\cdot\rangle$ is
evaluated only on tensors with no $Q_1$ mass.

The rest of the argument compares the score of $W^{\epsilon}$ on the whole target with its
score on a single term. Let $T^{\epsilon} = \sum_{j\leq r}R_j$ be a decomposition into uniform
rank-one tensors and let $P$ collect the terms with $\omega_1 > 0$, $H$ the rest.
Section~\ref{sec:deficit} bounds the $Q_3$ mass that $P$ can carry, by bounding
$\omega_3(R)/\omega_1(R)$ for a single term (Lemma~\ref{lem:ratio-uniform}); what is left
satisfies $\omega_3(H)\geq2\epsilon|Q_3|$ (Lemma~\ref{lem:deficit}) and hence
$\langle W^{\epsilon},H\rangle\geq1$ (Lemma~\ref{lem:witness-positive}).
Section~\ref{sec:kvs3} bounds the other side: a term of $H$ whose $Q_3$ mass is at least a
$1/\tau$ fraction of $2\epsilon|Q_3|$ has a residual component containing more than half the
vertices, and expansion then forces on it so much $Q_{k^*}$ mass that
$\langle W^{\epsilon},R_j\rangle\leq0$; every other term scores less than $4/\tau$.
Comparing the two gives $r\geq\tau/4$ in Section~\ref{sec:tseitin-main}.

\begin{remark}[Why arbitrary biases are excluded here]\label{rem:obstruction}
The restriction $p_e\in\{0,1,\tfrac12\}$ is essential to the argument of
Sections~\ref{sec:deficit}--\ref{sec:tseitin-main}, and not for convenience. For general $p_e\in(0,1)$ the ratio $\omega_3(R)/\omega_1(R)$ is \emph{unbounded}. Take $R$
whose residual graph has exactly two components: an odd one $C$, and an even one consisting of a
single free edge $e$ of bias $p_e$ whose two endpoints are satisfied when $e$ is unset and
defective when it is set. Defect counts add across components, and $C$ contributes an odd
number while $e$ contributes $0$ or $2$. Writing $\mu_j(C)$ for the weight of $C$ on its
$j$-defect patterns, as in the proof of Lemma~\ref{lem:ratio-uniform}, a total of one defect
forces $j_C = 1$ and $e$ unset, so $\omega_1(R) = \eta\,\mu_1(C)(1-p_e)$ while
$\omega_3(R) = \eta\bigl[\mu_3(C)(1-p_e) + \mu_1(C)p_e\bigr]$. Hence
$\omega_3(R)/\omega_1(R)\geq p_e/(1-p_e)\to\infty$ as $p_e\to1$, the weights of $C$
cancelling from the bound.
The contribution to $\omega_3$ of a term with one odd component can therefore not be controlled
by $\omega_1$ alone.
Section~\ref{sec:tseitin-refinement} lifts the restriction all the same, not by controlling
that ratio --- it really is unbounded --- but by refining the term into uniform ones and
charging each refinement separately.
\end{remark}

\subsubsection{The deficit after deleting the one-defect terms}\label{sec:deficit}

Deleting the terms with $\omega_1 > 0$ is of use only if they cannot carry away most of the
$Q_3$ mass with them. Their total $Q_1$ mass is fixed, at $(1-\epsilon)|Q_1|$, so what they can
carry is governed by the ratio $\omega_3(R)/\omega_1(R)$ of a single term. The extreme case is
\begin{equation}\label{eq:vertex-indicator}
   \chi(\TS_G) \;=\; \sum_{v\in V}\mathbf 1[v\text{ defective}] ,
\end{equation}
whose terms, refined into subcubes, achieve the largest such ratio. The next lemma computes it
and shows that no term does better.

\begin{lem}[Ratio bound at bounded bias]\label{lem:ratio-uniform}
Let $R$ be a uniform rank-one tensor with $\omega_1(R) > 0$; equivalently (Proposition
\ref{prop:oddcomp}) $G'(R)$ has exactly one odd component $C$, the remaining components
$D_1,\dots,D_s$ being even. Then
\[
  \frac{\omega_3(R)}{\omega_1(R)}
  \;=\;
  \frac{\binom{|C|}{3}}{\binom{|C|}{1}} + \sum_{i=1}^{s}\binom{|D_i|}{2}
  \;=\;
  \frac{(|C|-1)(|C|-2)}{6} + \sum_{i=1}^{s} \frac{|D_i|(|D_i|-1)}{2}
  \;\leq\;
  \frac{3|Q_3|}{|Q_1|},
\]
with equality iff $|C| = 1$ and there is a single even component (so $|D_1| = n-1$). More
generally, for every odd $k$,
\begin{equation}\label{eq:all-strata-ratio}
   \frac{\omega_k(R)}{\omega_1(R)} \;\leq\; \binom{n-1}{k-1} \;=\; \frac{k|Q_k|}{|Q_1|} .
\end{equation}
\end{lem}

\begin{proof}
\emph{Component factorisation.} For a component $X$ of $G'(R)$ let
$\mu_j(X) := 2^{-|E(X)|}|Q_j(X)|$ be its unnormalised weight on $j$-defect internal patterns.
By Lemma~\ref{lem:strata}(2) applied to $X$, each realisable internal defect set has
$2^{|E(X)|-|X|+1}$ preimages, so
\[
  \mu_j(X) \;=\; \frac{\binom{|X|}{j}}{2^{|X|-1}}
  \ \text{ if $j$ matches the parity of $X$},\qquad \mu_j(X)=0 \text{ otherwise.}
\]
By \eqref{eq:omega-uniform} and edge-disjointness of the components,
$\omega_j(R) = \eta\sum_{j_C+\sum_i j_{D_i}=j}\mu_{j_C}(C)\prod_i\mu_{j_{D_i}}(D_i)$, each
$j_X$ ranging over values of the parity of $X$.

\emph{The two weights.} With the single odd component $C$ and even components $D_i$, the only
way to total one defect is $j_C=1$ with all $j_{D_i}=0$; to total three, either $j_C=3$ and the
rest $0$, or $j_C=1$ and exactly one $j_{D_i}=2$. Hence
\[
  \omega_1 = \eta\,\mu_1(C)\!\prod_i \mu_0(D_i), \qquad
  \omega_3 = \eta\Bigl[\mu_3(C)\!\prod_i \mu_0(D_i)
              + \mu_1(C)\sum_i \mu_2(D_i)\!\prod_{j\neq i}\mu_0(D_j)\Bigr],
\]
and dividing (the $\eta$ and the $\prod_i\mu_0(D_i)$ cancel) gives the displayed identity.

\emph{Maximisation} over $|C|+\sum_i|D_i| = n$ with $|C|\geq1$, $|D_i|\geq1$. Since
$\binom{a+b}{2} = \binom a2 + \binom b2 + ab > \binom a2 + \binom b2$ for $a,b\geq1$, merging
two even components strictly increases $\sum_i\binom{|D_i|}{2}$; so the maximum has a
\emph{single} even component $D$ with $|D| = n-|C|$, and the objective becomes
$f(\kappa) = \frac{(\kappa-1)(\kappa-2)}{6} + \binom{n-\kappa}{2}$ for $\kappa := |C|\in[1,n]$.
Both terms are convex in $\kappa$ (leading coefficients $\tfrac16$ and $\tfrac12$, so
$f''=\tfrac43>0$), so $f$ attains its maximum on $[1,n]$ at an endpoint; since
$f(1) = \binom{n-1}{2}$ and $f(n) = \tfrac13 f(1)$, the maximum is $f(1)$, at $\kappa=1$. Both
reductions are strict, so equality holds iff $|C|=1$ and there is one even component. Finally,
by Lemma~\ref{lem:strata}(2), $3|Q_3|/|Q_1| = 3\binom n3/n = \tfrac{(n-1)(n-2)}{2} = f(1)$.

\emph{The all-strata form.} All realisable defect sets of $R$ have the same number of preimages,
so $\omega_k(R)/\omega_1(R) = |\mathcal F_k|/|C|$, where $\mathcal F_k$ is the family of
$k$-subsets of $V$ having odd intersection with $C$ and even intersection with every even
component, and where the denominator counts the $|C|$ admissible singleton defect sets. Every
$S\in\mathcal F_k$ meets $C$, so counting pairs $(v,S)$ with $v\in C\cap S$ and forgetting the
parity restrictions gives $|\mathcal F_k| \leq |C|\binom{n-1}{k-1}$. Lemma~\ref{lem:strata}
gives $|Q_k|/|Q_1| = \binom nk/n$ and $k\binom nk/n = \binom{n-1}{k-1}$.
\end{proof}

\begin{lem}[Deficit lemma]\label{lem:deficit}
Let $T^{\epsilon} = P + H$ be a decomposition into uniform rank-one tensors, where $P$ is the
sum of all terms with $\omega_1 > 0$ and $H$ the sum of the remaining terms. Then
$\omega_3(P) \leq 3(1-\epsilon)|Q_3|$ and consequently
\[
   \omega_3(H) \;\geq\; 2\epsilon\,|Q_3| .
\]
More generally $\omega_k(H) \geq \epsilon(k-1)|Q_k|$ for every odd $k\geq3$, and if $G-v$ is
connected for every vertex $v$ this is tight for all $k$ simultaneously.
\end{lem}

\begin{proof}
Every term of $H$ has zero $Q_1$ mass, so $\omega_1(P) = \omega_1(T^{\epsilon})$, and the
coefficient of $T^{\epsilon}$ at every $x\in Q_1$ is $k(x)-\epsilon = 1-\epsilon$, so
$\omega_1(P) = (1-\epsilon)|Q_1|$. Summing the all-strata bound \eqref{eq:all-strata-ratio} of
Lemma~\ref{lem:ratio-uniform} over the terms of $P$ gives
$\omega_k(P) \leq \frac{k|Q_k|}{|Q_1|}\omega_1(P) = (1-\epsilon)k|Q_k|$. Since
$\omega_k(T^{\epsilon}) = (k-\epsilon)|Q_k|$ and $H = T^{\epsilon}-P$,
\[
   \omega_k(H) \;\geq\; \bigl(k-\epsilon-(1-\epsilon)k\bigr)|Q_k| \;=\; \epsilon(k-1)|Q_k| ,
\]
which for $k=3$ is the displayed bound.

\emph{Tightness}, assuming $G-v$ connected for every $v$. Refine each
$\mathbf 1[v\text{ defective}]$ of \eqref{eq:vertex-indicator} into subcubes, one for every assignment of $\delta(v)$
violating the constraint at $v$, the other edges free, and decompose
$T^{\epsilon} = (1-\epsilon)T^{0}+\epsilon T^{1}$, with $T^{0} = \chi(\TS_G)$ the left side of
\eqref{eq:vertex-indicator} and $T^{1} = \chi(\TS_G)-\chi(1)$, as these subcubes scaled by
$1-\epsilon$ together with any decomposition of $\epsilon T^{1}$ into uniform terms. Every subcube of the first group
has positive $Q_1$ mass. Its free edges are exactly the edges of $G-v$, and the charges that
$G-v$ inherits once $\delta(v)$ is pinned have even total, because the pinned pattern violates
the constraint at $v$ while $\sum_u c(u)$ is odd; so $Q_0(G-v)\neq\emptyset$ by
Lemma~\ref{lem:strata}(2), and any such assignment of the free edges, taken together with the
pinned pattern, is an assignment of $G$ whose only defect is $v$, hence a point of $Q_1$ inside
the subcube. Every term of
the second group has $\omega_1 = 0$, since the coefficients of $T^{1}$ vanish on $Q_1$. The split
therefore returns the two groups as $P$ and $H$, and
$\omega_k(H) = \epsilon(k-1)|Q_k|$ for every odd $k$, which is the bound. Without the
hypothesis the first group may contain subcubes with no one-defect mass --- two copies of
$K_5$ each missing an edge, joined through a new vertex $v$ adjacent to the four endpoints, is
connected, bridgeless and $4$-regular while $G-v$ is not --- and those fall into $H$ instead,
so the decomposition no longer certifies equality.
\end{proof}

\begin{lem}\label{lem:witness-positive}
Let $P$ and $H$ be as in Lemma~\ref{lem:deficit} and let $W^{\epsilon}$ be the witness
\eqref{eq:witness-eps}, with its odd parameter $k^{*}\geq5$. Then
$\langle W^{\epsilon}, H\rangle \geq 4 - \tfrac{k^*-\epsilon}{k^*-1} \geq 1$.
\end{lem}

\begin{proof}
$H$ has no mass on $Q_1$, so only the strata $3$ and $k^{*}$ contribute. By
Lemma~\ref{lem:deficit}, $\omega_3(H)\geq2\epsilon|Q_3|$, and $H \leq T^{\epsilon}$ entrywise
gives $\omega_{k^*}(H)\leq(k^*-\epsilon)|Q_{k^*}|$. Hence
\[
   \langle W^{\epsilon}, H\rangle
   \;\geq\; \frac{2}{\epsilon|Q_3|}\cdot 2\epsilon|Q_3|
            \;-\; \frac{(k^*-\epsilon)|Q_{k^*}|}{(k^*-1)|Q_{k^*}|}
   \;=\; 4-\frac{k^*-\epsilon}{k^*-1} .
\]
For $k^{*}\geq5$ and $\epsilon\in(0,1]$ we have $\frac{k^*-\epsilon}{k^*-1}\leq\frac54$, so the
right side is at least $\tfrac{11}{4}\geq1$.
\end{proof}

\subsubsection{From a \texorpdfstring{$Q_3$}{Q3}-cover to a giant residual component}\label{sec:kvs3}

The heart of the argument is the estimate showing that a uniform term which contributes
substantially to $\omega_3$ must also contribute so much to $\omega_{k^*}$ that the witness
sees it as nonpositive. This is where expansion enters.

\begin{definition}[Pruning property]\label{def:pruning}
Let $\rho_0 \geq 0$. We say that $G$ has the \emph{$\rho_0$-pruning property} if for every
spanning subgraph $G'\subseteq G$ whose cycle-rank drop
$\rho := \dim Z(G)-\dim Z(G')$ is at most $\rho_0$, the graph $G'$ has a (necessarily
unique) component with more than $n/2$ vertices.
\end{definition}

Some hypothesis of this kind is needed: a chain of $b\geq2$ equally sized dense blocks joined
consecutively by two edges --- two, since a single joining edge would be a bridge --- has no
component above $n/2$ once the $2(b-1)$ connecting edges are deleted, and that deletion creates
$b-1$ new components, so its cycle-rank cost is only $2(b-1)-(b-1) = b-1$, a constant for
constant $b$. Such a graph therefore fails the $\rho_0$-pruning property for every
$\rho_0\geq b-1$; it cannot satisfy the property up to a linear cycle-rank drop.
The next lemma shows that the
property holds, with $\rho_0$ as large as a constant fraction of $n$, for expanders whose girth
exceeds $4/h$ --- for instance for the Ramanujan families of \cite{LubotzkyPhillipsSarnak88},
whose girth is $\Theta(\log n)$.

\begin{lem}[Pruning]\label{lem:pruning}
Let $G$ be connected, bridgeless and $d$-regular with $d\geq3$, of edge expansion $h>0$ and
girth $g > 4/h$, where $g$ is the length of a shortest cycle. Then $G$ has the
$\rho_0$-pruning property for every
$\rho_0 < \min\{h,2\}\,n/8$.
\end{lem}

\begin{proof}
Let $G'$ be spanning with components $X_1,\dots,X_p$ of sizes $s_1,\dots,s_p$, and suppose for
contradiction that $s_i\leq n/2$ for every $i$. Every edge of $G$ leaving $X_i$ is absent from
$G'$; writing $F$ for the set of edges of $G$ not in $G'$ we have
$|F| \geq \tfrac12\sum_i|\delta(X_i)|$, and
\[
\begin{aligned}
   \rho &= \dim Z(G)-\dim Z(G')
   =\bigl(|E|-n+1\bigr)-\bigl(|E|-|F|-n+p\bigr) \\
   &= |F|-p+1
   \;\geq\;\sum_{i=1}^{p}\Bigl(\tfrac12|\delta(X_i)|-1\Bigr)+1 .
\end{aligned}
\]
Since $G$ is bridgeless, $|\delta(X_i)|\geq2$ and every summand is nonnegative. Since
$s_i\leq n/2$, expansion gives $|\delta(X_i)|\geq h s_i$.

Call $X_i$ \emph{large} if $s_i\geq4/h$ and \emph{small} otherwise. For a large component
$\tfrac12|\delta(X_i)|-1\geq\tfrac h2 s_i-1\geq\tfrac h4 s_i$. For a small component we have
$s_i<4/h<g$, so $G[X_i]$ contains no cycle and, being connected, is a tree with $s_i-1$ edges;
$d$-regularity then gives $|\delta(X_i)| = ds_i-2(s_i-1)\geq s_i+2$, whence
$\tfrac12|\delta(X_i)|-1\geq\tfrac12 s_i$. Consequently
\[
   \rho \;\geq\; \frac h4\sum_{\text{large}}s_i
   \qquad\text{and}\qquad
   \rho \;\geq\; \frac12\sum_{\text{small}}s_i ,
\]
and since the two sums add up to $n$, one of them is at least $n/2$, so
\[
   \rho \;\geq\; \min\{h/4,\,1/2\}\cdot\frac n2 \;=\; \frac{\min\{h,2\}\,n}{8} \;>\; \rho_0 ,
\]
contradicting $\rho\leq\rho_0$. Hence some $s_i>n/2$; uniqueness is immediate, two components
of size $>n/2$ being impossible.
\end{proof}

\begin{lem}[Cost of separating a giant component]\label{lem:cutcost}
Let $G$ be connected and bridgeless with edge expansion $h>0$, let $B\subseteq V$ induce a
connected subgraph with $|B|>n/2$, and put $X := V\setminus B$. Then every $F\subseteq E$
containing $E(X,B)$ satisfies
\[
   \dim Z(G)-\dim Z(G-F) \;\geq\; \tfrac h2\,|X| .
\]
\end{lem}

\begin{proof}
Deleting edges only lowers $\dim Z$, so we may take $F = E(X,B)$. Let $X_1,\dots,X_q$ be the
components of $G[X]$; each has all its external edges going to $B$, and $B$ is connected, so
$G-F$ is the disjoint union of $B,X_1,\dots,X_q$ and
\[
   \dim Z(G)-\dim Z(G-F) \;=\; |F|-q \;=\; \sum_{i=1}^{q}\bigl(|E(X_i,B)|-1\bigr).
\]
Each $X_i$ has $|X_i|\leq|X|<n/2$, so expansion gives $|E(X_i,B)|\geq h|X_i|$, while
bridgelessness gives $|E(X_i,B)|\geq2$; hence
$|E(X_i,B)|-1\geq\tfrac12|E(X_i,B)|\geq\tfrac h2|X_i|$. Summing over $i$ gives the claim.
\end{proof}

\begin{lem}[Defect-ratio estimate]\label{lem:ratio}
Let $k^{*}$ be odd with $5\leq k^{*}\leq n/2$ and let $R$ be a uniform rank-one tensor whose
residual graph has exactly three odd components and a component $B$ with $|B|\geq n-n/k^{*}$.
Then
\[
   \frac{|Q_3|\cdot|Q_{k^*}(R)|}{|Q_{k^*}|\cdot|Q_3(R)|}
   \;\geq\; e^{-4}\,\frac{k^{*}(k^{*}-1)}{6} .
\]
\end{lem}

\begin{proof}
Write $C_1,C_2,C_3$ for the three odd components, $X := V\setminus B$ and $m := |B|$, so that
$|X| = n-m\leq n/k^{*}$ and $m\geq n-n/k^{*}\geq\tfrac45 n\geq n/2\geq k^{*}$, using
$k^{*}\geq5$ and $k^{*}\leq n/2$. At most one of
$C_1,C_2,C_3$ equals $B$, so $X$ contains at least two of them; without loss of generality
$C_1,\dots,C_t$ with $t\in\{2,3\}$ lie in $X$, and $B$ is either the remaining odd component
$C_3$ (if $t=2$) or an even component (if $t=3$).

\textbf{Case A: $B=C_3$.} We may place one defect
in each of $C_1, C_2$ and $k^{*}-2$ defects in $C_3$ --- a valid pattern, $k^{*}-2$ being odd
and matching the odd residual charge of $C_3$ --- and none in the even components. The
$3$-defect support points are exactly one defect per odd component, so the shared factors
$|Q_1(C_1)|, |Q_1(C_2)|$ and those of the even components cancel and Lemma~\ref{lem:strata}
gives
\[
  \frac{|Q_{k^*}(R)|}{|Q_3(R)|}
  \;\geq\;
  \frac{|Q_{k^*-2}(C_3)|}{|Q_1(C_3)|}
  \;=\;
  \frac{\binom{m}{k^*-2}}{m}
  \;=\;
  \frac{(m-1)(m-2)\cdots(m-k^*+3)}{(k^*-2)!}.
\]
Using $\binom n3/\binom{n}{k^*} = \frac{k^{*}!}{6}\cdot\frac{1}{(n-3)\cdots(n-k^*+1)}$ and
$k^{*}!/(k^{*}-2)! = k^{*}(k^{*}-1)$,
\[
  \frac{|Q_3|\cdot|Q_{k^*}(R)|}{|Q_{k^*}|\cdot|Q_3(R)|}
  \;\geq\;
  \frac{k^{*}(k^{*}-1)}{6}\cdot
  \frac{(m-1)\cdots(m-k^*+3)}{(n-3)\cdots(n-k^*+1)} .
\]
The second factor is a product of $k^{*}-3$ terms, each
$\frac{m-1-j}{n-3-j}\geq1-\frac{|X|}{n-k^*+1}\geq1-\frac{2|X|}{n}$ (using
$j\leq k^{*}-4$ and $k^{*}\leq n/2$), so it is at least
$(1-\tfrac{2|X|}{n})^{k^*}\geq e^{-4|X|k^{*}/n}\geq e^{-4}$, since $|X|\leq n/k^{*}$ and
$1-u\geq e^{-2u}$ for $u\in[0,\tfrac12]$.

\textbf{Case B: $B$ even.} All three odd components lie in $X$; placing one defect in each
$C_i$ and $k^{*}-3$ defects in $B$ (valid, $k^{*}-3$ being even) and none elsewhere gives
$|Q_{k^*}(R)|/|Q_3(R)|\geq|Q_{k^*-3}(B)|/|Q_0(B)| = \binom{m}{k^*-3}$, whence
\[
  \frac{|Q_3|\cdot|Q_{k^*}(R)|}{|Q_{k^*}|\cdot|Q_3(R)|}
  \;\geq\;
  \frac{k^{*}(k^{*}-1)(k^{*}-2)}{6}\cdot
  \frac{m(m-1)\cdots(m-k^*+4)}{(n-3)(n-4)\cdots(n-k^*+1)},
\]
a product of $k^{*}-3$ ratios each at least $1-\tfrac{2|X|}{n}$ by the same estimate, so at
least $e^{-4}k^{*}(k^{*}-1)(k^{*}-2)/6$ --- stronger than needed.
\end{proof}

\begin{lem}[$Q_3$ mass forces a giant residual component]\label{lem:kvs3}
Let $G$ be connected, bridgeless, $d$-regular with $d\geq3$ and edge expansion $h>0$, let
$\epsilon\in(0,1]$ and $\tau\geq2$, and put
\[
   K \;:=\; \log_2\tau+\log_2\frac{3-\epsilon}{2\epsilon},
   \qquad
   \Lambda \;:=\; \frac{2K}{h}.
\]
Assume $G$ has the $K$-pruning property. Then every uniform rank-one tensor $R$ with
$R\leq T^{\epsilon}$ entrywise, $\omega_1(R)=0$ and $\omega_3(R)\geq 2\epsilon|Q_3|/\tau$ has
exactly three odd residual components, and $G'(R)$ has a component $B$ with
$|V\setminus B|\leq\Lambda$.
\end{lem}

\begin{proof}
By Proposition~\ref{prop:oddcomp}, $\omega_1(R)=0$ gives $\ell\geq3$, while $\omega_3(R)>0$
forces the minimum defect count $\ell\leq3$. So $G'(R)$ has exactly three odd components
$C_1,C_2,C_3$, the remaining components $\{D_j\}$ being even.

\medskip
\noindent\emph{Step 1: the cycle-rank drop is at most $K$.}
By \eqref{eq:omega-uniform} every support point of $R$ has the same value, and
$[\mathbf x^{x}]T^{\epsilon}=3-\epsilon$ for $x\in Q_3$, so $R\leq T^{\epsilon}$ gives
$\eta\,2^{-|E(G'(R))|}\leq3-\epsilon$ and hence $\omega_3(R)\leq(3-\epsilon)|Q_3(R)|$.
Counting the $3$-defect support points by components (one defect in each $C_i$, none in the
$D_j$) and using Lemma~\ref{lem:strata} within each component,
\[
  |Q_3(R)| \;=\; |C_1||C_2||C_3|\cdot 2^{\dim Z(G'(R))}
  \;=\; |Q_3|\cdot\frac{|C_1||C_2||C_3|}{\binom n3}\cdot 2^{-\rho},
\]
where
\[
  \rho \;:=\; \dim Z(G) - \dim Z(G'(R)) \;\geq\; 0,
\]
where we also used $|Q_3| = \binom n3 2^{\dim Z(G)}$ and $Z(G'(R))\subseteq Z(G)$. Since
$|C_1||C_2||C_3|\leq\binom n3$, combining with the hypothesis
$\omega_3(R)\geq2\epsilon|Q_3|/\tau$ gives
\[
  \frac{2\epsilon|Q_3|}{\tau} \;\leq\; (3-\epsilon)|Q_3|\,2^{-\rho},
  \qquad\text{i.e.}\qquad
  \rho \;\leq\; \log_2\tau+\log_2\frac{3-\epsilon}{2\epsilon} \;=\; K .
\]

\medskip
\noindent\emph{Step 2: a giant residual component, and the ratio.}
Since $\rho\leq K$, the $K$-pruning property supplies a unique component
$B$ of $G'(R)$ with $|B|>n/2$. Set $X := V\setminus B$ and $F := E(X,B)$. No edge of $G'(R)$
crosses $F$, so $Z(G'(R))\subseteq Z(G-F)$ and Lemma~\ref{lem:cutcost} gives
\[
  \tfrac h2|X| \;\leq\; \dim Z(G)-\dim Z(G-F) \;\leq\; \rho \;\leq\; K,
  \qquad\text{that is}\qquad
  |X| \;\leq\; \frac{2K}{h} \;=\; \Lambda .
\]
This is the assertion.
\end{proof}

\begin{lem}[Local sign of the Tseitin witness]\label{lem:local-sign}
Let $k^{*}\geq24e^{4}/\epsilon$ be odd with $k^{*}\leq n/2$, and let $R\leq T^{\epsilon}$ be a
uniform rank-one term with $\omega_1(R) = 0$. Then $\langle W^{\epsilon},R\rangle\leq0$ in
each of the following two cases: the residual graph of $R$ has at least five odd components; or
it has exactly three, and some component $B$ satisfies
\begin{equation}\label{eq:local-sign-condition}
   |V\setminus B| \;\leq\; n/k^{*}.
\end{equation}
\end{lem}

\begin{proof}
If there are at least five odd components then $\omega_3(R) = 0$ by
Proposition~\ref{prop:oddcomp}, while the only other finite nonzero coefficient of
$W^{\epsilon}$ is negative.

Suppose there are exactly three odd components. Then \eqref{eq:local-sign-condition} is the
hypothesis $|B|\geq n-n/k^{*}$ of Lemma~\ref{lem:ratio}, which gives
\[
   \frac{|Q_3|\,|Q_{k^*}(R)|}{|Q_{k^*}|\,|Q_3(R)|}
   \;\geq\;
   e^{-4}\,\frac{k^{*}(k^{*}-1)}{6} ,
\]
and since $R$ is uniform its masses have the same ratio as the corresponding support sizes by
\eqref{eq:omega-uniform}, so
\[
   \langle W^{\epsilon},R\rangle
   \;\leq\; \frac{\omega_3(R)}{|Q_3|}
            \Bigl(\frac{2}{\epsilon}-\frac{e^{-4}k^{*}}{6}\Bigr)
   \;\leq\; 0
\]
by $k^{*}\geq24e^{4}/\epsilon$.
\end{proof}

\subsubsection{The bounded-bias lower bound}\label{sec:tseitin-main}

\begin{theorem}[Bounded-bias Tseitin lower bound]\label{thm:tseitin-lb-eps}
There is an absolute constant $c_0 > 0$ such that the following holds. Let $G=(V,E)$ be
connected, bridgeless and $d$-regular with $d\geq3$, on $n$ vertices, with edge expansion
$h>0$; write $\bar h := \min\{h,1\}$; let the Tseitin charge have odd total parity; and let
$\epsilon\in(0,1]$. Let $\tau\geq2$ satisfy
\begin{equation}\label{eq:tau-condition}
   \log_2\tau \;+\; \log_2\frac{2}{\epsilon} \;\leq\; c_0\,\epsilon\,\bar h\,n ,
\end{equation}
and assume $G$ has the $K$-pruning property for
$K = \log_2\tau+\log_2\tfrac{3-\epsilon}{2\epsilon}$, which holds in particular when the girth
of $G$ exceeds $4/h$. Then every decomposition
$T^{\epsilon} = \sum_{j=1}^{r}R_j$ into uniform (bounded-bias) rank-one tensors has
$r\geq\tau/4$. In particular, if the girth exceeds $4/h$ and
$\epsilon\bar h n\geq C\log_2(2/\epsilon)$ for a sufficiently large absolute constant $C$, then
\[
   \rktu\bigl(\chi(\TS_G)-\epsilon\cdot\chi(1)\bigr) \;=\; 2^{\Omega(\epsilon \bar h n)} ,
\]
and hence every $\AMCb^{\epsilon}$ proof of $\TS_G$ has this lower bound. For bounded degree,
$h\leq d$ gives $\bar h = \Theta(h)$, and the bound reads
$2^{\Omega(\epsilon h n)}$.
\end{theorem}

\begin{proof}
Take $c_0 := 1/(200\,e^{4})$ and let $k^{*}$ be the least odd integer with
$k^{*}\geq 24e^{4}/\epsilon$; then $k^{*}\leq 25e^{4}/\epsilon$. Since $\tau\geq2$ and
$(3-\epsilon)/(2\epsilon)\geq1$ for $\epsilon\leq1$, we have $K\geq1$, and
\eqref{eq:tau-condition} gives
$K \leq\log_2\tau+\log_2\frac2\epsilon\leq c_0\epsilon\bar hn$. In particular
$K\leq c_0\bar hn<\min\{h,2\}n/8$, since $c_0<\tfrac18$ and
$\bar h = \min\{h,1\}\leq\min\{h,2\}$, so Lemma~\ref{lem:pruning} supplies the $K$-pruning
property whenever the girth exceeds $4/h$. Hence
$\Lambda := 2K/\bar h \leq 2c_0\epsilon n$ and $\Lambda\geq 2K/\bar h\geq2$, so
\[
   k^{*}\Lambda \;\leq\; \frac{25e^{4}}{\epsilon}\cdot2c_0\epsilon n
   \;=\; 50e^{4}c_0\,n \;=\; \frac n4 \;\leq\; n ,
   \qquad\text{and}\qquad
   k^{*} \;\leq\; \frac{k^{*}\Lambda}{2} \;\leq\; \frac n8 \;\leq\; \frac n2 ,
\]
so $k^{*}$ and $\Lambda$ satisfy the hypotheses of Lemma~\ref{lem:kvs3} --- note that the
$\Lambda$ appearing there is $2K/h \leq 2K/\bar h$, so the bounds just derived apply a fortiori
--- and we use the witness $W^{\epsilon}$ of~\eqref{eq:witness-eps}.

\textbf{Step 1: eliminate the one-defect terms.}
Let $P$ be the sum of those $R_j$ with $\omega_1(R_j)>0$ and $H$ the sum of the rest, so
$T^{\epsilon}=P+H$. By Lemma~\ref{lem:witness-positive}, $\langle W^{\epsilon},H\rangle\geq1$;
the functional is applied to $H$ only, as \eqref{eq:witness-eps} records.

\textbf{Step 2: bound each remaining term.}
Every $R_j$ occurring in $H$ satisfies $\omega_1(R_j)=0$ and $R_j\leq T^{\epsilon}$ entrywise.
Write $\ell_j$ for its number of odd residual components; by Proposition~\ref{prop:oddcomp},
$\ell_j\geq3$ and $\ell_j$ is odd. We distinguish three cases.
\begin{itemize}
\item $\ell_j\geq5$. Then $\langle W^{\epsilon},R_j\rangle\leq0$ by
  Lemma~\ref{lem:local-sign}.
\item $\ell_j=3$ and $\omega_3(R_j)\geq2\epsilon|Q_3|/\tau$. Lemma~\ref{lem:kvs3} supplies a
  component $B$ of $G'(R_j)$ with $|V\setminus B|\leq\Lambda\leq n/k^{*}$, so again
  $\langle W^{\epsilon},R_j\rangle\leq0$ by Lemma~\ref{lem:local-sign}.
\item $\ell_j=3$ and $\omega_3(R_j)<2\epsilon|Q_3|/\tau$. Then
  $\langle W^{\epsilon},R_j\rangle\leq W^{\epsilon}_3\,\omega_3(R_j)
  <\frac{2}{\epsilon|Q_3|}\cdot\frac{2\epsilon|Q_3|}{\tau}=\frac4\tau$.
\end{itemize}

\textbf{Step 3: count.}
Let $N$ be the number of terms of the third kind. All other terms contribute at most $0$, so
$1\leq\langle W^{\epsilon},H\rangle\leq N\cdot\frac4\tau$, giving $N\geq\tau/4$ and hence
$r\geq N\geq\tau/4$.

Finally, taking $\tau$ as large as \eqref{eq:tau-condition} permits gives
$\log_2 r\geq c_0\epsilon\bar hn-\log_2\frac2\epsilon-2$, which is $\Omega(\epsilon\bar hn)$
whenever $\epsilon\bar hn\geq C\log_2(2/\epsilon)$ for a suitable constant $C$.
Under the girth assumption the required pruning property holds for this choice of $\tau$,
which proves the final assertion.
\end{proof}

\subsubsection{Removing the bias restriction by weighted refinement}
\label{sec:tseitin-refinement}

By Remark~\ref{rem:obstruction} the defect-ratio lemma does not apply to an arbitrary-bias
term. We refine each term exactly into uniform ones, where it does apply, and charge the
refinements by weight rather than by number: those the witness can score on fix many edges and
carry little weight, and the total weight of a term is capped by entrywise domination.

For $q\in[-1,1]$ let
\[
   \nu_q(x) \;:=\; \frac{1+q(-1)^{x}}{2},\qquad x\in\{0,1\},
\]
and let $b(q) = 0$ if $q\geq0$ and $b(q) = 1$ if $q<0$. Thus $\nu_q$ is the Bernoulli
distribution with bias $p = (1-q)/2$, and $b(q)$ is a modal bit: $\nu_q(b(q)) = (1+|q|)/2$ is
the larger of the two values. We write $\delta_b$ for the point mass at $b\in\{0,1\}$.

\begin{lem}[Canonical bounded-bias refinement]\label{lem:refinement}
For $q\in[-1,1]$, putting $a = |q|$,
\begin{equation}\label{eq:one-edge-refinement}
   \nu_q \;=\; a\,\delta_{b(q)} + (1-a)\,\nu_0 .
\end{equation}
Consequently, for $q = (q_e)_{e\in E}$ put $a_e = |q_e|$ and, for $F\subseteq E$,
\[
   \alpha_q(F) \;:=\; \prod_{e\in F}a_e\prod_{e\notin F}(1-a_e).
\]
If $\mu_q = \bigotimes_{e\in E}\nu_{q_e}$, and $\mu_{q,F}$ is the uniform distribution which
fixes every $e\in F$ to $b(q_e)$ and leaves every $e\notin F$ unbiased, then
\begin{equation}\label{eq:refinement}
   \mu_q \;=\; \sum_{F\subseteq E}\alpha_q(F)\,\mu_{q,F},
   \qquad
   \sum_{F\subseteq E}\alpha_q(F)\,2^{|F|} \;=\; \prod_{e\in E}(1+a_e).
\end{equation}
Terms of zero weight in the first sum are omitted.
\end{lem}

\begin{proof}
If $q\geq0$ then $\nu_q = q\delta_0+(1-q)\nu_0$; if $q\leq0$ then
$\nu_q = (-q)\delta_1+(1+q)\nu_0$. This is \eqref{eq:one-edge-refinement}. Tensoring these
identities gives the first identity in \eqref{eq:refinement}, and the second is
\[
   \sum_{F}\prod_{e\in F}(2a_e)\prod_{e\notin F}(1-a_e)
   \;=\;\prod_{e}(1-a_e+2a_e)\;=\;\prod_{e}(1+a_e). \qedhere
\]
\end{proof}

The next observation is the weight-sensitive ingredient. Write $m := |E|$ and put
\[
   \Lambda_0 \;:=\; \sum_{x}T^{\epsilon}(x) \;=\; 2^{m-1}(n-2\epsilon),
\]
the sum being $2^{m}(n/2-\epsilon)$ because flipping one edge at $v$ toggles the defect at $v$,
so each vertex is defective on exactly half the edge cube.

\begin{lem}[Modal domination]\label{lem:modal-domination}
Suppose
\[
   T^{\epsilon} \;=\; \sum_{j=1}^{r}\eta_j\,\mu_{q_j},
   \qquad \lambda_j := \eta_j/\Lambda_0,
\]
is an arbitrary-bias decomposition. Put $a_{j,e} = |q_{j,e}|$ and
$Z_j = \prod_{e}(1+a_{j,e})$. Then
\begin{equation}\label{eq:modal-domination}
   \lambda_jZ_j \;\leq\; M_\epsilon \;:=\; \frac{2(n-\epsilon)}{n-2\epsilon}
   \qquad\text{for every }j.
\end{equation}
\end{lem}

\begin{proof}
At the modal assignment $x^{*}_j = (b(q_{j,e}))_{e}$ each factor contributes
$\nu_{q_{j,e}}(b(q_{j,e})) = (1+a_{j,e})/2$, so $\mu_{q_j}(x^{*}_j) = 2^{-m}Z_j$. Every summand is nonnegative, so
$\eta_j\mu_{q_j}(x^{*}_j)\leq T^{\epsilon}(x^{*}_j)\leq n-\epsilon$. Substituting the value of
$\Lambda_0$ and rearranging gives \eqref{eq:modal-domination}.
\end{proof}

\begin{theorem}[Arbitrary-bias Tseitin lower bound]\label{thm:tseitin-lb-arbitrary}
Let $G = (V,E)$ be connected, bridgeless and $d$-regular with $d\geq3$, on $n$ vertices, with
edge expansion $h>0$ and girth greater than $4/h$. Put $\bar h = \min\{h,1\}$, and let the
Tseitin charge have odd total parity. If $0<\epsilon\leq1$ and
\begin{equation}\label{eq:arbitrary-bias-eps-condition}
   \epsilon n \;\geq\; 50e^{4},
\end{equation}
then
\begin{equation}\label{eq:arbitrary-bias-bound}
   \rkt\bigl(\chi(\TS_G)-\epsilon\cdot\chi(1)\bigr)
   \;\geq\;
   \frac{\epsilon}{2(n-\epsilon)}\,2^{\epsilon\bar hn/(50e^{4})} .
\end{equation}
The same bound holds for the size of every $\SHAMC^{\epsilon}$ proof of $\TS_G$, that is, every
general depth-three $\AMC^{\epsilon}$ proof.
\end{theorem}

\begin{remark}
The girth hypothesis enters only through Lemma~\ref{lem:pruning}: what the proof uses is the
$K$-pruning property for $K = \epsilon\bar hn/(50e^{4})$, supplied by Lemma~\ref{lem:pruning}
as in Theorem~\ref{thm:tseitin-lb-eps}, where $K$ is chosen differently.
\end{remark}

\begin{proof}
Let $k^{*}$ be the least odd integer at least $24e^{4}/\epsilon$, and use the witness
$W^{\epsilon}$ of \eqref{eq:witness-eps} with this choice of $k^{*}$. Then
$k^{*}\leq25e^{4}/\epsilon$, and \eqref{eq:arbitrary-bias-eps-condition} gives $k^{*}\leq n/2$.
Set
\begin{equation}\label{eq:X0-t}
   X_0 \;:=\; \frac{\epsilon n}{25e^{4}},
   \qquad
   K \;:=\; \frac{\epsilon\bar hn}{50e^{4}} .
\end{equation}
Thus $k^{*}X_0\leq n$.

Consider an arbitrary-bias decomposition
\begin{equation}\label{eq:arbitrary-decomposition}
   T^{\epsilon} \;=\; \sum_{j=1}^{r}\eta_j\,\mu_{q_j} .
\end{equation}
Apply Lemma~\ref{lem:refinement} to every summand and write
\[
   R_{j,F} \;:=\; \eta_j\alpha_j(F)\,\mu_{j,F},
   \qquad \alpha_j(F) := \alpha_{q_j}(F).
\]
This is an exact decomposition of $T^{\epsilon}$ into uniform rank-one terms. Put into $P$ all
refinements with positive one-defect mass and into $H$ all remaining refinements, and write
$H = \sum_{j=1}^{r}H_j$, where $H_j$ is the sum of the refinements of the $j$-th original term
which occur in $H$. Lemmas~\ref{lem:deficit} and~\ref{lem:witness-positive} ask only that
$T^{\epsilon} = P+H$ be a decomposition into uniform terms split by one-defect mass, and say
nothing about how many terms there are; they therefore apply verbatim to the refinement and give
\begin{equation}\label{eq:refined-witness-positive}
   \langle W^{\epsilon},H\rangle \;\geq\; 1 .
\end{equation}
In particular $P$ needs no further attention: \eqref{eq:refined-witness-positive} is a bound on
$H$ alone, and it is Lemma~\ref{lem:deficit} that keeps the $Q_3$ mass of $H$ large however
large $P$ is and however many refinements it contains. Since
$\langle W^{\epsilon},H\rangle = \sum_{j\leq r}\langle W^{\epsilon},H_j\rangle$, a bound
$\langle W^{\epsilon},H_j\rangle\leq c$ holding for each $j$ turns
\eqref{eq:refined-witness-positive} into $1\leq rc$, a lower bound on the number $r$ of
\emph{original} terms; the number of refinements never enters.

Fix an original term $j$. Every refinement of it which occurs in $H$ is entrywise at most
$T^{\epsilon}$, because it is a nonnegative summand of the exact refined decomposition, and it
has an odd number $\ell\geq3$ of odd residual components. By Lemma~\ref{lem:local-sign} its
witness pairing is nonpositive if $\ell\geq5$, or if $\ell = 3$ and the complement of its
largest residual component has size at most $X_0$. Call all remaining refinements \emph{bad}.
Since $\ell$ is odd, these two cases leave only $\ell = 3$: a bad refinement has exactly three
odd components and either has no component larger than $n/2$, or has a unique largest component
$B$, with $X = V\setminus B$ satisfying $|X|>X_0$.

We claim that every bad refinement, with fixed-edge set $F$, has cycle-rank drop
\begin{equation}\label{eq:bad-rank-drop}
   \rho(F) \;:=\; \dim Z(G)-\dim Z(G-F) \;\geq\; K .
\end{equation}
Since $\epsilon,\bar h\leq1$ and $50e^{4}>8$, we have $K<\min\{h,2\}n/8$, so
Lemma~\ref{lem:pruning} gives $G$ the $K$-pruning property. If $G-F$ has no component larger
than $n/2$, the contrapositive of that property gives $\rho(F)>K$ at once. If it has a
component $B$ larger than $n/2$, put
$X = V\setminus B$; all edges between $X$ and $B$ belong to $F$, so Lemma~\ref{lem:cutcost}
gives
\[
   \rho(F) \;\geq\; \frac h2|X|
   \;>\; \frac{h\epsilon n}{50e^{4}}
   \;\geq\; K .
\]
This proves \eqref{eq:bad-rank-drop}.

Suppose now that $F$ is a bad refinement and its three odd components are $C_1,C_2,C_3$. Put
\[
   g(F) \;:=\; \frac{|C_1||C_2||C_3|}{\binom n3} \;\leq\; 1 .
\]
Write $Q_3(j,F) := Q_3\cap\mathrm{supp}\,\mu_{j,F}$. The one-defect-per-odd-component count and
Lemma~\ref{lem:strata} give
\begin{equation}\label{eq:refined-Q3-count}
   \frac{|Q_3(j,F)|}{|Q_3|} \;=\; 2^{-\rho(F)}g(F).
\end{equation}
Using $\eta_j = \Lambda_0\lambda_j$, the fact that $\mu_{j,F}$ is uniform on the subcube
fixing the $|F|$ edges of $F$ and so has coefficient $2^{-(m-|F|)}$ at each of its
$2^{m-|F|}$ support points, and \eqref{eq:refined-Q3-count}, we obtain the exact identity
\begin{equation}\label{eq:refined-omega3}
   \frac{\omega_3(R_{j,F})}{|Q_3|}
   \;=\; \frac{n-2\epsilon}{2}\,\lambda_j\alpha_j(F)\,2^{|F|-\rho(F)}g(F).
\end{equation}

Let $H^{\mathrm{bad}}_j$ be the sum of the bad refinements of the $j$-th original term. Summing
\eqref{eq:refined-omega3}, using \eqref{eq:bad-rank-drop}, $g(F)\leq1$, the second identity in
\eqref{eq:refinement}, and Lemma~\ref{lem:modal-domination}, gives
\[
   \frac{\omega_3(H^{\mathrm{bad}}_j)}{|Q_3|}
   \;\leq\; \frac{n-2\epsilon}{2}\,\lambda_j2^{-K}\sum_{F}\alpha_j(F)2^{|F|}
   \;=\; \frac{n-2\epsilon}{2}\,\lambda_jZ_j2^{-K}
   \;\leq\; \frac{n-2\epsilon}{2}\,M_\epsilon2^{-K} .
\]
All nonbad refinements in $H_j$ pair nonpositively with the witness, and the negative
$k^{*}$-stratum can only decrease the pairing of a bad refinement. Therefore
\begin{equation}\label{eq:refined-per-term}
   \langle W^{\epsilon},H_j\rangle
   \;\leq\; \frac{2}{\epsilon|Q_3|}\,\omega_3(H^{\mathrm{bad}}_j)
   \;\leq\; \frac{2(n-\epsilon)}{\epsilon}\,2^{-K} .
\end{equation}
Finally sum \eqref{eq:refined-per-term} over the $r$ original terms and combine it with
\eqref{eq:refined-witness-positive}:
\[
   1 \;\leq\; \langle W^{\epsilon},H\rangle
     \;\leq\; r\,\frac{2(n-\epsilon)}{\epsilon}\,2^{-K} .
\]
Rearranging and substituting the definition of $K$ gives \eqref{eq:arbitrary-bias-bound}.

For the proof-system conclusion, Corollary~\ref{lem:autosml} turns every general depth-three
monotone expression computing $T^{\epsilon}$ into an arbitrary-bias tensor decomposition with
at most as many rank-one terms as the proof has size.
\end{proof}

\begin{corollary}[Depth-three regimes]\label{cor:tseitin-depth-three-regimes}
Let $G$ range over an explicit family of fixed-degree, bridgeless expanders with
$h = \Omega(1)$ and girth greater than $4/h$, and give each graph any explicitly chosen odd
charge. Then every $\SHAMC^{\epsilon}$ proof of $\TS_G$
has size at least
\[
   \frac{\epsilon}{2(n-\epsilon)}\,2^{\epsilon\bar hn/(50e^{4})} .
\]
In particular:
\begin{itemize}
  \item if $\epsilon = \omega(\log n/n)$, the size is $2^{\Omega(\epsilon hn)} = n^{\omega(1)}$;
  \item if $\epsilon = C\log_2 n/n$, then the explicit estimate in
        Theorem~\ref{thm:tseitin-lb-arbitrary} is $n^{C\bar h/(50e^{4})-2+o(1)}$;
  \item for constant $\epsilon$, the size is $2^{\Omega(n)}$.
\end{itemize}
Such families exist. For distinct primes $p,q\equiv1\pmod4$ with $p$ a quadratic residue
mod $q$, the graphs $X^{p,q}$ of Lubotzky, Phillips and Sarnak
\cite{LubotzkyPhillipsSarnak88} are $(p+1)$-regular on $n = q(q^{2}-1)/2$ vertices, Ramanujan,
so that every nontrivial eigenvalue is at most $2\sqrt p$, and of girth exceeding
$\tfrac23\log_p n$. The spectral bound $h\geq(d-\lambda)/2$
\cite{HooryLinialWigderson06} gives $h\geq(\sqrt p-1)^{2}/2$, which for $p\geq5$ exceeds
$\tfrac12$ and so also makes these graphs bridgeless; and their girth, being
$\Theta(\log n)$, exceeds the constant $4/h$ for all large $n$.
\end{corollary}

\subsubsection{A polynomial-size proof at exponentially small shift}\label{sec:tiny-shift}

The bounds above cannot be extended to all $\epsilon>0$: below an
exponentially small threshold the Tseitin target has an explicit arbitrary-bias decomposition
with $O_d(n^{3})$ terms.

Work in edge-sign variables $z_e\in\{-1,1\}$ and write
$s_v(z) = (-1)^{c(v)}\prod_{e\ni v}z_e$, so that $B_v := (1-s_v)/2$ is the defect indicator at
$v$ and $T^{\epsilon} = \sum_{v\in V}B_v-\epsilon$. By Lemma~\ref{lem:pm-rank-one} the rank-one terms
of~\eqref{eq:prodmix} are the nonnegative products $\prod_{e\in E}g_e(z_e)$, so the least number
of those summing to $T^{\epsilon}$ is $\rkt(T^{\epsilon})$.

\begin{lem}[Telescoping the global inconsistency]\label{lem:telescope}
Order $V = \{v_1,\dots,v_n\}$ and put $H_0 := 1$ and $H_i := \prod_{j\leq i}s_{v_j}$. Then
$\sum_{i=1}^{n}B_{v_i}H_{i-1} = 1$, and consequently
\begin{equation}\label{eq:telescope}
   T^{\epsilon} \;=\; \sum_{i=1}^{n}B_{v_i}\bigl(1-\epsilon H_{i-1}\bigr) .
\end{equation}
Each $H_{i-1}$ is $\pm$ the parity of the cut $\delta(\{v_1,\dots,v_{i-1}\})$.
\end{lem}

\begin{proof}
Odd total charge gives $H_n = (-1)^{\sum_vc(v)}\prod_{e}z_e^{2} = -1$, and
$2B_{v_i}H_{i-1} = (1-s_{v_i})H_{i-1} = H_{i-1}-H_i$, so the sum telescopes to
$(H_0-H_n)/2 = 1$. Substituting $\epsilon = \epsilon\sum_iB_{v_i}H_{i-1}$ into
$T^{\epsilon} = \sum_iB_{v_i}-\epsilon$ gives~\eqref{eq:telescope}. In $H_{i-1}$ every edge
internal to $\{v_1,\dots,v_{i-1}\}$ occurs twice and cancels.
\end{proof}

Everything therefore reduces to decomposing a very weakly biased parity.

\begin{lem}[Weak parity has quadratic product rank]\label{lem:weak-parity}
For $s\geq2$ put $\ell := s-1$ and
\[
   \delta_s \;:=\; \frac{\ell!}{s^{2}\,2^{\ell}\,\ell^{2\ell}} \;=\; 2^{-\Theta(s\log s)} .
\]
If $|\theta|\leq\delta_s$ then $1+\theta\prod_{i=1}^{s}z_i$ is a sum of exactly $s^{2}$
nonnegative products. The construction is a finite-difference moment-matching argument in the
spirit of the mixtures of product distributions of \cite{ChenMoitra18}.
\end{lem}

\begin{proof}
For $j = 0,\dots,\ell$ set $t_j := j/\ell^{2}$ and
$a_j := (-1)^{\ell-j}\binom{\ell}{j}2^{-\ell}$, let
$\Pi_{j,k} := \prod_{i=k}^{\ell}(1+t_jz_i)$ for $1\leq k\leq\ell+1$, so that
$\Pi_{j,\ell+1} = 1$, and put
\[
   P_j \;:=\; \Pi_{j,1},
   \qquad
   R_{j,k} \;:=\; (1-kt_jz_k)\,\Pi_{j,k+1}
   \qquad (1\leq k\leq\ell) .
\]
Since $\Pi_{j,k} = (1+t_jz_k)\Pi_{j,k+1}$ one has
$t_jz_k\Pi_{j,k+1} = \Pi_{j,k}-\Pi_{j,k+1}$, so
$R_{j,k} = (k+1)\Pi_{j,k+1}-k\Pi_{j,k}$ and $P_j+\sum_{k=1}^{\ell}R_{j,k} = \ell+1 = s$. Expanding
$P_j = \sum_{S\subseteq[\ell]}t_j^{|S|}z_S$, the coefficient of $z_S$ in $\sum_ja_jP_j$ is
$\sum_ja_jt_j^{|S|}$, and substituting $t_j = j/\ell^{2}$ gives, for every $p\geq0$,
\[
   \sum_{j=0}^{\ell}a_jt_j^{p}
   \;=\; \frac{1}{2^{\ell}\ell^{2p}}\sum_{j=0}^{\ell}(-1)^{\ell-j}\binom{\ell}{j}j^{p}
   \;=\; \frac{\Delta^{\ell}[x^{p}](0)}{2^{\ell}\ell^{2p}},
\]
where $\Delta f(x) := f(x+1)-f(x)$ is the forward difference with unit step and
$\Delta^{\ell}$ its $\ell$-fold iterate; writing $\Delta = S-I$ for the shift
$(Sf)(x) = f(x+1)$ and expanding the binomial in the commuting operators $S,I$ gives
$\Delta^{\ell}f(0) = \sum_{j=0}^{\ell}(-1)^{\ell-j}\binom{\ell}{j}f(j)$.
Applied to $f = \sum_{i\leq m}c_ix^{i}$ of degree $m\geq1$, the operator $\Delta$ gives
$\sum_{i\leq m}c_i\bigl[(x+1)^{i}-x^{i}\bigr]$, in which each bracket has degree $i-1$;
so $\Delta f$ has degree exactly $m-1$ with leading coefficient $c_mm$, and $\Delta$
annihilates constants. Iterating $\ell$ times therefore annihilates every polynomial of
degree $<\ell$ and returns the constant $c_{\ell}\ell!$ on one of degree $\ell$: in
particular $\Delta^{\ell}[x^{p}](0) = 0$ for $p<\ell$ and $=\ell!$ for $p = \ell$. Hence only $S = [\ell]$ survives and
$\sum_ja_jP_j = \frac{\ell!}{2^{\ell}\ell^{2\ell}}\prod_{i\leq\ell}z_i$. Combining the two, for
every $u\in[-1,1]$,
\begin{equation}\label{eq:weak-parity-identity}
   \frac{1}{s^{2}}\sum_{j=0}^{\ell}
   \Bigl[(1+ua_jz_s)P_j+\sum_{k=1}^{\ell}R_{j,k}\Bigr]
   \;=\; 1+u\,\delta_s\prod_{i=1}^{s}z_i ,
\end{equation}
a sum of $s(\ell+1) = s^{2}$ products. Every one-coordinate factor is nonnegative on
$\{-1,1\}$, since $t_j\leq1/\ell$, $kt_j\leq1$ and $|ua_j|\leq1$. Take $u = \theta/\delta_s$.
\end{proof}

\begin{theorem}[Tiny-shift upper bound]\label{thm:tseitin-tiny-shift}
Let $G$ have maximum degree $d$ and odd total Tseitin charge, put $m := |E| \geq 2$ and
$\epsilon_0(m) := \delta_m = 2^{-\Theta(m\log m)}$. Then
\[
   \rkt\bigl(\chi(\TS_G)-\epsilon\cdot\chi(1)\bigr) \;\leq\; 2^{d-1}\,|V|\,|E|^{2}
   \qquad\text{for every }0<\epsilon\leq\epsilon_0(m),
\]
which for fixed degree is $O(n^{3})$ product terms, hence an $\SHAMC^{\epsilon}$ proof of
$\TS_G$ of that size.
\end{theorem}

\begin{proof}
In \eqref{eq:telescope} the $i$-th factor $1-\epsilon H_{i-1}$ is $1+\theta\prod_{e\in S}z_e$
with $S = \delta(\{v_1,\dots,v_{i-1}\})$ and $|\theta| = \epsilon$. Here $|S|\leq m$, and
$\delta_s$ decreases in $s$, so the hypothesis $\epsilon\leq\delta_m$ gives $\epsilon\leq\delta_{|S|}$. For $|S|\geq2$ therefore
Lemma~\ref{lem:weak-parity} writes that factor as a sum of $|S|^{2}\leq m^{2}$ nonnegative
products in the variables $z_e$, $e\in S$ (with $g_e\equiv1$ off $S$). The two degenerate cuts are handled directly and more
cheaply: $S = \emptyset$, which occurs at $i = 1$, leaves the constant $1-\epsilon>0$, a single
nonnegative product, and $|S| = 1$ leaves $1+\theta z_e$ with $|\theta|\leq1$, again a single
one. Each $B_{v_i}$ is the sum of the $2^{\deg v_i-1}$ locally violating complete
patterns $\pi$ on $\delta(v_i)$, and each such pattern is itself the nonnegative product
$\prod_{e\in\delta(v_i)}\mathbf 1[z_e = \pi_e]$. Nonnegative products are closed under
multiplication, since at each edge the two factors combine into the single nonnegative
one-coordinate function $g_e(z_e)\mathbf 1[z_e = \pi_e]$; so every product obtained above is
by Lemma~\ref{lem:pm-rank-one} a rank-one set-multilinear term. Summing over $i$
gives at most $n2^{d-1}m^{2}$ products.
\end{proof}

Two consequences. The first is a separation inside depth three.

\begin{corollary}[Bounded and arbitrary bias separate at tiny shift]\label{cor:tiny-shift-separation}
Let $G$ be a constant-degree expander and $\epsilon\leq2^{-\Theta(n\log n)}$. Every $\AMCb$
proof of $\TS_G$ has size $2^{\Omega(n)}$ by Corollary~\ref{cor:all-eps-amcb}, while $\SHAMC$
has one of size $O(n^{3})$. The two fragments are therefore exponentially separated.
\end{corollary}

The second concerns the quantifier over $\epsilon$.

\begin{remark}[Tseitin cannot settle the $\epsilon$ quantifier]\label{rem:tseitin-not-all-eps}
By Theorem~\ref{thm:tseitin-tiny-shift} no superpolynomial lower bound on $\SHAMC$ proof size
for $\TS_G$ holds uniformly in $\epsilon>0$, whatever the method; and by
Theorem~\ref{thm:mu0-upper} no superpolynomial lower bound on $\AMC^{\epsilon}$ proof size holds
for $\TS_G$ at any $\epsilon\in(0,1]$ once the depth is unrestricted, nor at any fixed
mixed-product depth. An unconditional $\AMC$ lower
bound must therefore be sought on a different formula. Section~\ref{sec:selector-lift} shows how
little has to change: an index gadget of constant size applied to $\TS_G$ itself is hard at
every $\epsilon\geq2^{-\Omega(N)}$. What the
Tseitin bounds of this section do settle is the behaviour above $\epsilon = \Omega(\log n/n)$,
and the true threshold between that scale and $2^{-\Theta(n\log n)}$ is open.
\end{remark}

\subsection{A proof of mixed-product depth zero at every shift}\label{sec:mu0-upper}

Both bounds of Section~\ref{sec:tseitin-lb} are bounds on $\rkt$, that is on depth three, and
Section~\ref{sec:tiny-shift} shows that even there they cannot reach every $\epsilon$. Once the
depth is not restricted the picture changes completely: $\TS_G$ has polynomial-size proofs at
\emph{every} $\epsilon\in(0,1]$, and they can be arranged to have mixed-product depth zero with
respect to any bipartition of the edges. The ingredient is that the parities which the
telescoping of Lemma~\ref{lem:telescope} leaves behind have small monotone set-multilinear
formulas; what Lemma~\ref{lem:weak-parity} had to buy with an exponentially small $\theta$ is
free as soon as logarithmic depth is allowed.

\begin{lem}[Parity polynomials]\label{lem:parity-formula}
For $S\subseteq E$ and $b\in\{0,1\}$ put
\[
   P^{b}_{S} \;:=\;
   \sum_{\substack{\sigma\in\{0,1\}^{S}\\ \bigoplus_{e\in S}\sigma_e = b}}\ \prod_{e\in S}x_e^{\sigma_e},
\]
the set-multilinear tensor of the indicator of $\bigoplus_{e\in S}\sigma_e = b$, with one
variable block $\{x_e^{0},x_e^{1}\}$ per edge $e\in S$. Each of $P^{0}_{S}$ and $P^{1}_{S}$
is computed by a monotone syntactically set-multilinear formula with $O(|S|^{2})$ leaves and
depth $O(\log|S|)$.
\end{lem}

\begin{proof}
For a partition $S = A\sqcup B$,
\begin{equation}\label{eq:parity-split}
   P^{0}_{S} \;=\; P^{0}_{A}P^{0}_{B}+P^{1}_{A}P^{1}_{B},
   \qquad
   P^{1}_{S} \;=\; P^{0}_{A}P^{1}_{B}+P^{1}_{A}P^{0}_{B} .
\end{equation}
Halving $S$ repeatedly therefore gives both polynomials, starting from $x_e^{0},x_e^{1}$ at
$|S| = 1$. Writing $L(s)$ for the total number of leaves of the two formulas when $|S| = s$, one has
$L(1) = 2$ and $L(s)\leq4L(s/2)$, so
$L(s) = O(s^{2})$, and the depth grows by $O(1)$ per halving.
\end{proof}

\begin{theorem}[Polynomial size at mixed-product depth zero, for every shift]\label{thm:mu0-upper}
Let $G$ have maximum degree $d$ and odd total Tseitin charge, and put $m := |E|\geq2$. Fix a
bipartition of the edge blocks into $\mathcal X\sqcup\mathcal Y$. Then for every
$\epsilon\in(0,1]$ the target $T^{\epsilon} = \chi(\TS_G)-\epsilon\cdot\chi(1)$ is computed by a
monotone syntactically set-multilinear formula with $O(2^{d}nm^{2})$ leaves, of depth
$O(d+\log m)$, whose mixed-product depth with respect to that bipartition is zero.
\end{theorem}

\begin{proof}
Write $D_i := \delta(\{v_1,\dots,v_{i-1}\})$ and $C_{i-1} := \sum_{j<i}c(v_j)$, so that
$H_{i-1} = (-1)^{C_{i-1}}\prod_{e\in D_i}z_e$ by Lemma~\ref{lem:telescope}. Expand $B_{v_i}$
into the $2^{\deg v_i-1}$ local assignments $\pi$ of $\delta(v_i)$ that violate the charge at
$v_i$. Fixing such a $\pi$ fixes the coordinates of $D_i\cap\delta(v_i)$. Put
$S_i := D_i\setminus\delta(v_i)$ and
$b(i,\pi) := C_{i-1}+\sum_{e\in D_i\cap\delta(v_i)}\pi_e \bmod 2$; the factor
$1-\epsilon H_{i-1}$ then equals $1-\epsilon$ at the assignments of $S_i$ of parity
$b(i,\pi)$ and $1+\epsilon$ at the others. The term indexed by
$(i,\pi)$ in the resulting expansion of \eqref{eq:telescope} thus involves $\delta(v_i)$,
$S_i$ and the remaining blocks separately, so it is a product of tensors:
\begin{equation}\label{eq:mu0-formula}
   T^{\epsilon}
   \;=\;
   \sum_{i=1}^{n}\ \sum_{\pi}\
   \Bigl(\prod_{e\in\delta(v_i)}x_e^{\pi_e}\Bigr)
   \Bigl((1-\epsilon)P^{\,b(i,\pi)}_{S_i}+(1+\epsilon)P^{\,1-b(i,\pi)}_{S_i}\Bigr)
   \prod_{e\notin\delta(v_i)\cup S_i}\bigl(x_e^{0}+x_e^{1}\bigr),
\end{equation}
with nonnegative coefficients because $\epsilon\leq1$, and syntactically set-multilinear
because sums have children of equal block support and products of disjoint block support. Of its
at most $n2^{d-1}$ summands each costs $O(m^{2})$ leaves and depth $O(\log m)$ by
Lemma~\ref{lem:parity-formula}, whence the stated size and depth $O(d+\log m)$.

For the mixed-product depth put $A := S_i\cap\mathcal X$ and $B := S_i\cap\mathcal Y$ and
apply \eqref{eq:parity-split} to both parity polynomials of the $(i,\pi)$ summand of
\eqref{eq:mu0-formula}: its bracket becomes
$\sum_{b_1,b_2}c_{b_1b_2}P^{b_1}_{A}P^{b_2}_{B}$, one term for each of the four pairs, with
$c_{b_1b_2} = 1-\epsilon$ when $b_1\oplus b_2 = b(i,\pi)$ and $1+\epsilon$ otherwise. Every
fixed literal $x_e^{\pi_e}$ and every free block $x_e^{0}+x_e^{1}$ involves a single block, so
lies on one side; distributing these over the four terms and computing $P^{b_1}_{A}$,
$P^{b_2}_{B}$ inside their own sides by Lemma~\ref{lem:parity-formula} writes the summand as a
sum of four products $c_{b_1b_2}F\cdot G$, with $F$ built from variables of blocks in
$\mathcal X$ alone and $G$ from blocks in $\mathcal Y$ alone. The
support of a gate is the union of the blocks occurring below it, so no gate inside $F$ or inside
$G$ is mixed; and the top gate of each term multiplies the two one-sided polynomials $F$ and
$G$. No product gate therefore has two mixed children, so the mixed-product depth is zero.
\end{proof}

For fixed degree the formula of Theorem~\ref{thm:mu0-upper} has $O(n^{3})$ leaves and depth
$O(\log n)$. Taking it together with the circuits of
Section~\ref{sec:encoding} for $\chi(\TS_G)$ and $\chi(1)$ yields one monotone circuit, still
of size $O(n^{3})$ and depth $O(\log n)$, which computes the three polynomials $\chi(1)$,
$\chi(\TS_G)$ and $\chi(\TS_G)-\epsilon\cdot\chi(1)$ simultaneously. By
Theorem~\ref{thm:AMC-size}(b), $\chi(1)\amcle\chi(\TS_G)$ then has an $\AMC$ derivation of
size $O(n^3)$. The depth bound refers to the displayed certificate formula.

\begin{corollary}[The flattening bound is empty for Tseitin]\label{cor:tseitin-flat-easy}
For fixed degree and every $\epsilon\in(0,1]$, every bipartite flattening of $T^{\epsilon}$ has
nonnegative rank $O(n^{3})$. So \eqref{eq:method} gives no superpolynomial bound for $\TS_G$ at
any mixed-product depth, and in particular the exponential bounds of
Section~\ref{sec:tseitin-lb} cannot be obtained from a flattening.
\end{corollary}

\begin{proof}
Theorem~\ref{thm:mu0-upper} supplies, for each bipartition, a formula of mixed-product depth
zero with $L = O(n^{3})$ leaves, so Theorem~\ref{thm:mixed-depth-flattening} gives
$\rkp(\Flat(T^{\epsilon}))\leq L^{2^{0}} = O(n^{3})$.
\end{proof}

\begin{remark}[Depth three versus mixed-product depth zero]\label{rem:mu0-vs-depth3}
Depth three implies mixed-product depth zero (Remark~\ref{rem:t0}), and the converse now fails
exponentially: on a constant-degree expander at constant $\epsilon$,
Theorem~\ref{thm:tseitin-lb-arbitrary} gives $\rkt(T^{\epsilon}) = 2^{\Omega(n)}$ while
Theorem~\ref{thm:mu0-upper} gives a formula of mixed-product depth zero with $O(n^{3})$ leaves.
At mixed-product depth zero the flattening bound \eqref{eq:method} is therefore strictly weaker
than the tensor rank bound, and mixed-product depth organises a hierarchy of its own, not the
hierarchy of ordinary depth: the formula above has depth $O(\log n)$ and mixed-product depth
zero simultaneously. The
$\epsilon$-sensitive Tseitin bounds measure tensor rank, which is depth three, and nothing
beyond it.
\end{remark}

\subsection{Lifting Sherali--Adams degree to \texorpdfstring{$\SHAMC$}{Sigma Pi Sigma-AMC} size at exponentially small shift}\label{sec:selector-lift}

Remark~\ref{rem:tseitin-not-all-eps} leaves the quantifier over $\epsilon$ to be settled on a
formula other than Tseitin. We give one whose admissible range reaches exponentially small
shifts, and it is obtained from Tseitin by an index gadget. The statement is about $\rkt$, hence about depth-three proof size, and only about that:
unlike Section~\ref{sec:polytopes} it yields no bipartite flattening bound and therefore does
not feed Theorem~\ref{thm:mixed-depth-flattening}.

\paragraph{Instances, Fourier notation and local expectation functionals.}
This subsection and Section~\ref{sec:csp-dnf} both use a Tseitin base instance that fools a
Sherali--Adams relaxation, and the objects involved are fixed here once, for both.

A \emph{normalized Boolean Max-$k$-CSP instance} on $m$ variables is identified with the function
$I:\{-1,1\}^{m}\to[0,1]$ sending an assignment to the \emph{fraction} of its constraints that the
assignment satisfies; $\mathrm{opt}(I) := \max_xI(x)$, so that $I$ is unsatisfiable exactly when
$\mathrm{opt}(I)<1$.

For a finite coordinate set $V$ and $\alpha\subseteq V$ write $x^{\alpha}:=\prod_{i\in\alpha}x_i$,
with $x^{\emptyset}\equiv1$. These $2^{|V|}$ monomials are an orthonormal basis of the real functions on $\{-1,1\}^{V}$ for the inner product
$\langle f,g\rangle=\mathbb E[fg]$, every expectation here and below being over a uniform
$x\in\{-1,1\}^{V}$. Every such $g$ therefore has a
unique expansion $g=\sum_{\alpha\subseteq V}\widehat g(\alpha)\,x^{\alpha}$ with coefficients
$\widehat g(\alpha)=\mathbb E[g\,x^{\alpha}]$, its \emph{Fourier coefficients}. The \emph{Fourier
degree} of $g$ is the largest $|\alpha|$ with $\widehat g(\alpha)\neq0$, and
\[
   g^{\leq d}\;:=\;\sum_{|\alpha|\leq d}\widehat g(\alpha)\,x^{\alpha}
\]
is the \emph{degree-$d$ truncation} of $g$. We write
$\binom{m}{\leq d} := \sum_{j\leq d}\binom mj$ for the number of monomials of degree at most
$d$ on $m$ coordinates.

A \emph{degree-$d$ local expectation functional} is a linear functional $\widetilde{\mathbb E}$ on
the real functions on $\{-1,1\}^{m}$ of Fourier degree at most $d$ with
$\widetilde{\mathbb E}\,1 = 1$ which is nonnegative on every nonnegative function depending on at
most $d$ of the coordinates. Such a functional is a relaxed substitute for an average over genuine
assignments: it need not come from a distribution, but it cannot be told apart from one by looking
at $d$ coordinates at a time. The tilde separates it from the genuine expectation $\mathbb E$
under the uniform measure, which keeps that meaning throughout; in particular the Fourier
coefficients $\widehat g(\alpha)$ above are taken against $\mathbb E$ and not against
$\widetilde{\mathbb E}$. A \emph{$d$-junta} is a function depending on at most $d$
coordinates. Only nonnegative ones occur below, as in the LP framework of
\cite{ChanLeeRaghavendraSteurer16}, and \emph{$d$-junta} always means a nonnegative one here.
These are not the conical juntas of Section~\ref{sec:juntas}.

These functionals are the degree-$d$ Sherali--Adams solutions, which is what connects the
selector lift below and the planted family of Section~\ref{sec:csp-dnf} to the
Sherali--Adams hierarchy. A $d$-junta $f$ depending on the $d$-set $T$ is
$\sum_{\sigma}f(\sigma)\mathbf 1[x|_T=\sigma]$ with nonnegative coefficients, and each
indicator $\mathbf 1[x|_T=\sigma]$ is a product of at most $d$ literals, hence of degree at most
$d$. Nonnegativity on the $d$-juntas therefore says exactly that
$p_T(\sigma) := \widetilde{\mathbb E}\,\mathbf 1[x|_T=\sigma]\geq0$ for every $d$-set $T$ and
every $\sigma$. As these indicators sum to $1$ and $\widetilde{\mathbb E}\,1 = 1$, each $p_T$ is
a probability distribution on $\{-1,1\}^{T}$, whose moments are the numbers
$\widetilde{\mathbb E}\,x^{\alpha}$ with $\alpha\subseteq T$. Distributions belonging to
different $d$-sets agree on the overlap, but need not be the marginals of any single distribution
on all the coordinates.
Finally, $\SA_d(I)$ is the degree-$d$ Sherali--Adams value of $I$, the maximum of
$\widetilde{\mathbb E}I$ over the degree-$d$ local expectation functionals, so that $\SA_d(I)=1$
says the degree-$d$ relaxation is fooled into valuing $I$ at one.

\paragraph{The base.}
Let $I_0$ be an unsatisfiable normalized Max-$k$-CSP instance on $y_1,\dots,y_m\in\{-1,1\}$ with
$L$ constraints, let $B_\ell$ be the indicator that the $\ell$-th constraint is violated, and put
$\mathcal V := \sum_{\ell\leq L}B_\ell = L\,(1-I_0)$, the number of violated constraints, so that
$1\leq\mathcal V\leq L$. Fix $k\leq d<m$ and assume $\SA_d(I_0) = 1$, witnessed by a degree-$d$ local
expectation functional $\widetilde{\mathbb E}$; in terms of $\mathcal V$ this reads
\[
   \widetilde{\mathbb E}1 = 1, \qquad \widetilde{\mathbb E}\mathcal V = 0 .
\]
\paragraph{The lift.}
Fix $b\geq1$ and put $q := 2^{b}$. As in Section~\ref{sec:tiny-shift} we use $\pm1$ variables,
and the lifted formula has two kinds of them: for each base variable $u$ there are $b$
\emph{selector bits}, taking values in $\{-1,1\}$, whose joint value is read as an address
$a_u\in[q]$; and $q$ \emph{table bits} $z_{u,1},\dots,z_{u,q}\in\{-1,1\}$ private to $u$. Thus
there are
\[
   N \;=\; m(q+b)
\]
variables in all. An assignment to them is a pair $(a,z)$, where $a = (a_1,\dots,a_m)\in[q]^{m}$
collects the address of every base variable as read off its own selector bits, and it
\emph{induces} the base assignment
\[
   y_u \;:=\; z_{u,a_u} \;\in\; \{-1,1\},
\]
the entry of $u$'s table at $u$'s own address; $y$ is a function of the lifted assignment, not a
variable of the lifted formula. Equivalently $y_u = \mathrm{IND}_q(a_u,z_u)$ for the index
gadget $\mathrm{IND}_q(x,y) := y_x$ on $[q]\times\{-1,1\}^{q}$, the address $x = a_u$ being
given in binary by $u$'s $b$ selector bits; the construction below is thus the composition
$I_0\circ\mathrm{IND}_q^{\,m}$, one gadget per base variable.

For each constraint with scope $\{u_1,\dots,u_t\}$, each tuple of addresses
$(a_{u_1},\dots,a_{u_t})\in[q]^{t}$ and each pattern
$(\sigma_1,\dots,\sigma_t)\in\{-1,1\}^{t}$ violating that constraint, take the conjunction which
fixes the $tb$ selector bits of the scope to those addresses and fixes the $t$ addressed table
bits $z_{u_1,a_{u_1}},\dots,z_{u_t,a_{u_t}}$ to that pattern, and let $A$ be the disjunction of
all of them. Such a conjunction holds at $(a,z)$ exactly when the scope has those addresses and
the induced $y$ falsifies that constraint through that pattern; so each violated base constraint
contributes exactly one satisfied conjunction and each satisfied one none, giving
\begin{equation}\label{eq:selector-count}
   c_A(a,z) \;=\; \mathcal V(y) \;=\; \mathcal V\bigl(z_{1,a_1},\dots,z_{m,a_m}\bigr) .
\end{equation}
Hence $A$ is a tautology, with at most $L2^{k}q^{k}$ terms of width at most $k(b+1)$.

Fixing the selector bits to spell a given $a$ restricts the lifted instance, by
\eqref{eq:selector-count}, to $I_0$ evaluated on the $m$ addressed table bits
$z_{1,a_1},\dots,z_{m,a_m}$.
The lift is therefore $q^{m}$ copies of $I_0$, one for each address tuple and none of them
distinguished, and a purported decomposition of $c_A-\epsilon$ has to serve all of them at once;
the proof below selects a copy only after seeing the decomposition.

\begin{theorem}[Selector-lift lower bound]\label{thm:selector-lift}
With $I_0$, $\widetilde{\mathbb E}$, $k$, $d$, $q$ and $A$ as above, for every
$\epsilon>0$ for which $c_A-\epsilon$ is nonnegative,
\begin{equation}\label{eq:selector-lift}
   \rkt\bigl(\chi(A)-\epsilon\cdot\chi(1)\bigr) \;\geq\; \frac{\epsilon\,q^{\,d+1}}{L\,\binom{m}{\leq d}} .
\end{equation}
More generally the same bound holds for the number of terms in any identity
$c_A(a,z)-\epsilon = \sum_{j=1}^{r}\lambda_j(a)Q_j(z)$ in which each $\lambda_j$ is an arbitrary
nonnegative function of all the selector bits and each $Q_j$ a nonnegative product over the
individual table bits.
\end{theorem}

\begin{proof}
We prove the general form of the statement, for an identity
$c_A(a,z)-\epsilon = \sum_{j\leq r}\lambda_j(a)Q_j(z)$ with each $\lambda_j\geq0$ arbitrary on
the selector bits and each $Q_j$ a nonnegative product over the table bits; a tensor
decomposition into $r$ rank-one set-multilinear terms is the special case in which each
$\lambda_j$ is itself a product over the selector bits. All coordinates are $\pm1$-valued and
each $Q_j$ is written in the corresponding form: discard identically zero terms and normalise each table factor to have uniform average one,
absorbing its average into $\lambda_j$, so that
\[
   Q_j(z) \;=\; \prod_{u=1}^{m}\prod_{t=1}^{q}\bigl(1+\beta_{j,u,t}z_{u,t}\bigr),
   \qquad |\beta_{j,u,t}|\leq1 .
\]
Every nonnegative product over the table bits can be written this way, by
Lemma~\ref{lem:pm-rank-one}, the scalar $\eta'$ of \eqref{eq:pm-rank-one} being what has been
absorbed into $\lambda_j$, and $Q_j$ is $2^{mq}$ times a product distribution, in the bias
parametrisation of Lemma~\ref{lem:refinement}. Write $h_{j,u,t} := |\beta_{j,u,t}|$ and let
$Z_j := \max_{z}Q_j(z) = \prod_{u,t}(1+h_{j,u,t})$ be the value of $Q_j$ at its modal
assignment $z_{u,t} = \operatorname{sgn}\beta_{j,u,t}$.
All summands are nonnegative and $c_A-\epsilon\leq L$, so evaluating at a modal table assignment
for $Q_j$ gives
\begin{equation}\label{eq:selector-modal}
   \lambda_j(a)\,Z_j \;\leq\; L \qquad\text{for every }j\text{ and every }a ,
\end{equation}
an estimate uniform in the selector assignment; the selector may therefore be chosen later.

\emph{Localising at a selector.} Fix $a\in[q]^{m}$, that is, pass to one copy. Write $y_u := z_{u,a_u}$ for the addressed bits. Now average the identity
$c_A(a,z)-\epsilon = \sum_j\lambda_j(a)Q_j(z)$ over the unaddressed table bits, independently
and uniformly. On the right each factor $1+\beta_{j,u,t}z_{u,t}$ with
$t\neq a_u$ averages to $1$, since $\mathbb E\,z_{u,t} = 0$; only the $m$ addressed factors
survive. On the left, by \eqref{eq:selector-count}, $c_A-\epsilon$ already depends on the
addressed bits alone, and equals $\mathcal V(y)-\epsilon$ there. Hence
\begin{equation}\label{eq:selector-localised}
   \mathcal V(y)-\epsilon \;=\; \sum_{j=1}^{r}\lambda_j(a)\,Q_{j,a}(y),
   \qquad Q_{j,a}(y) := \prod_{u=1}^{m}\bigl(1+\beta_{j,u,a_u}y_u\bigr) .
\end{equation}
The base functional $\widetilde{\mathbb E}$ is defined only in Fourier degree at most $d$; extend
it to every function on $\{-1,1\}^{m}$ by
\[
   \Lambda(g) \;:=\; \widetilde{\mathbb E}\bigl(g^{\leq d}\bigr).
\]
Then $\Lambda$ is linear and agrees with $\widetilde{\mathbb E}$ in degree at most $d$, so
$\Lambda(\mathcal V-\epsilon) = \widetilde{\mathbb E}\mathcal V-\epsilon = -\epsilon$ because $\deg\mathcal V\leq k\leq d$.
Applying $\Lambda$ to \eqref{eq:selector-localised} therefore gives
$\sum_j\lambda_j(a)\Lambda(Q_{j,a}) = -\epsilon$.

\emph{Testing one localised product.} For $F\subseteq[m]$ and $\sigma\in\{-1,1\}^{F}$ put
$\rho_{F,\sigma}(y) := \prod_{u\in F}(1+\sigma_uy_u) = 2^{|F|}\mathbf 1[y_F = \sigma]$, which
is nonnegative, supported on the subcube $y_F = \sigma$ and of average
$\mathbb E\,\rho_{F,\sigma} = 1$. If $|F|\leq d$ then
$\Lambda(\rho_{F,\sigma})\geq0$ by local nonnegativity. For arbitrary $F$, expanding the product gives
$\rho_{F,\sigma} = \sum_{\alpha\subseteq F}\sigma^{\alpha}y^{\alpha}$ with
$\sigma^{\alpha} := \prod_{u\in\alpha}\sigma_u\in\{-1,1\}$, so
\[
   \Lambda(\rho_{F,\sigma})
   \;=\; \sum_{\alpha\subseteq F,\ |\alpha|\leq d}\sigma^{\alpha}\,
   \widetilde{\mathbb E}\,y^{\alpha},
\]
a sum of at most $\binom{m}{\leq d}$ terms whatever $|F|$ is. Each term is at most one in absolute value:
$1\pm x^{\alpha}$ is a nonnegative function of $|\alpha|$ coordinates, so
\begin{equation}\label{eq:lef-moment}
   \bigl|\widetilde{\mathbb E}\,x^{\alpha}\bigr|\;\leq\;\widetilde{\mathbb E}\,1\;=\;1
   \qquad (|\alpha|\leq d) .
\end{equation}
Hence $|\Lambda(\rho_{F,\sigma})|\leq \binom{m}{\leq d}$.

For the fixed $j$ and $a$ of this step abbreviate $\beta_u := \beta_{j,u,a_u}$,
$\sigma_u := \operatorname{sgn}(\beta_u)$ and $h_u := h_{j,u,a_u} = |\beta_u|$, so that
$\beta_u = h_u\sigma_u$ and hence
\[
   1+\beta_uy_u \;=\; (1-h_u) \;+\; h_u\bigl(1+\sigma_uy_u\bigr) .
\]
Substitute this in each of the $m$ factors of $Q_{j,a}$ and expand the product. Each resulting
term is indexed by the set $F\subseteq[m]$ of coordinates at which the second summand was taken,
and equals $\prod_{u\in F}h_u\prod_{u\notin F}(1-h_u)$ times $\rho_{F,\sigma_F}$, where
$\sigma_F := (\sigma_u)_{u\in F}$. These coefficients are the probabilities of the random set
$F$ containing each $u$ independently with probability $h_u$, so
\[
   Q_{j,a} \;=\; \mathbb E_{F}\bigl[\rho_{F,\sigma_F}\bigr] .
\]
A coordinate with $\beta_u = 0$ has $h_u = 0$ and so never enters $F$, which makes the choice
of $\sigma_u$ there immaterial. Apply
$\Lambda$ and split this expectation at $|F| = d$: the terms with $|F|\leq d$ contribute
nonnegatively, and each term with $|F|>d$ is at least $-\binom{m}{\leq d}$. Writing
$D := d+1$ and $\tau_j(a) := \Pr\bigl[|F|\geq D\bigr]$,
\begin{equation}\label{eq:selector-tail}
   \Lambda(Q_{j,a}) \;\geq\; -\binom{m}{\leq d}\,\tau_j(a) .
\end{equation}

\emph{Choosing a common selector.} Bound $\tau_j(a)$ first. For an integer $s\geq0$ the
binomial coefficient $\binom sD$ is $0$ when $s<D$ and at least $1$ when $s\geq D$, so
$\binom{|F|}{D}$ dominates the indicator of the event $|F|\geq D$; and $\binom{|F|}{D}$ counts
the $D$-element subsets of $F$. Hence
\[
   \tau_j(a) \;=\; \Pr\bigl[|F|\geq D\bigr]
   \;\leq\; \mathbb E_{F}\binom{|F|}{D}
   \;=\; \sum_{U\subseteq[m],\,|U| = D}\Pr\bigl[\,U\subseteq F\,\bigr]
   \;=\; \sum_{U\subseteq[m],\,|U| = D}\ \prod_{u\in U}h_u ,
\]
the coordinates entering $F$ independently. Now let $a$ vary: draw each $a_u$ independently and
uniformly from $[q]$. Since $h_u = h_{j,u,a_u}$ depends on $a_u$ we write it out in full, and
averaging, with independence used once more,
\[
   \mathbb E_a\,\tau_j(a)
   \;\leq\; \sum_{|U| = D}\ \prod_{u\in U}\frac{1}{q}\sum_{t\leq q}h_{j,u,t}
   \;=\; q^{-D}\!\!\sum_{|U| = D}\ \prod_{u\in U}\Bigl(\sum_{t\leq q}h_{j,u,t}\Bigr)
   \;\leq\; q^{-D}\prod_{u,t}(1+h_{j,u,t}) \;=\; q^{-D}Z_j ,
\]
the last inequality because expanding the middle sum gives distinct squarefree monomials --- one
factor $h_{j,u,t}$ from each of $D$ distinct blocks $u$ --- each occurring with coefficient one
in the expansion of $\prod_{u,t}(1+h_{j,u,t})$, all of whose monomials are nonnegative. Dividing by $Z_j$ and summing over $j$ gives
$\mathbb E_a\sum_j\tau_j(a)/Z_j\leq rq^{-D}$, so some selector assignment $a^{*}$ --- depending
on the whole purported decomposition --- satisfies
\begin{equation}\label{eq:selector-choice}
   \sum_{j=1}^{r}\frac{\tau_j(a^{*})}{Z_j} \;\leq\; r\,q^{-D} .
\end{equation}

\emph{Contradiction at the selected copy.} At $a = a^{*}$, combining
\eqref{eq:selector-localised}, \eqref{eq:selector-tail}, \eqref{eq:selector-modal} and
\eqref{eq:selector-choice} in that order,
\[
\begin{aligned}
   -\epsilon &= \sum_{j}\lambda_j(a^{*})\Lambda(Q_{j,a^{*}})
   \;\geq\; -\binom{m}{\leq d}\sum_{j}\lambda_j(a^{*})\tau_j(a^{*})
   \\
   &\geq -L\,\binom{m}{\leq d}\sum_{j}\frac{\tau_j(a^{*})}{Z_j}
   \;\geq\; -L\,\binom{m}{\leq d}\,r\,q^{-D} ,
\end{aligned}
\]
which rearranges to \eqref{eq:selector-lift}. A nonnegative rank-one tensor over the raw Boolean
variables is in particular of the displayed form, so the bound applies to $\rkt$ and, by
Corollary~\ref{lem:autosml}, to the size of every $\SHAMC^{\epsilon}$ proof of $A$.
\end{proof}

The base we use is Tseitin itself.

\begin{corollary}[Lifted Tseitin is hard at exponentially small shift]\label{cor:exp-small-shift}
Fix an explicit family of connected $d_G$-regular expanders with $d_G = O(1)$ and edge expansion
$h = \Omega(1)$, and give each member an explicitly computable odd charge, for instance a single
charged vertex. There are constants $c>0$ and $w$,
depending only on $d_G$ and $h$, and an explicit family of tautological DNFs
$A_N := \mathrm{Ind}_q(\TS_G)$ on $N$ Boolean variables with $O(N)$ terms of width at most $w$,
such that for all sufficiently large members
\[
   2^{-cN}\;\leq\;\epsilon\;\leq\;1
   \qquad\Longrightarrow\qquad
   \rkt\bigl(\chi(A_N)-\epsilon\cdot\chi(1)\bigr)\;\geq\;2^{cN} .
\]
Every depth-three $\AMC^{\epsilon}$ proof of $A_N$ therefore has at least $2^{cN}$ product terms
throughout that range of shifts.
\end{corollary}

\begin{proof}
Take $I_0 := \TS_G$ as base: $m = |E| = d_Gn/2$ variables, $L = n$ constraints of arity $d_G$,
and unsatisfiable. By \cite{Grigoriev01} its sum-of-squares refutation degree is $\Omega(n)$,
hence so is its Sherali--Adams refutation degree, and linear programming duality for
Sherali--Adams turns the latter into a degree-$d$ local expectation functional with
$d = \alpha m$ for a constant $\alpha = \alpha(d_G,h)>0$. Each $B_\ell$ is a nonnegative
$d_G$-junta and $\widetilde{\mathbb E}\mathcal V = 0$, so $\widetilde{\mathbb E}B_\ell = 0$ for every
$\ell$: on each vertex's scope the local distribution satisfies that vertex's parity with
probability one. (Each nonnegative $j$-junta is a nonnegative
combination of subcube indicators, which are idempotent and hence squares of degree-$j$
polynomials; a degree-$2d$ sum-of-squares solution therefore restricts to a degree-$d$ local
expectation functional, and the factor $2$ is absorbed into $\alpha$.)

Put $b := \lceil4/\alpha\rceil$, $q := 2^{b}$, $\kappa := q+b$ and $c := \kappa^{-1}$, all
constants determined by $d_G$ and $h$. The lift has $N = \kappa m$ variables, at most
$L2^{d_G}q^{d_G} = O(N)$ terms of width at most $w := d_G(b+1)$, and is constructible in
polynomial time. Since $\binom{m}{\leq d}\leq2^{m}$ and $b(d+1)\geq4m$, \eqref{eq:selector-lift} gives
$r\geq\epsilon2^{b(d+1)}/(L\,\binom{m}{\leq d})\geq\epsilon2^{3m}/L$; here $L = n\leq m\leq2^{m}$, so
$r\geq\epsilon2^{2m}$, and $\epsilon\geq2^{-m}$ gives $r\geq2^{m} = 2^{cN}$.
\end{proof}

\begin{remark}[Limits of the construction]\label{rem:selector-scope}
The argument acts on products over the
individual table bits, and a general nonnegative column factor of a bipartite flattening need
not factor over them, so it gives nothing for $\rkp$ and nothing at mixed-product depth above
zero. For $\TS_G$ itself no such bound exists at all, by Theorem~\ref{thm:mu0-upper}; whether
$\mathrm{Ind}_q(\TS_G)$ is likewise easy at mixed-product depth zero we do not know. Section~\ref{sec:csp-dnf} runs the same skeleton with copies of the same Tseitin base planted in a shared
variable set instead of private tables, and trades the reach in $\epsilon$ for exactly those two
properties; the comparison is set out there.
\end{remark}

\section{Lower bounds at bounded mixed-product depth}\label{sec:polytopes}

We now feed the flattening bound \eqref{eq:method}. What it asks for is a DNF tautology $A$, a
bipartition of the blocks of $\chi(A)$, and a submatrix of
$\flt{X_I}{X_{\ol I}}{\chi(A)} - \epsilon\cdot J_{X_I \mid X_{\ol I}}$ of large nonnegative
rank; monotonicity of $\rkp$ under submatrices then bounds the leaves of every monotone formula
of mixed-product depth $t$ computing $\chi(A)-\epsilon\cdot\chi(1)$, and at $t=0$ gives the
$\SHAMC^{\epsilon}$ proof size bound \eqref{eq:flat-lb}. Every such submatrix has the form
$M - \epsilon\cdot J$ with $M$ a submatrix of the unperturbed flattening and $J$ the all-ones
matrix of matching dimensions.

We isolate two ways of producing one, differing in where the rank lower bound comes from. In
Recipe I the rows are a family of valid inequalities of a combinatorial polytope and the rank is
its approximate extension complexity; in Recipe II (Section~\ref{sec:csp-dnf}) the rows are the
instances of a planted constraint satisfaction family and the rank comes from a
Sherali--Adams integrality gap, no polytope being named. The two have disjoint hypotheses and
different admissible ranges of $\epsilon$, and share only the encoding lemma,
Proposition~\ref{prop:local-witness-dnf}, so we state each interface once and present the
constructions as instances.

Recipe I is stated in the language of approximate extension complexity, which we fix here. Let
$P$ be a polytope with vertex set $V$ and let $\mc F$ be a family of inequalities
$\ol a_f \cdot \ol x - b \geq 0$ valid at every vertex. Write $P^{\mc F}_{\delta}$ for the
outer relaxation cut out by $\ol a_f \cdot \ol x - b + \delta \geq 0$, $f \in \mc F$, so that
$P^{\mc F}_{0}$ is the body cut out by $\mc F$ itself. Let $S$ be the slack matrix of $\mc F$ at
$V$ and put $M := S + b\cdot J$, so that $M-\epsilon\cdot J = S+(b-\epsilon)J$ is the slack
matrix at $V$ of the relaxed family defining $P^{\mc F}_{b-\epsilon}$. Then
\begin{equation}\label{eq:approx-xc}
   \mathsf{xc}\bigl(P,P^{\mc F}_{b-\epsilon}\bigr)
   \;\leq\;\rkp(M - \epsilon\cdot J)\;\leq\;
   \mathsf{xc}\bigl(P,P^{\mc F}_{b-\epsilon}\bigr)+1 ,
   \qquad
   \mathsf{xc}(P,P') := \min\{\mathsf{xc}(Q) : P\subseteq Q\subseteq P'\} :
\end{equation}
a nonnegative factorization of the perturbed slack matrix is an extended formulation of a body
squeezed between $P$ and the relaxation, and conversely up to one further term, the constant
slack such a formulation may need \cite[Lem.~3.4]{Hrubes20}, taken there from
\cite{BraunFioriniPokuttaSteurer12}. Equality can fail: for
$P = \{1\}\subset\mathbb R$ with the single inequality $x\geq1$ and $0<\epsilon<1$ the perturbed
slack matrix is $[\,1-\epsilon\,]$, of nonnegative rank one, while $Q = P$ is cut out by an
equality and needs no inequality at all. Recipe I uses only the left inequality, and $Q$ is
allowed to be an unbounded polyhedron, as the outer body of Section~\ref{sec:pm-lb} is.

The construction behind the first interface is this. Specialise the above to a combinatorial
setting: a finite ground set $E$, a family $\mc H$ of subsets of $E$ whose indicator vectors
$\chi_H$ are the vertices of $P$, and for $\mc F$ a family of covering inequalities
$\lta a_f,\ol x\rta\geq b$ --- covering in the usual sense of nonnegative coefficients and the
$\geq$ direction, and restricted further so that every $a_f$ lies in $\{0,1\}^{E}$ and the right
hand side $b$ is the same throughout the family. For such an $a_f$ the left hand side
$\lta a_f,\chi_H\rta$ is just the overlap count $|\supp(a_f)\cap H|$, and the slack of $f$ at
$\chi_H$ is that count less $b$. So we want a DNF whose valid slice --- the submatrix of its
flattening carried by the valid encodings, in the sense of
Proposition~\ref{prop:local-witness-dnf} --- is the overlap count $|\supp(a_f)\cap H|$: its
perturbation is then the slack matrix of a nested pair of bodies, up to the shift by $b$, and
\eqref{eq:approx-xc} applies to it. Proposition~\ref{prop:local-witness-dnf} turns any list of
witness conjunctions into such a DNF, the valid slice counting how many of them a valid pair
satisfies. We choose the list element by element: a group of conjunctions for each $e \in E$,
arranged so that exactly one member of the group fires when $e$ lies in $\supp(a_f)$ and in $H$
at once, and none fires otherwise. One of the two memberships is free: the encoding of a member of $\mc H$ includes its
indicator vector, so $e \in H$ is recorded by a single variable. The other must be read off
whatever encoding $\mc F$ carries, by a test of constant size --- and read \emph{without
multiplicity}, since a second firing for the same $e$ would inflate the slice. Invalid encodings
on either side are absorbed by the two invalidity DNFs.

\begin{prop}[Recipe I: covering polytopes]\label{prop:recipe-polytope}
Let $E$ be a finite ground set, $\mc H$ a family of subsets of $E$, and $\mc F$ a family of
inequalities $\lta a_f, \ol x\rta \geq b$ with $a_f \in \{0,1\}^{E}$ and a common right hand
side $b \in \mathbb Z_{>0}$, all valid at every $\chi_H$, $H \in \mc H$; put
$P := \mathsf{conv}\{\chi_H : H \in \mc H\}$. Encode the members of $\mc F$ over a set $X$ of
Boolean variables and those of $\mc H$ over a disjoint set $Y$ containing their indicator bits,
$\gamma_e \in Y$ recording $e \in H$. Either encoding may carry auxiliary variables beyond
these, in which case their consistency with the encoded object is part of validity. Assume:
\begin{enumerate}
\item[(I.1)] for every $e \in E$ there are conjunctions $\rho_{e,1}, \dots, \rho_{e,c}$ of
literals in $X$, with $c = O(1)$, of which exactly one holds at the encoding of $f$ when
$e \in \supp(a_f)$ and none holds otherwise;
\item[(I.2)] invalid encodings are recognised exactly by DNFs $I_R(X)$ and $I_C(Y)$ with
polynomially many terms.
\end{enumerate}
Then
\[
   A \;:=\; I_R \vee I_C \vee \bigvee_{e \in E}\ \bigvee_{j=1}^{c}
   \bigl(\rho_{e,j} \wedge \gamma_e\bigr)
\]
is a tautological DNF with $|I_R| + |I_C| + c|E|$ terms whose valid slice is
$M(f,H) = |\supp(a_f) \cap H|$, so that $M = S + b\cdot J$ for the slack matrix $S$ of $\mc F$ at
$\mc H$; and every monotone formula of mixed-product depth at most $t$ computing
$\chi(A) - \epsilon\cdot\chi(1)$ has at least
\[
   \Bigl(\min\bigl\{\,\mathsf{xc}(Q)\ :\ P \subseteq Q \subseteq P^{\mc F}_{b-\epsilon}\,\bigr\}\Bigr)^{1/2^{t}}
\]
leaves.
\end{prop}

\begin{proof}
On a valid pair $(f,H)$ the two invalidity DNFs are unsatisfied, and by \textup{(I.1)} exactly one
witness term $\rho_{e,j}\wedge\gamma_e$ is satisfied for each $e \in \supp(a_f)\cap H$ and none
for any other $e$; the number of satisfied terms is therefore
$|\supp(a_f)\cap H| = \lta a_f, \chi_H\rta \geq b \geq 1$. Proposition~\ref{prop:local-witness-dnf}
now gives the term count, the tautology, and the valid slice $M$. Finally
$M - \epsilon\cdot J = S + (b-\epsilon)J$ is the slack matrix of the defining inequalities of
$P^{\mc F}_{b-\epsilon}$ evaluated at the points $\chi_H$, so the left inequality of
\eqref{eq:approx-xc} bounds its nonnegative rank below by the displayed minimum, and
Corollary~\ref{cor:local-witness-mixed-depth} converts that into the leaf bound.
\end{proof}

The right hand side matters. With $b = 1$ and $\epsilon \in (0,1)$ the relaxation
$P^{\mc F}_{b-\epsilon}$ loosens each inequality by less than a unit; with $b>1$ it loosens by
more, so the same range of $\epsilon$ demands a bound against a weaker relaxation. Odd cut
inequalities have $b=1$; subtour elimination constraints, for instance, have $b=2$.

Sections~\ref{sec:pm-polytopes}--\ref{sec:pm-lb} instantiate Recipe I with $E$ the edge set of
$K_{2n}$, $\mc H$ the perfect matchings or the odd factors, and $\mc F$ the odd cut
inequalities $\ol x(\delta(U)) \geq 1$. Section~\ref{sec:csp-dnf} states its own interface,
Proposition~\ref{prop:recipe-csp}, which shares only
Proposition~\ref{prop:local-witness-dnf} with this one.

\subsection{Matching, perfect matching and odd factor polytopes}\label{sec:pm-polytopes}

Let $G = (V, E)$ be a graph. The \emph{matching polytope} is
\[
   P_M(G) = \mathsf{conv}\{\chi_M \in \{0,1\}^E \subset \mathbb{R}^E \, |\, M \subseteq E
   \text{ is a matching}\},
\]
which is described by $x_e \geq 0$ for $e \in E$, $x(\delta(v)) \leq 1$ for $v \in V$, and
$x(E(U)) \leq \frac{|U| - 1}{2}$ for $U \subseteq V$ with $|U|$ odd.

If $|V|$ is even, then the \emph{perfect matching polytope} $P_{\PM}(G)$ is
the face of $P_M(G)$ cut out by $x(\delta(v)) = 1$, $v\in V$:
\[
   P_{\PM}(G) = \mathsf{conv}\{\chi_M \in \{0,1\}^E \subset \mathbb{R}^E \, |\, M \subseteq E
   \text{ is a perfect matching}\},
\]
which is described by
\begin{enumerate}
\item\label{itm:eq_cond} $x_e \geq 0$, $e \in E$, together with $x(\delta(v)) = 1$, $v \in V$;
\item $x(\delta(U)) \geq 1$, $U \subseteq V$, $|U|$ odd.
\end{enumerate}
For an arbitrary graph neither list is irredundant, and the equalities are not facets.

An \emph{odd factor} of $G$ is a spanning subgraph in which every vertex has odd degree;
equivalently, a $T$-join for $T = V$. This forces $|V|$ to be even. The \emph{odd factor
polytope} is
\[
   P_{\OF}(G) \;=\; \mathsf{conv}\{\chi_H \in \{0,1\}^E : H \subseteq E \text{ is an odd factor of } G\}.
\]
The odd cut inequalities $x(\delta(U)) \geq 1$, $|U|$ odd, are valid for it --- this is checked
in Section~\ref{sec:pm-lb} --- but they do not describe it, even together with
$0\leq x_e\leq1$: in $K_4$ the indicator of a four-cycle satisfies all of them, while every
vertex of the four-cycle has even degree, so it is not an odd factor and, being integral, not a
convex combination of odd factors either. What those inequalities together with $x\geq0$ do
describe, by Edmonds and Johnson \cite{EdmondsJohnson73}, is the dominant
$P_{\OF}(G)+\mathbb R^{E}_{\geq0}$. Nothing below needs a facet description: only the validity of
the odd cut inequalities is used, and that every perfect matching is a $1$-regular odd factor,
so that $P_{\PM}(G)$ is the face of $P_{\OF}(G)$ cut out by $x(\delta(v)) = 1$ for all $v\in V$.

Write $S_{\PM(G)}$ and $S_{\OF(G)}$ for the \emph{odd cut slack matrices} of the two bodies:
the slacks of the odd cut inequalities at the vertices, a submatrix of the slack matrix of any
description containing them. They share the same rows --- indexed by odd
cuts $\delta(U)$, $|U|$ odd --- and the same entry rule
\[
   S(U, H) \;=\; |\delta(U) \cap H| - 1,
\]
their columns ranging respectively over perfect matchings and over all odd
factors of $G$. Since every perfect matching is an odd factor, $S_{\PM(G)}$
is a column submatrix of $S_{\OF(G)}$; by monotonicity of $\rkp$ under
submatrices, $\rkp(S_{\OF(G)}) \geq \rkp(S_{\PM(G)})$.

For the complete graph $K_{2n}$ --- the vertex count must be even for a perfect matching
to exist, and we fix it as $2n$ throughout Sections~\ref{sec:pm-dnfs} and~\ref{sec:pm-lb} and
Appendix~\ref{sec:threshold}
--- we denote the corresponding polytopes simply by $P_{\PM}$, $P_{\OF}$. By Rothvo\ss{} \cite{Rothvoss17},
$\rkp(S_{\PM}) = 2^{\Omega(n)}$ and therefore $\rkp(S_{\OF}) = 2^{\Omega(n)}$.
Writing $M_{\PM}$, $M_{\OF}$ for the corresponding edge-count matrices, so that
$S_{\PM} = M_{\PM} - J$ and $S_{\OF} = M_{\OF} - J$, the same column-submatrix inclusion and
Braun and Pokutta's \cite{BraunPokutta15} $\epsilon$-perturbed bound give
\begin{equation}\label{eq:BP-import}
   \rkp(M_{\OF} - \epsilon\cdot J) \;\geq\; \rkp(M_{\PM} - \epsilon\cdot J) \;=\; 2^{\Omega(n)}
\end{equation}
for every constant $\epsilon \in (0,1)$, remaining superpolynomial down to
$\epsilon = c_{0}\log\log n/\log n$ for any fixed $c_{0}>8$. Appendix~\ref{sec:threshold} makes
the constants explicit and establishes this sufficient range.

\subsection{Perfect matching and odd factor DNFs}\label{sec:pm-dnfs}

We define a DNF tautology $A_{\PM}(\ol x, \ol y, \ol z)$ phrasing ``for
every odd cut $\ol x$, every perfect matching $\ol y$ intersects $\delta(\ol x)$''
($\ol z$ are auxiliary variables). Specifically,
$A_{\PM}(\ol x, \ol y, \ol z)$ is the disjunction of the
DNFs $\neg B_{odd}(\ol x, \ol z)$, $\neg C_{match}(\ol y)$,
$\neg C_{V\text{-}cover}(\ol y)$ and $D_{intersects}(\ol x, \ol y)$, saying that either
$\ol x$ is not odd, or $\ol y$ is not a matching, or $\ol y$ does not cover $V$,
or $\ol y$ intersects $\delta(\ol x)$, respectively. These DNFs are in
turn defined as follows.

\begin{enumerate}
\item $\neg B_{odd}(\ol x, \ol z)$: ``$\ol x$ is not odd''. The standard encoding
of $x_1\oplus \dots \oplus x_{2n} = 0$ as a DNF: extra Boolean variables
$\ol z = \{z_{x_1\oplus\dots\oplus x_i}\}_{i > 1}$ (we write $z_{x_1}\equiv x_1$) represent
values of partial XORs, and the terms are
$\{z_{x_1\oplus\dots\oplus x_i}^a\wedge x_{i+1}^b \wedge z_{x_1\oplus\dots\oplus x_{i+1}}^{\ol{a\oplus b}}\}_{a,b\in \{0,1\},\, i \in [2n-1]}$
together with $\neg z_{x_1\oplus\dots\oplus x_{2n}}$, where $\ol c := 1-c$. A term of the first
kind is satisfied exactly when the guessed partial XOR at step $i+1$ is \emph{inconsistent} with the
one at step $i$, so $\neg B_{odd}$ is satisfied precisely when $\ol z$ misrepresents the
partial XORs of $\ol x$ or when $\ol x$ has even parity. This is what makes an odd $\ol x$
with a consistent $\ol z$ --- the rows used in the lower bound below --- contribute nothing
to $\chi(\neg B_{odd})$.
\item $\neg C_{match}(\ol y)$: ``$\ol y$ is not a matching''. Terms
$\{y_{i, j} \wedge y_{j, k}\}$ for $j \in [2n]$ and $i < k$ in $[2n]\setminus\{j\}$,
one for each pair of distinct edges sharing the vertex $j$. Distinctness of $i$ and $k$ is
needed: the term with $i = k$ is $y_{i,j}$ itself, which is satisfied by every nonempty
matching, so it would contribute to the coefficient at a valid pair and the perturbed valid
slice computed in Section~\ref{sec:pm-lb} would not be $M_{\PM} - \epsilon\cdot J$.
\item $\neg C_{V\text{-}cover}(\ol y)$: ``$\ol y$ does not cover $V$''. Terms
$\{\bigwedge_{j\in [2n]\setminus \{i\}} \neg y_{i, j}\}_{i \in [2n]}$.
\item $D_{intersects}(\ol x, \ol y)$: ``$\ol y$ intersects $\delta(\ol x)$''. Terms
$\{x_i \wedge y_{i, j} \wedge \neg x_j\}$.
\end{enumerate}

The tautology $A_{\OF}(\ol x, \ol y, \ol z, \ol w)$ asserts the analogous
principle for odd factors: ``for every odd cut $\ol x$, every odd factor
$\ol y$ intersects $\delta(\ol x)$.'' It is a tautology by the parity argument
$\sum_{v \in U}\deg_H(v) = 2|E_H(U)| + |\delta(U)\cap H|$: an odd number of odd summands is
odd, so $|\delta(U) \cap H|$ is odd and in particular positive.
$A_{\OF}$ is the disjunction of
$\neg B_{odd}(\ol x, \ol z)$, $\neg C_{oddfactor}(\ol y, \ol w)$ and
$D_{intersects}(\ol x, \ol y)$, saying that either $\ol x$ is not odd, or
some vertex has even degree in $\ol y$, or $\ol y$ intersects $\delta(\ol x)$.
The disjuncts $\neg B_{odd}$ and $D_{intersects}$ are as in $A_{\PM}$ above,
and

\begin{enumerate}
\item[$(2')$] $\neg C_{oddfactor}(\ol y, \ol w)$: the disjunction over
$v \in V$ of the standard encoding of $\bigoplus_{j : \{v, j\} \in E} y_{v, j} = 0$
as a DNF; for each $v$, extra Boolean variables
$\ol w_v = \{w_{v, i}\}_{i > 1}$ represent the partial XORs of the incident
edge variables $y_{v, j_1}, y_{v, j_2}, \dots$ (with $j_1, j_2, \dots$ an
enumeration of the neighbours of $v$), and the terms have the same shape
as those of $\neg B_{odd}$ with $x$ replaced by these incidence variables.
\end{enumerate}

Compared to $A_{\PM}$, the matching DNF $\neg C_{match}$ is dropped
entirely and the covering DNF $\neg C_{V\text{-}cover}$ is replaced by
a union of $|V|$ parity gadgets structurally identical to $\neg B_{odd}$; the
resulting DNF $A_{\OF}$ is a disjunction of parity/intersection gadgets of a single
shape, giving its encoding a more uniform block structure than $A_{\PM}$. In
view of the previous subsection, $A_{\OF}$ yields the same $\SHAMC$ lower bounds as
$A_{\PM}$. Both DNFs have polynomially many terms.

\subsection{Lower bounds for \texorpdfstring{$A_{\PM}$ and $A_{\OF}$}{A-PM and A-OF}}\label{sec:pm-lb}

Both bounds are instances of Recipe I, and it remains only to name its data and check its
hypotheses. Take $E$ the edge set of $K_{2n}$, $\mc H$ the perfect matchings, respectively the
odd factors, and $\mc F$ the odd cut inequalities $\ol x(\delta(U)) \geq 1$ indexed by the odd
$U \subseteq V$, so that $a_f = \chi_{\delta(U)}$ and $b = 1$, and
$P = \mathsf{conv}\{\chi_H : H \in \mc H\}$ is $P_{\PM}$, respectively $P_{\OF}$. Members of
$\mc F$ are encoded over $X := \ol x \cup \ol z$, with $\ol x$ the indicator of $U$ and $\ol z$
the partial XORs, and members of $\mc H$ over $Y := \ol y$, respectively $\ol y \cup \ol w$;
this is the split of the variables that induces the flattening
$\flt{U_x}{U_y}{\chi(A) - \epsilon\cdot\chi(1)}$ used below, with $U_x := X$ and $U_y := Y$.

\emph{The inequalities of $\mc F$ hold at every $\chi_H$.} No perfect matching lies inside a set
of odd size, so a perfect matching $M$ has an edge leaving $U$ and $|\delta(U)\cap M| \geq 1$.
For an odd factor $H$, the identity
$\sum_{v\in U}\deg_H(v) = 2|E_H(U)| + |\delta(U)\cap H|$ exhibits $|\delta(U)\cap H|$ as a sum of
$|U|$ odd numbers minus an even number; as $|U|$ is odd, $|\delta(U)\cap H|$ is odd and in
particular positive.

\emph{Condition \textup{(I.1)}.} For an edge $e = \{i,j\}$ put $\rho_{e,1} := x_i \wedge \neg x_j$
and $\rho_{e,2} := x_j \wedge \neg x_i$, so $c = 2$. They are contradictory, and their
disjunction holds at the encoding of $U$ exactly when $e$ has one endpoint in $U$ and one
outside, that is exactly when $e \in \delta(U) = \supp(a_f)$. Take $\gamma_e := y_{i,j}$, which
holds exactly when $e$ belongs to the encoded $H$. The resulting witness terms
$\rho_{e,j} \wedge \gamma_e$ are precisely the terms $x_i \wedge y_{i,j} \wedge \neg x_j$ of
$D_{intersects}$.

\emph{Condition \textup{(I.2)}.} Invalid rows are recognised by $I_R := \neg B_{odd}$: its terms
of the first kind are satisfied exactly when $\ol z$ misrepresents some partial XOR of $\ol x$,
and $\neg z_{x_1\oplus\dots\oplus x_{2n}}$ exactly when the recorded total parity is even, so
$\neg B_{odd}$ holds precisely at those $(\ol x, \ol z)$ that do not encode an odd $U$ with its
partial XORs. Invalid columns are recognised by
$I_C := \neg C_{match} \vee \neg C_{V\text{-}cover}$, since $\ol y$ fails to encode a perfect
matching exactly when two of its edges share a vertex or some vertex is left uncovered; for
$A_{\OF}$ by $I_C := \neg C_{oddfactor}$, whose parity gadgets are satisfied exactly when
$\ol w$ misrepresents a partial XOR at some vertex or some vertex has even degree in $\ol y$.
Both have polynomially many terms.

The DNF that Recipe I builds from this data --- the disjunction of the two invalidity DNFs with
the witness terms --- is therefore exactly $A_{\PM}$, respectively $A_{\OF}$, of
Section~\ref{sec:pm-dnfs}. So both are tautologies, and their valid slice is the ground set
overlap count $M(U, H) = |\delta(U) \cap H|$: the valid rows biject with the odd subsets of $V$
and the valid columns with the perfect matchings, respectively the odd factors, which are exactly
the row and column indices of $M_{\PM}$, respectively $M_{\OF}$. The perturbed valid slice is
$M_{\PM} - \epsilon\cdot J$, respectively $M_{\OF} - \epsilon\cdot J$.
Recipe I now gives the following, connecting proof size directly to the nonnegative rank of the
odd cut slack matrices of $P_{\PM}$ and $P_{\OF}$.

\begin{theorem}\label{thm:pm-of-lb}\label{thm:matching-mixed-depth}
Let $\Phi$ be a monotone formula, of mixed-product depth at most
$t$ relative to the partition $(U_x, U_y)$ above, computing
$\chi(A_{\PM}) - \epsilon\cdot\chi(1)$ or $\chi(A_{\OF}) - \epsilon\cdot\chi(1)$. Then its
number of leaves is at least
\[
   \rkp\bigl(M_{\PM} - \epsilon\cdot J\bigr)^{1/2^{t}},
\]
where $M_{\PM} - \epsilon\cdot J$ is the slack matrix, at the odd cut inequalities $\mc F$, of
the pair $(P_{\PM}, P^{\mc F}_{\PM, 1 - \epsilon})$, and correspondingly for $A_{\OF}$.
\end{theorem}

\begin{proof}
The valid slice identified above is $M_{\PM}-\epsilon\cdot J$, respectively
$M_{\OF}-\epsilon\cdot J$. Write $F$ for the polynomial computed by $\Phi$ and $L$ for its
leaf count. Monotonicity under submatrices and Theorem~\ref{thm:mixed-depth-flattening} give
\[
   \rkp(M_{\PM}-\epsilon\cdot J)
   \;\leq\;\rkp(\Flat(F))\;\leq\;L^{2^t}.
\]
For the odd factor target, the same argument applies because the perfect matching matrix is
a column submatrix of the odd factor matrix. Taking the $2^{-t}$ power proves the claim.
Since $b=1$, the slack-matrix interpretation is the relaxation obtained by loosening each odd
cut inequality by $1-\epsilon$.
\end{proof}

\begin{corollary}\label{cor:pm-of-lb}
For constant $\epsilon \in (0,1)$ the leaf count is at least $2^{\Omega(n/2^{t})}$. For
$\epsilon \geq c_{0}\log\log n/\log n$ with $c_{0}>8$ fixed it is at least
$2^{n^{1-8/c_{0}+o(1)}/2^{t}}$, and in particular superpolynomial for every constant $t$. At
$t=0$, both $\SHAMC^{\epsilon}$ and $\AMCb^{\epsilon}$ satisfy the exponential bound in the
constant-shift regime and the second bound in the vanishing-shift regime.
\end{corollary}

\begin{proof}
Insert $\rkp(M_{\PM}-\epsilon\cdot J)=2^{\Omega(n)}$ from~\eqref{eq:BP-import} for constant
$\epsilon$; in the vanishing regime \eqref{eq:bp-boundary} gives
$\log\rkp(M_{\PM}-\epsilon\cdot J)\geq n^{1-8/c_0+o(1)}$, and raising the rank to the power
$2^{-t}$ divides this logarithm by $2^{t}$. The case $t=0$ is
Remark~\ref{rem:t0}.
\end{proof}

By Proposition~\ref{prop:circuit-sharing-obstruction} the restriction to formulas cannot be
removed as stated.

Braun and Pokutta state their per-chunk constant only as a positive constant depending on $k$
and on their perturbation parameter. Since the threshold
$\epsilon\geq c_{0}\log\log n/\log n$, $c_0>8$, rests on uniformity in $k$, Appendix~\ref{sec:threshold} unpacks their proof, makes the dependence on both parameters
explicit, and proves the stated sufficient threshold.

\subsection{An explicit DNF hard at inverse-polynomial shift}\label{sec:csp-dnf}

Sections~\ref{sec:pm-polytopes}--\ref{sec:pm-lb} took the DNF from a combinatorial polytope and
drew the rank bound from its approximate extension complexity. Here we change the DNF family
instead, and obtain a much smaller admissible shift: an explicit polynomial-size tautological DNF whose
perturbed flattening has superpolynomial nonnegative rank for every inverse-polynomial
$\epsilon \geq K^{-\gamma}$ with $\gamma < 1/4$ fixed. The witness conjunctions of
Proposition~\ref{prop:local-witness-dnf} record a violated constraint of a planted Tseitin
copy on a constant-degree expander, and the rank lower bound comes from a Sherali--Adams integrality gap rather than
from the extension complexity of a fixed body.

The skeleton and the Tseitin base are those already used in Section~\ref{sec:selector-lift}: plant many copies of a
base instance whose degree-$d$ Sherali--Adams relaxation is fooled, so that the copy to be tested
is chosen only after the purported decomposition, then test that decomposition with the local
expectation functional witnessing the failure. One design choice differs,
and everything else follows from it. There each base variable carried a private table, so every
address tuple already held a copy and no work was needed to find one; here the copies are planted
into a \emph{shared} pool of $N$ variables, and a copy the factors cannot see must first be
located. That costs an imported random-restriction lemma, whose $N^{-1/4}$ is what caps the
admissible shift at $K^{-\gamma}$ with $\gamma<1/4$ rather than the $2^{-\Omega(N)}$ reached
there. What it buys is the converse: the argument runs on marginals of arbitrary nonnegative
factors rather than on products over private bits, so it bounds $\rkp$ and survives mixed-product
depth above zero, neither of which the lift does.

The
mechanism is that of Chan, Lee, Raghavendra and Steurer \cite{ChanLeeRaghavendraSteurer16}; what
is added here is the dependence of the admissible shift on all parameters, which is what fixes
the range of $\epsilon$. The only input taken from their work is a random-restriction
lemma, quoted below in the form we use.

\paragraph{The matrix.} Fix an unsatisfiable constraint satisfaction instance $I_0$ on $m$
variables, writing $I_0(x)$ for the fraction of its constraints that $x$ satisfies, and let
$I_\phi$ be the copy of $I_0$ planted along an injection $\phi:[m]\hookrightarrow[N]$ --- one
copy for each injection. The matrix whose nonnegative rank is bounded is
\[
   A^\eta_{\phi,x} \;=\; 1-\eta-I_\phi(x), \qquad \phi \text{ an injection},
   \quad x \in \{-1,1\}^{N} ,
\]
the slack at the assignment $x$ of the valid inequality $I_\phi(x)\leq1-\eta$. It is nonnegative
exactly because planting does not change the optimum, so that every copy has
$\mathrm{opt}(I_\phi)=\mathrm{opt}(I_0)\leq1-\eta$; and by the factorization
theorem a nonnegative factorization of it of rank $T$ is a linear program with $T$ inequalities
certifying $I_\phi\leq1-\eta$ for all planted copies simultaneously.
Proposition~\ref{prop:quantitative-planted-csp-rank} says that no such program of size $N^{d/2}$
exists once $I_0$ fools the degree-$d$ Sherali--Adams relaxation, and
Section~\ref{sec:csp-interface} realises $A^\eta$ as the perturbed valid slice of a DNF.

Read geometrically, $A^\eta$ is again a slack matrix of a nested pair of convex bodies, as in
Section~\ref{sec:pm-lb} and Appendix~\ref{sec:threshold}. Lift an assignment $x$ to the vector $v(x)$ whose
coordinates are the indicators $\mathbf 1[x|_T=\sigma]$, one for each $k$-subset
$T\subseteq[N]$ and each pattern $\sigma\in\{-1,1\}^{T}$. Every constraint of $I_\phi$ is a sum
of such indicators on its own scope, so $I_\phi$ is linear in this lift:
$I_\phi(x)=\lta a_\phi,v(x)\rta$ with $a_\phi\geq0$. The inner body is the convex hull of the $v(x)$ and the outer one is cut out by
the inequalities $\lta a_\phi,y\rta\leq1-\eta$, one per planted copy, the two being nested
precisely because $\mathrm{opt}(I_0)<1-\eta$. Nothing below appeals to this reading, and in
particular no polytope is named: the rows here are the instances of a family, not the facets of
one fixed body, and it is the family that the argument consumes.

\paragraph{Why the witness cannot be fixed in advance.} The matrices produced by
Proposition~\ref{prop:local-witness-dnf} are entrywise strictly positive, so by
Section~\ref{sec:mixed-depth} a usable witness must be signed; and a signed witness fixed in
advance would have to defeat every admissible rank-one matrix at once.

The argument below does not fix one. It supplies instead a whole family of tests, one for each
$m$-element set $S\subseteq[N]$ --- each useless on its own, certifying nothing against a general
nonnegative function and only something against functions that are nearly juntas on $S$ --- and
a counting principle, Lemma~\ref{lem:clrs-restriction}, guaranteeing that any family of at most
$N^{d/2}$ nonnegative functions of mean one and bounded sup-norm has a \emph{single} $S$ on whose coordinates all of their
marginals are nearly juntas. The assumed factorization then selects the test that defeats it.
Where the lift of Section~\ref{sec:selector-lift} located a copy by averaging over selector
assignments, here it is this counting principle that does so.

\subsubsection{The planted family and its rank}\label{sec:planting}

Normalized Max-$k$-CSP instances, $\mathrm{opt}$, the Fourier notation, the degree-$d$ local
expectation functionals, the $d$-juntas and $\SA_d$ are those of
Section~\ref{sec:selector-lift}, and are used here without further comment. One object is new.
For an injection $\phi:[m]\hookrightarrow[N]$, the \emph{planted} copy $I_\phi$ of an instance
$I$ on $m$ variables is the instance on $N$ variables obtained by renaming the variable $u$ as
$\phi(u)$ and leaving the other $N-m$ variables unconstrained, so that
$I_\phi(x) = I(x_{\phi(1)},\dots,x_{\phi(m)})$ for $x \in \{-1,1\}^{N}$, and
$\mathrm{opt}(I_\phi)=\mathrm{opt}(I)$. Finally we put
\begin{equation}\label{eq:delta-def}
   \Delta(N,m,d)\;:=\;\binom{m}{\leq d}\left(4d\sqrt{m\log_2N}\;N^{-1/4}+N^{-d/2}\right).
\end{equation}

\begin{prop}[Quantitative planting bound]\label{prop:quantitative-planted-csp-rank}
Let $I_0$ be a normalized Boolean Max-$k$-CSP instance on $m$ variables, where
$3\leq m$ and $k \leq d \leq m$, such that
\[
   \mathrm{opt}(I_0)\le s<1,
   \qquad \SA_d(I_0)=1 .
\]
For $N\ge4m$ let $\Phi$ be the set of injections $[m]\hookrightarrow[N]$, and for
$\phi \in \Phi$ let $I_\phi$ be the copy of $I_0$ planted according to $\phi$. For
$0<\eta<1-s$ define the nonnegative matrix
\[
   A^{\eta}_{\phi,x}\;=\;1-\eta-I_\phi(x),
   \qquad \phi\in\Phi,\quad x\in\{-1,1\}^{N}.
\]
If $\eta>\Delta(N,m,d)$, with $\Delta$ as in \eqref{eq:delta-def}, then
$\rkp(A^{\eta})>N^{d/2}$.
\end{prop}

The mechanism is this. Suppose $A^{\eta}$ had a nonnegative factorization
$A^{\eta}_{\phi,x}=\sum_{i\leq R}\lambda_i(\phi)q_i(x)$ with $R\leq N^{d/2}$, each column factor
normalized to mean one and its mean absorbed into $\lambda_i$, so that the $q_i$ are densities.
The proof localizes
its factors to a set of $m$ coordinates, evaluates the identity at the row $\phi$ that plants
$I_0$ on exactly those coordinates, and applies the functional witnessing $\SA_d(I_0)=1$. The left side is then the
function $1-\eta-I_0$ of those $m$ coordinates, on which the functional returns
$(1-\eta)-\widetilde{\mathbb E}I_0 = -\eta$: the failure of the degree-$d$ Sherali--Adams
relaxation to certify $I_0\leq1-\eta$. On the right it can be made only slightly negative, and
the two terms of $\Delta(N,m,d)$ are the two reasons why.

The functional is nonnegative on $d$-juntas, so a factor whose marginal on those $m$ coordinates
is a $d$-junta costs nothing at all, and the whole question is how far the factors are from that.
Call a factor \emph{flat} if $\|q_i\|_\infty\leq N^{d}$. The lemma we import from
\cite{ChanLeeRaghavendraSteurer16} produces a single $m$-set on which the marginal of every flat
factor splits as a $d$-junta plus an error whose Fourier coefficients of degree at most $d$ are
all bounded by $4d\sqrt{m\log_2N}\,N^{-1/4}$. Its higher coefficients are left unbounded, and
need not be bounded: the functional reaches the error only through its degree-$d$ truncation,
which is a combination of the $\binom{m}{\leq d}$ monomials of degree at most $d$ with exactly the
coefficients the lemma controls. On each of those monomials the functional has value at most one
in absolute value by \eqref{eq:lef-moment}, so the error costs at most $\binom{m}{\leq d}$ times that bound;
and averaging the identity over $x$ gives
$\sum_i\lambda_i(\phi) = \mathbb E_xA^{\eta}_{\phi,x}\leq1$, the $q_i$ having mean one and
$A^{\eta}$ being at most one. So this is the first term of $\Delta(N,m,d)$.

The lemma applies only to flat factors, and the rest are held down by their coefficients instead.
Each summand lies pointwise below $A^{\eta}\leq1$, so $\lambda_i(\phi)\|q_i\|_\infty\leq1$ and a
factor failing flatness has $\lambda_i(\phi)<N^{-d}$. Each factor enters the identity multiplied by its
coefficient, so by linearity such a factor contributes $\lambda_i(\phi)$ times the functional's
value on its marginal, hence at most $N^{-d}$ times that value; and $\binom{m}{\leq d}$ bounds it for every
factor. It comes from the same expansion: what the functional is applied to is the factor's marginal on the chosen $m$-set,
which is again a density; a density $q$ has
$|\widehat q(\alpha)| = |\mathbb E[q\,x^{\alpha}]| \leq \mathbb Eq = 1$ for every $\alpha$; and
only the coefficients of degree at most $d$ reach the functional, of which there are $\binom{m}{\leq d}$.
There are at most $N^{d/2}$ factors in all, so together these cost at most
$\binom{m}{\leq d}\,N^{d/2}N^{-d}$ --- the second term. The right hand side is therefore at least
$-\Delta(N,m,d)$, while the left is $-\eta$; so $\eta>\Delta(N,m,d)$ contradicts the assumed
factorization, and none with $R\leq N^{d/2}$ exists.

The imported lemma is the only place where the ambient dimension $N$ enters, and its $N^{-1/4}$,
visible in \eqref{eq:delta-def}, is what fixes the ceiling $1/4$ in
Theorem~\ref{thm:explicit-csp-dnf-small-shift}. Appendix~\ref{sec:planting-proof} states the
lemma and gives the proof.

\subsubsection{From rank bounds to hard DNFs}\label{sec:csp-interface}\label{sec:csp-theorem}

It remains to convert the rank bound into a statement about DNFs. That conversion is the only
step of Section~\ref{sec:csp-dnf} in which the planting plays no part: it is the encoding of
Proposition~\ref{prop:local-witness-dnf} applied to the planted family, and nothing more.
Proposition~\ref{prop:local-witness-dnf} is also the only thing shared with
Proposition~\ref{prop:recipe-polytope}; the interface below names no polytope, and its row
structure is demanded by the route rather than incidental to it.

\begin{prop}[Recipe II: planted Max-$k$-CSP instances]\label{prop:recipe-csp}
Let $I_0$ be a normalized Boolean Max-$k$-CSP instance on $m$ variables with $L$ constraints and
$\mathrm{opt}(I_0)<1$, and let $N\ge4m$. There is a tautological DNF $A_{m,N}(Z,X)$ in
$K:=(m+1)N$ variables, constructible from $I_0$ in time polynomial in its output size with at most
\begin{equation}\label{eq:recipe-csp-size}
   m+m\binom N2+\binom m2N+2^{k}LN^{k}\;=\;O\bigl(LN^{k}+mN^{2}\bigr)
\end{equation}
terms and maximum term width $N$, every witness term having width $2k$, whose valid slice in the
$Z\mid X$ flattening is the violated constraint count
\[
   V_{\phi,x}\;=\;\#\{\text{constraints of }I_\phi\text{ violated by }x\}
   \;=\;L\bigl(1-I_\phi(x)\bigr),
\]
the rows ranging over the injections $\phi:[m]\hookrightarrow[N]$ and the columns over all
assignments $x\in\{-1,1\}^{N}$. If moreover $\SA_d(I_0)=1$ for some $k\le d\le m$ and $\epsilon$
satisfies
\begin{equation}\label{eq:recipe-csp-condition}
   \Delta(N,m,d)\;<\;\frac{\epsilon}{L}\;<\;1-\mathrm{opt}(I_0),
\end{equation}
then $\rkp(V-\epsilon\cdot J)>N^{d/2}$, and every monotone formula
of mixed-product depth at most $t$ computing $\chi(A_{m,N})-\epsilon\cdot\chi(1)$ has more than
$N^{d/2^{t+1}}$ leaves.
\end{prop}

\begin{proof}
Introduce Boolean variables $z_{u,a}$ for $u\in[m]$, $a\in[N]$, and $x_a$ for $a\in[N]$; a valid $Z$
sets exactly one $z_{u,a}=1$ for each $u$, these locations being distinct, and so encodes an
injection $\phi:[m]\hookrightarrow[N]$ by $\phi(u) := a$. The DNF $A_{m,N}$ is:

\begin{itemize}
\item \emph{Invalid-injection terms.} One term $\bigwedge_{a=1}^{N}\neg z_{u,a}$ for every $u$
(``$u$ has no location''), one term $z_{u,a}\wedge z_{u,b}$ for every $u$ and $a<b$ (``$u$ has
two locations''), and one term $z_{u,a}\wedge z_{v,a}$ for every $u<v$ and every $a$ (``two base
variables collide'').
\item \emph{Violated-constraint terms.} For every constraint of $I_0$, on base variables
$u_1,\dots,u_k$ say, every tuple of distinct locations $a_1,\dots,a_k\in[N]$, and every one of
the at most $2^{k}$ local assignments violating that constraint, the term fixing
$z_{u_1,a_1}=\dots=z_{u_k,a_k}=1$ together with the corresponding literals of
$x_{a_1},\dots,x_{a_k}$.
\end{itemize}

This is an instance of Proposition~\ref{prop:local-witness-dnf} with valid rows $R$ the $Z$
encoding injections, valid columns $C$ all of $\{0,1\}^{N}$, and witness conjunctions the
violated-constraint terms: on a valid pair $(\phi,x)$ all invalidity terms vanish, and for each
constraint of $I_\phi$ violated by $x$ exactly one of its local terms is satisfied, while no
local term of a constraint satisfied by $x$ is. So $A_{m,N}$ is a tautology as soon as every planted copy is
unsatisfiable, which holds because $\mathrm{opt}(I_\phi)=\mathrm{opt}(I_0)<1$, the valid slice is
$V$ as displayed, and the term count and widths are as stated.

Assume now \eqref{eq:recipe-csp-condition} and put $\eta:=\epsilon/L$. Entrywise
\begin{equation}\label{eq:recipe-validslice}
   \bigl(V-\epsilon\cdot J\bigr)_{\phi,x}\;=\;L\bigl(1-\eta-I_\phi(x)\bigr),
\end{equation}
so $V-\epsilon\cdot J$ is a positive multiple of the matrix $A^{\eta}$ of
Proposition~\ref{prop:quantitative-planted-csp-rank}: the right hand inequality of
\eqref{eq:recipe-csp-condition} supplies its hypothesis
$0<\eta<1-s$ with $s=\mathrm{opt}(I_0)$, and the left hand inequality is its hypothesis
$\eta>\Delta(N,m,d)$. Hence $\rkp(V-\epsilon\cdot J)>N^{d/2}$, positive scaling not affecting
$\rkp$. The leaf bound is Corollary~\ref{cor:local-witness-mixed-depth}.
\end{proof}

For the base instance we reuse Tseitin from Corollary~\ref{cor:exp-small-shift}.

\begin{theorem}[Explicit inverse-polynomial-shift DNF]\label{thm:explicit-csp-dnf-small-shift}
For every fixed $0<\gamma_0<1/4$, there is an explicit infinite family of tautological DNFs
$A_K$ on $K$ variables, with $K^{O(1)}$ terms and constructible in deterministic polynomial
time in the output size, such that for all sufficiently large members of the family and every
\[
   K^{-\gamma_0}\;\le\;\epsilon\;<\;1,
\]
the flattening of $\chi(A_K)-\epsilon\cdot\chi(1)$ has nonnegative rank
\[
   K^{\Omega(\log K)} .
\]
Consequently, by Theorem~\ref{thm:mixed-depth-flattening}, every
monotone formula of mixed-product depth at most $t$ computing this $\AMC$ target has at least
\[
   \exp\!\left(\Omega\!\left(\frac{(\log K)^2}{2^{t}}\right)\right)
\]
leaves; in particular the bound is superpolynomial for every constant $t$.
The family and the implicit constants may depend on $\gamma_0$.
\end{theorem}

\begin{proof}
Fix $0<\gamma_0<1/4$. We instantiate Proposition~\ref{prop:recipe-csp} with the Tseitin parity constraints on the
explicit connected $d_G$-regular expanders of Corollary~\ref{cor:exp-small-shift}, with the same
odd charges. Let $I_0$ be the normalized instance, with $n:=|V(G)|$,
$m:=|E(G)|=d_Gn/2$, $L:=n$ and arity $k:=d_G=O(1)$. Then
\[
   \mathrm{opt}(I_0)=1-\frac1n
   \qquad\text{and}\qquad
   \SA_d(I_0)=1
\]
for $d=\lfloor\alpha m\rfloor$ and a constant $\alpha>0$. The first identity follows from
Lemma~\ref{lem:strata}: every assignment has at least one defect, and single-defect assignments
exist because $G$ is connected. The second is the Sherali--Adams gap established in the proof
of Corollary~\ref{cor:exp-small-shift}.

Fix a constant $C>4/(1-4\gamma_0)$, set $N=\lceil2^{Cm}\rceil$, and let $A_K:=A_{m,N}$ be the DNF supplied by
Proposition~\ref{prop:recipe-csp}, on $K=(m+1)N$ variables. By \eqref{eq:recipe-csp-size} with
$k=d_G$ it has $O(LN^{d_G}+mN^{2})=K^{O(1)}$ terms.

It remains to verify \eqref{eq:recipe-csp-condition} throughout the stated range of $\epsilon$.
Its right hand inequality is exactly $\epsilon<1$, since $\eta=\epsilon/L=\epsilon/n$ and
$1-\mathrm{opt}(I_0)=1/n$. For the left hand one, $\binom{m}{\leq d}\le2^{m}$ and so
\[
   \Delta(N,m,d)\;\le\;2^{m}\,\mathrm{poly}(m,\log N)\,N^{-1/4}\;+\;2^{m}N^{-d/2}.
\]
Here $\log_2N=Cm+O(1)$, so the first term is
$\mathrm{poly}(m)\,2^{m(1-C/4)}$ and the second is $2^{m-C\alpha m^{2}/2}$; also
$\log K=\log N+O(\log m)=(1+o(1))\log N$, so
$\epsilon/L\geq K^{-\gamma_0}/L=2^{-C\gamma_0m(1+o(1))}$. Since $C(1/4-\gamma_0)>1$, the
inequality $\epsilon/L>\Delta(N,m,d)$ holds uniformly for $\epsilon\ge K^{-\gamma_0}$ and all
sufficiently large members of the family. Proposition~\ref{prop:recipe-csp} therefore gives
$\rkp(V-\epsilon\cdot J)>N^{d/2}$, and $\rkp$ is monotone under submatrices.

Finally, $\log N=\Theta(m)$ gives $\log K=\Theta(m)$, so $d=\Theta(m)=\Theta(\log K)$ and
$\log N=\Theta(\log K)$; hence
\[
   \log\rkp \;\geq\; (d/2)\log N \;=\; \Theta\bigl((\log K)^{2}\bigr),
\]
which is the stated rank $K^{\Omega(\log K)}$. After pruning zero gates,
Lemma~\ref{lem:autosml-depth} permits applying Theorem~\ref{thm:mixed-depth-flattening} to any
monotone formula for the target. Taking $R=N^{d/2}$ in its leaf bound $R^{1/2^t}$ gives the
claimed formula lower bound, since $\log R=\Theta((\log K)^2)$.
\end{proof}

\section{Conclusion and open problems}\label{sec:conclusion}

Our lower bounds for arbitrary rank-one summands hold over prescribed ranges of shifts and
for restricted classes of certificates, either depth three or bounded mixed-product depth.
These are independent restrictions. For the subcube fragment, the Sherali--Adams equivalence
also gives lower bounds uniform in $\epsilon$.

\paragraph{1. A bound uniform in \texorpdfstring{$\epsilon$}{epsilon} at depth three.} An unconditional $\SHAMC$ lower
bound asks for an explicit tautological DNF family $A_N$ with
\[
   \min_{\epsilon>0}\rkt\bigl(\chi(A_N)-\epsilon\cdot\chi(1)\bigr)\;=\;N^{\omega(1)} ,
\]
the minimum being over all admissible shifts at once. All three routes of this paper stop short
of it. The flattening recipes of Section~\ref{sec:polytopes} need $\epsilon\geq K^{-\gamma}$ with
$\gamma<1/4$; the selector lift of Section~\ref{sec:selector-lift} reaches
$\epsilon\geq2^{-cN}$ for a constant $c>0$ and says nothing at $\epsilon=2^{-N^{2}}$; and the Tseitin bounds of
Section~\ref{sec:tseitin-lb} cannot be pushed to every shift, since
Theorem~\ref{thm:tseitin-tiny-shift} gives $\TS_G$ proofs of size $O(n^{3})$ at
$\epsilon=2^{-\Theta(n\log n)}$. The lifted family $\mathrm{Ind}_q(\TS_G)$ and the planted family $A_{m,N}$ of
Section~\ref{sec:csp-dnf} are candidates; Theorem~\ref{thm:tseitin-tiny-shift} does not apply
to either.

\paragraph{2. Unbounded fan-in depth four.} No superpolynomial lower bound is available for
$\AMC^{\epsilon}$ proofs given by a monotone formula of ordinary depth four with unbounded
fan-in, at any $\epsilon$, and the flattening bound does not apply to one, since bracketing such
a formula into binary gates can make its mixed-product depth grow with $n$
(Figure~\ref{fig:bracketing}). Even a bound at depth four and constant $\epsilon$ would be new.

\paragraph{3. The Tseitin threshold.} For a constant-degree expander,
Theorem~\ref{thm:tseitin-lb-arbitrary} gives $\rkt(T^{\epsilon})\geq2^{\Omega(\epsilon n)}$
when $\epsilon=\omega(\log n/n)$, hence an exponential bound at constant shift and a
superpolynomial bound throughout that range. The rank is $O(n^{3})$ at
$\epsilon=2^{-\Theta(n\log n)}$ by Theorem~\ref{thm:tseitin-tiny-shift}. Where
between the two the transition occurs is open, as is whether the quadratic product rank of
Lemma~\ref{lem:weak-parity} is optimal for a weakly biased parity.

\paragraph{4. Small shifts through the planting route.} The ceiling $\gamma<1/4$ of
Theorem~\ref{thm:explicit-csp-dnf-small-shift} comes from the sampling scale in
Lemma~\ref{lem:clrs-restriction}, not from the encoding; a sharper planting estimate would
enlarge the admissible range by the same exponent. Reaching $\epsilon=K^{-\omega(1)}$ by this
route would need a Sherali--Adams certificate lower bound at superpolynomially small gap.

\paragraph{5. Flattenings of the lifted families.} The selector lift bounds $\rkt$ but gives
nothing for $\rkp$ of a flattening, and whether $\mathrm{Ind}_q(\TS_G)$ is easy at
mixed-product depth zero --- as $\TS_G$ itself is, by Theorem~\ref{thm:mu0-upper} --- we do not
know. An answer either way would say which of the two features of Section~\ref{sec:csp-dnf},
the reach in $\epsilon$ or the applicability above depth three, is intrinsic to the shared-pool
planting.

\paragraph{6. Which systems does \texorpdfstring{$\AMC$}{AMC} simulate?} The simulation established here already
holds at depth three, where the bounded-bias fragment is equivalent to Sherali--Adams up to clause width
(Corollary~\ref{cor:amcb-sa-equivalent}). Whether $\AMC$, at depth three or at any depth,
p-simulates Cutting Planes, sums-of-squares or Polynomial Calculus is open, as is whether any
standard system p-simulates $\AMC$.

\paragraph{7. The bounded-bias fragment.} The equivalence with Sherali--Adams gives lower
bounds for $\AMCb$ at every shift (Corollary~\ref{cor:all-eps-amcb}). At
$\epsilon=2^{-\Theta(n\log n)}$, $\SHAMC$ has exponentially smaller proofs of Tseitin
(Corollary~\ref{cor:tiny-shift-separation}). It remains open whether the fragments separate at
constant $\epsilon$.

\appendix

\section{The \texorpdfstring{$\epsilon$}{epsilon} threshold in the Braun--Pokutta bound}\label{sec:threshold}

This appendix justifies the claim of Section~\ref{sec:pm-lb}: the bound of Braun and Pokutta
\cite{BraunPokutta15} makes $\rkp(M - \epsilon\cdot J)$ at least $2^{n^{\Omega(1)}}$ for every
$\epsilon \geq c_{0}\log\log n/\log n$ with $c_{0}>8$ fixed. We establish this sufficient
range without claiming that the constant $8$ is optimal. They state their per-chunk constant and the auxiliary constant
$B$ only as positive constants depending on $k$ and on the perturbation. Their perturbation
parameter is called $\epsilon$ there and is our $\delta = 1-\epsilon$ below, and their goodness
threshold, called $\delta$ there, is our $t$. Since the
threshold computation rests on uniformity in $k$, we unpack their proof
\cite[Proposition~3.2, Sections~3.3 and~3.4]{BraunPokutta15} and make the dependence on the
chunk parameter $k$ and the perturbation $\delta$ explicit. Throughout, $M$ is the matrix
indexed by perfect matchings (columns) and odd cuts $\delta(U)$ (rows) of $K_{2n}$, with entry
$|\delta(U) \cap M|$, and $J$ is the all-ones matrix; logarithms are base~$2$.

\paragraph{Reduction to the Braun--Pokutta slack matrix.}
Write $S^{+\delta}$ for the perturbed slack matrix of \cite[Section~3]{BraunPokutta15}, whose
entry for a matching $M$ and an odd cut $U$ is
\[
   S^{+\delta}_{M,U} \;=\; \frac{|\delta(U) \cap M| - 1 + \delta}{2}, \qquad 0 \le \delta < 1 .
\]
Since $M - \epsilon\cdot J$ has entries $|\delta(U)\cap M| - \epsilon = 2\,S^{+(1-\epsilon)}_{M,U}$,
we have $M - \epsilon\cdot J = 2\,(S^{+(1-\epsilon)})^{\mathsf T}$, so the relevant Braun--Pokutta parameter is
$\delta = 1-\epsilon$ and $\rkp(M - \epsilon\cdot J) = \rkp\bigl(S^{+(1-\epsilon)}\bigr)$ (the
factor $2$ and transposition are immaterial for nonnegative rank).

\paragraph{The bound and its chunk structure.}
Their key estimate is \cite[Proposition~3.2]{BraunPokutta15}: for an odd integer $k \ge 7$ and
$0 \le \delta < 1 - 4/k$, partition the vertices of $K_{2n}$ minus a reserved $3$-matching $H$
($6$ vertices) into $m$ chunks $T_1, \dots, T_m$ of $2(k-3)$ vertices each, plus a remainder
chunk $L$ of fewer than $2(k-3)$ vertices, so that
\[
   m \;=\; \frac{2n - 6 - |L|}{2(k-3)} \;\ge\; \frac{n-3}{k-3} - 1 \;=\; \Theta(n/k),
\]
and bound the common information $C[\,\cdot\,]$ of $S^{+\delta}$
\cite[Definition~2.9]{BraunPokutta15} from below chunk by chunk:
\[
   \log \rkp\bigl(S^{+\delta}\bigr) \;\ge\; C\bigl[M;U \mid T,N,H,E\bigr] \;\ge\; c_{k,\delta}\cdot m ,
\]
using $\log \rkp \ge C[\,\cdot\,]$ \cite[Lemma~2.11]{BraunPokutta15}. Under $\delta = 1-\epsilon$
the admissibility constraint $\delta < 1 - 4/k$ reads
\[
   \epsilon \;>\; \frac{4}{k} .
\]

\paragraph{The per-chunk constant, explicitly.}
Unwrapping the proof of \cite[Proposition~3.2]{BraunPokutta15} in
\cite[Section~3.3]{BraunPokutta15} gives the following admissible choice of constant:
\begin{equation}\label{eq:ckdelta-exact}
   c_{k,\delta} \;=\; B_{k,\delta}\,(2\log e)\,t^{2}\,
   \min\Bigl\{\,
      \frac{\Pr[N=0 \mid E]\;\alpha^{2}}{\Pr[U = C(H) \mid E_{0}]},\;
      \frac{\Pr[N=1 \mid E]/4}{\Pr[\delta(C)\cap M = H \mid E_{0}]}
   \,\Bigr\},
   \qquad \alpha \;=\; \frac{1}{(2k-7)!!},
\end{equation}
where $\alpha$ is the reciprocal of the number of perfect matchings on $2k - 6$ vertices, $0<t<1$
is a free parameter of their proof (the goodness threshold of
\cite[Definition~3.3]{BraunPokutta15}, denoted $\delta$ there), the squares
$(t\alpha)^{2}$ and $(t/2)^{2}$ are the gaps fed into Pinsker's inequality
\cite[Lemma~2.2]{BraunPokutta15}, and $C$, $D$, $H$, $N$, $E_{0}$, $E$ are the objects of the
local (single-chunk) case of their proof on $K_{2k}$: $C \sqcup D$ is a bipartition of the vertex
set into two $k$-sets, $H$ a uniformly random $3$-matching between $C$ and $D$, $N$ a fair coin,
and $E_{0} = \{U \in \{C, C(H)\},\ H \subseteq M\}$,
$E = E_{0} \wedge \{U = C(H) \text{ if } N = 0;\ \delta(C)\cap M = H \text{ if } N = 1\}$.
The constant $B_{k,\delta}$, a lower bound on the probability of the pair $(M, U)$ being
``bad'' for the seed, is computed in \cite[Section~3.4]{BraunPokutta15}:
\begin{equation}\label{eq:Bkdelta-exact}
   B_{k,\delta} \;=\;
   \Bigl(\,2 - (3k-4)(1-\beta) - \frac{6(k-4)}{k(k-1)}\,\Bigr)
   \frac{\gamma A}{k\,\beta\,\Pr[E_{0}]},
\end{equation}
with
\[
   \beta \;=\; \Bigl(\frac{1-t}{1+t}\Bigr)^{2}, \qquad
   A \;=\; \bigl((k-4)!!\bigr)^{2}, \qquad
   \gamma \;=\; \frac{1}{(k/2 - 1 + \delta)\,(2k-1)!!\,2^{2k-1}},
\]
$A$ counting the perfect matchings of $K_{2k}$ that cross the cut $(C, D)$ exactly along a fixed
$3$-matching, and $\gamma$ being the normalization of the distribution induced by $S^{+\delta}$
on $K_{2k}$.

These expressions evaluate as follows. Every perfect matching $M' \supseteq H$ meets
$\delta(C(H))$ in exactly the three edges of $H$, and meets $\delta(C)$ in between $3$ and $k$
edges; the matchings with $\delta(C) \cap M' = H$ are exactly the $A$ matchings counted above.
Hence
\[
   \Pr[E \mid N = 0] \;=\; (2+\delta)\,\gamma\,(2k-7)!!, \qquad
   \Pr[E \mid N = 1] \;=\; 2\,(2+\delta)\,\gamma\,\bigl((k-4)!!\bigr)^{2},
\]
\[
   2\,(2+\delta)\,\gamma\,(2k-7)!! \;\le\; \Pr[E_{0}] \;\le\; (k+1+2\delta)\,\gamma\,(2k-7)!! .
\]
Consequently $\Pr[N=0\mid E] \ge 1/2$, the second branch of the minimum
in~\eqref{eq:ckdelta-exact} is $\ge 1/4 \gg \alpha^{2}$, so the minimum is always the first
branch, and $\gamma$ cancels against $\Pr[E_{0}]$. The parenthesized factor
in~\eqref{eq:Bkdelta-exact} never exceeds $2$ and is positive only when $1 - \beta = O(1/k)$,
i.e.\ $t = O(1/k)$; the choice $t = 1/(12k)$ makes it $\ge 1/7$ (and $\beta \ge 20/21$) for
every odd $k \ge 7$ and every $\delta \in [0, 1-4/k]$. Fixing this choice of $t$, uniformly in
$\delta \in [0, 1-4/k]$,
\begin{equation}\label{eq:ckdelta-twosided}
   \frac{1}{1500\,k^{4}}\cdot\frac{\bigl((k-4)!!\bigr)^{2}}{\bigl((2k-7)!!\bigr)^{3}}
   \;\le\; c_{k,\delta} \;\le\;
   \frac{\bigl((k-4)!!\bigr)^{2}}{\bigl((2k-7)!!\bigr)^{3}},
   \qquad\text{i.e.}\qquad
   c_{k,\delta} \;=\; 2^{-2k\log k \;+\; O(k)} .
\end{equation}
Two features matter for the threshold computation: the dominant term is the squared gap
$\alpha^{2} = \bigl((2k-7)!!\bigr)^{-2}$ contributed by Pinsker's inequality, and the
perturbation $\delta$ enters this choice of $c_{k,\delta}$ only through the sufficient restriction
$\delta < 1 - 4/k$ and through polynomial-in-$k$ factors (the term $k/2 - 1 + \delta$ in
$\gamma$, and $\Pr[E_{0}]$).

\paragraph{Optimizing the trade-off.}
For $k=o(n)$, combining~\eqref{eq:ckdelta-twosided} with $m = \Theta(n/k)$ gives
\begin{equation}\label{eq:bp-optimised}
   \log \rkp(M - \epsilon\cdot J) \;\ge\; c_{k,\,1-\epsilon}\cdot m
   \;=\; \frac{n}{k}\,2^{-2k\log k + O(k)},
   \qquad k > 4/\epsilon .
\end{equation}
Choose the smallest odd $k\geq7$ with $k>4/\epsilon$, so
$k = \Theta(\max\{1, 1/\epsilon\})$. For constant $\epsilon \in (0,1)$ this gives
$\log\rkp(M - \epsilon\cdot J) = \Omega(n)$. In general put
$\epsilon = c_{0}\log\log n/\log n$ with $c_{0}$ constant, and let $k$ be the smallest odd
integer above $4/\epsilon$; then $k = n^{o(1)}$, $\log k = (1+o(1))\log\log n$ and
$2k\log k = (8/c_{0}+o(1))\log n$, so that \eqref{eq:bp-optimised} reads
\begin{equation}\label{eq:bp-boundary}
   \log \rkp(M - \epsilon\cdot J) \;\ge\; n^{\,1-8/c_{0}+o(1)} .
\end{equation}
For $c_{0} > 8$ the exponent is positive and the conclusion is
$\rkp(M - \epsilon\cdot J)\geq2^{n^{\Omega(1)}}$, far beyond superpolynomial; for $c_{0} < 8$ the
right hand side is $o(1)$ and nothing is asserted. Superpolynomiality itself needs only
$c_{k,\,1-\epsilon}\cdot m = \omega(\log n)$, so the argument delivers it throughout
$\epsilon\geq c_{0}\log\log n/\log n$ for every fixed $c_{0} > 8$, and a fortiori whenever
$\epsilon = \omega(\log\log n/\log n)$.

\paragraph{Scope of the constant.}
The restriction $\delta<1-4/k$ is sufficient for the estimate used above. Before substituting
this restriction, \cite[Section~3.4]{BraunPokutta15} gives the badness lower bound
\[
   \left(2+\delta-
      \frac{6(k-1+\delta)/(k(k-1))+3\delta}{\beta}\right)
   \frac{\gamma A}{\Pr[E_0]}.
\]
At $\delta=1-\epsilon$, its parenthesized factor tends, as $\beta\to1$, to
$2\epsilon-6(k-\epsilon)/(k(k-1))$. This is positive whenever
$\epsilon>3k/(k(k-1)+3)$, which permits smaller chunks than $k>4/\epsilon$.
Thus the sufficient constant $8$ above is not an optimality claim for the full
Braun--Pokutta argument. We do not optimize the sharper estimate here.

\section{Proof of the quantitative planting bound}\label{sec:planting-proof}

We prove Proposition~\ref{prop:quantitative-planted-csp-rank}, in the notation fixed in
Sections~\ref{sec:selector-lift} and~\ref{sec:planting}. The one estimate we import is the
following.

\begin{lem}[Random restriction, {\cite[Lem.~3.5]{ChanLeeRaghavendraSteurer16}}, in the form used
here]\label{lem:clrs-restriction}
Let $3\leq m$, let $d\leq m$, let $N\geq4m$, and put
$\gamma:=4d\sqrt{m\log_2N}\,N^{-1/4}$. Let $\mathcal Q$ be a family of at most $N^{d/2}$
densities $q:\{-1,1\}^{N}\to\mathbb R_{\geq0}$ with $\|q\|_\infty\leq N^{d}$. Then there is a single $m$-element set $S\subseteq[N]$ such that
the marginal on $S$ of every $q\in\mathcal Q$ can be written as $c+e$, where $c$ is a $d$-junta
on $\{-1,1\}^{S}$ and the error $e:\{-1,1\}^{S}\to\mathbb R$, which is of no prescribed sign,
has all of its low-degree Fourier coefficients small:
\[
   |\widehat e(\alpha)|\;\leq\;\gamma\qquad\text{for every }\alpha\subseteq S
   \text{ with }|\alpha|\leq d .
\]
\end{lem}

Here a \emph{density} means a nonnegative function of mean one, and the \emph{marginal} on $S$ of
a function $q$ on $\{-1,1\}^{N}$ is
$(\pi_Sq)(y):=\mathbb E[q(x)\mid x|_S=y]$, which is again a density and which fixes every
function already depending only on the coordinates in $S$. The hypothesis on an individual
density is stated in \cite{ChanLeeRaghavendraSteurer16} as a lower bound on the entropy of the
associated measure; the sup-norm bound above is the sufficient condition through which it is
applied, since $\|q\|_\infty\leq N^{d}$ forces that entropy to be at least $N-d\log_2N$. The
cardinality
bound $N^{d/2}$ is a hypothesis of the lemma and not an artefact of the application, and within
that range the error $\gamma$ does not depend on the size of the family at all.

\begin{proof}[Proof of Proposition~\ref{prop:quantitative-planted-csp-rank}]
Write $\gamma$ as in Lemma~\ref{lem:clrs-restriction}, take the uniform measure
$\mu$ on $\mathcal X:=\{-1,1\}^{N}$, and abbreviate $A:=A^{\eta}$, $A_\phi:=A(\phi,\cdot)$. The
matrix is nonnegative because $\mathrm{opt}(I_\phi)=\mathrm{opt}(I_0)\le s<1-\eta$, and
$I_\phi\geq0$ with $\eta>0$ gives $A\leq1$ pointwise, hence also $\int A_\phi\,d\mu\leq1$.
As $\SA_d(I_0)$ is a maximum over degree-$d$ local expectation functionals, fix one,
$\widetilde{\mathbb E}$, attaining it, so $\widetilde{\mathbb E}I_0=1$; for a function $g$ on an
$m$-element coordinate set put
\[
   \Lambda(g)\;:=\;\widetilde{\mathbb E}\bigl(g^{\leq d}\bigr) ,
\]
a linear functional defined on \emph{all} functions there, the truncation being what extends
$\widetilde{\mathbb E}$ past its degree bound. Fix for each $m$-set $S\subseteq[N]$ a bijection
$[m]\to S$ and use it to transport $\widetilde{\mathbb E}$, and hence $\Lambda$, to functions on
$\{-1,1\}^{S}$.

Suppose, for contradiction, that $A$ has a nonnegative factorization
\begin{equation}\label{eq:loc-fact}
   A(\phi,x)\;=\;\sum_{i\leq R}\lambda_i(\phi)\,q_i(x),
   \qquad \lambda_i:\Phi\to\mathbb R_{\geq0},\quad q_i:\mathcal X\to\mathbb R_{\geq0},
\end{equation}
with $R\leq N^{d/2}$. Discard the factors with $\lambda_i\equiv0$ or $q_i\equiv0$, which changes
nothing and only decreases $R$; as $\mu$ has full support every surviving $q_i$ has
$\int q_i\,d\mu>0$, so absorbing that integral into $\lambda_i$ makes each $q_i$ a density.
Integrating \eqref{eq:loc-fact} in $x$ and using the pointwise bound on $A$,
\begin{equation}\label{eq:loc-mass}
   \sum_{i\leq R}\lambda_i(\phi)\;=\;\int A_\phi\,d\mu\;\leq\;1 \qquad\text{for every }\phi ,
\end{equation}
and since all summands of \eqref{eq:loc-fact} are nonnegative each is bounded pointwise by $A$,
so
\begin{equation}\label{eq:loc-spike}
   \lambda_i(\phi)\,\|q_i\|_\infty\;=\;\|\lambda_i(\phi)q_i\|_\infty\;\leq\;\|A_\phi\|_\infty
   \;\leq\;1 \qquad\text{for every }i,\phi .
\end{equation}

\emph{Choosing the coordinate set.} Split the factors at the level $N^{d}$ into
$\mathcal G:=\{i:\|q_i\|_\infty\leq N^{d}\}$ and the \emph{spiky} ones $\mathcal B$, for which
\eqref{eq:loc-spike} gives $\lambda_i(\phi)<N^{-d}$ at every row. Now $\{q_i\}_{i\in\mathcal G}$
is a family of at most $R\leq N^{d/2}$ densities of sup-norm at most $N^{d}$, which is exactly
the hypothesis of Lemma~\ref{lem:clrs-restriction} at these $N,m,d$; the standing assumption
$N\geq4m$ is that lemma's. It supplies a single $m$-set $S\subseteq[N]$ and splittings
$\pi_Sq_i=c_i+e_i$, $i\in\mathcal G$, with $c_i$ a $d$-junta on $\{-1,1\}^{S}$ and
$|\widehat{e_i}(\alpha)|\leq\gamma$ for $|\alpha|\leq d$.

\emph{The hard row on that set.} Let $\phi_S\in\Phi$ be the chosen bijection $[m]\to S$,
regarded as an injection into $[N]$. Every $m$-set is the image of such a row, which is the whole
point of indexing the rows by all injections. Since $A_{\phi_S}=1-\eta-I_{\phi_S}$ depends only
on the coordinates in $S$, marginalizing \eqref{eq:loc-fact} at $\phi_S$ gives
\[
   1-\eta-I_0\;=\;\pi_SA_{\phi_S}\;=\;\sum_{i\leq R}\lambda_i(\phi_S)\,\pi_Sq_i
\]
under the identification, $\pi_S$ being linear. Apply $\Lambda$. On the left, $I_0$ is an average
of indicators of $k$-ary constraints and so has Fourier degree at most $k\leq d$; the function
$1-\eta-I_0$ therefore equals its own degree-$d$ truncation and
\[
   \Lambda(1-\eta-I_0)\;=\;(1-\eta)\widetilde{\mathbb E}1-\widetilde{\mathbb E}I_0\;=\;-\eta .
\]
On the right, split the sum at $\mathcal G$ and $\mathcal B$ and bound the three pieces.
A $d$-junta has Fourier degree at most $d$, so $\Lambda(c_i)=\widetilde{\mathbb E}(c_i)\geq0$ by
the defining nonnegativity of a local expectation functional. Expanding
$\Lambda(e_i)=\sum_{|\alpha|\leq d}\widehat{e_i}(\alpha)\,\widetilde{\mathbb E}\,x^{\alpha}$ and
using \eqref{eq:lef-moment} together with the bound from the lemma gives
$|\Lambda(e_i)|\leq \binom{m}{\leq d}\gamma$; the same expansion for a density $q$, whose Fourier
coefficients satisfy $|\widehat q(\alpha)|\leq\mathbb E|q|=1$, gives
$|\Lambda(\pi_Sq)|\leq \binom{m}{\leq d}$. Hence, using \eqref{eq:loc-mass} for the middle sum and
$\lambda_i(\phi_S)<N^{-d}$ with $|\mathcal B|\leq N^{d/2}$ for the last,
\[
\begin{aligned}
   -\eta &= \sum_{i\in\mathcal G}\lambda_i(\phi_S)\Lambda(c_i)
   \;+\;\sum_{i\in\mathcal G}\lambda_i(\phi_S)\Lambda(e_i)
   \;+\;\sum_{i\in\mathcal B}\lambda_i(\phi_S)\Lambda(\pi_Sq_i)
   \\
   &\geq 0\;-\;\binom{m}{\leq d}\gamma\;-\;N^{d/2}N^{-d}\binom{m}{\leq d} ,
\end{aligned}
\]
whose right hand side is $-\Delta(N,m,d)$ by \eqref{eq:delta-def}. So $\eta\leq\Delta(N,m,d)$,
contrary to hypothesis.
\end{proof}

\bibliographystyle{plain}
\bibliography{references}

\end{document}